\pdfoutput=1
\documentclass[10pt]{article}
\usepackage[margin=0.78in]{geometry}
\usepackage{amsmath,amssymb,amsthm,mathtools}
\usepackage{microtype}
\usepackage{graphicx,booktabs}
\usepackage{placeins}
\usepackage{floatrow}
\usepackage{enumitem}
\usepackage{comment}
\usepackage[dvipsnames]{xcolor}
\definecolor{darkblue}{rgb}{0.0,0.0,0.55}
\usepackage{hyperref}
\hypersetup{
     colorlinks = true,
     citecolor = Maroon,
     urlcolor = darkblue,
     linkcolor = darkblue
}
\usepackage[round]{natbib}
\usepackage{mathtools}
\usepackage{lmodern}
\setlist{nosep,leftmargin=1.3em}
\newcommand{\E}{\mathbb{E}}

\usepackage{xspace}

\newtheorem{theorem}{Theorem}
	\newtheorem{lemma}{Lemma}
	\newtheorem{proposition}{Proposition}

    \newtheorem{corollary}{Corollary}
	\newtheorem{assumption}{Assumption}
	\newtheorem{example}{Example}
	\newtheorem{remark}{Remark}
	\newtheorem*{example*}{Example}
\usepackage{tikz}
\usetikzlibrary{calc,arrows.meta,positioning}

\definecolor{srwink}{HTML}{1D1D1F}     %
\definecolor{srwmuted}{HTML}{7A7A80}   %
\definecolor{srwrule}{HTML}{E4E4E8}    %
\definecolor{srwcodeblue}{HTML}{2F5D8C}%
\definecolor{srwagent}{HTML}{8B1E4F}   %
\definecolor{srwbot}{HTML}{2E3440}     %
\definecolor{srwother}{HTML}{A9A9B0}   %

\tikzset{
  srw flow/.style={-{Stealth[length=1.8mm,width=1.4mm]},line width=0.6pt,draw=srwink},
  srw feed/.style={-{Stealth[length=1.5mm,width=1.1mm]},line width=0.4pt,draw=srwmuted!80},
  srw txt/.style={font=\scriptsize,text=srwink,align=center},
  srw note/.style={font=\scriptsize\itshape,text=srwmuted,align=center},
  srw lab/.style={font=\scriptsize\bfseries,text=#1},
}

\newcommand{\srwRobot}[2]{%
  \begin{scope}[shift={#1}]
    \draw[srwbot,line width={0.9pt*#2}] (0,{0.5*#2}) -- (0,{0.66*#2});
    \fill[srwbot] (0,{0.68*#2}) circle ({0.05*#2});
    \fill[srwbot,rounded corners={0.16cm*#2}] ({-0.525*#2},{-0.5*#2}) rectangle ({0.525*#2},{0.5*#2});
    \fill[white] ({-0.21*#2},{0.075*#2}) circle ({0.075*#2});
    \fill[white] ({0.21*#2},{0.075*#2}) circle ({0.075*#2});
  \end{scope}}

\newcommand{\srwDataset}[6]{%
  \begin{scope}[shift={#1}]
    \fill[white] (0,0) rectangle (#2,{#3*#4});
    \foreach \r in {#6}{\fill[#5!22] (0,{(\r-1)*#4}) rectangle (#2,{\r*#4});}
    \foreach \r in {1,...,#3}{\draw[#5!35,line width=0.25pt] (0,{\r*#4}) -- (#2,{\r*#4});}
    \draw[#5!35,line width=0.25pt] ({#2/3},0) -- ({#2/3},{#3*#4})
                                   ({2*#2/3},0) -- ({2*#2/3},{#3*#4});
    \draw[#5!80,line width=0.45pt] (0,0) rectangle (#2,{#3*#4});
  \end{scope}}

\newcommand{\srwHit}[1]{\fill[srwink] #1 circle (0.07);}
\newcommand{\srwMiss}[1]{\draw[srwother,fill=white,line width=0.5pt] #1 circle (0.065);}

\title{\textbf{Inference After Coding Agents}\thanks{W\"uthrich is also affiliated with CESifo. We thank participants at the University of Virginia's Human+Machine Intelligence seminar for helpful comments. We acknowledge the use of the frontier versions of the GPT and Claude models (Apr--Nov 2026) and Grammarly for assistance with mathematical proofs, coding, editing, and proofreading. The usual disclaimer applies. Van Dijcke gratefully acknowledges a gift from Schmidt Sciences and start-up research support from the University of Virginia. Part of this research was conducted while Van Dijcke was affiliated with the University of Michigan. This research was supported in part through computational resources and services provided by Advanced Research Computing at the University of Michigan, Ann Arbor. A replication package for this paper is available at \url{https://github.com/Davidvandijcke/inference_after_coding_agents_replication}.}}
\author{Florian Gunsilius\thanks{Department of Economics, Emory University. Email: \url{fgunsil@emory.edu}.}\quad \quad David Van Dijcke\thanks{Corresponding author. Department of Economics, University of Virginia. Email: \url{eju5az@virginia.edu}.}\quad \quad Kaspar W\"uthrich\thanks{Department of Economics, University of Michigan. Email: \url{kasparwu@umich.edu}.}}
\date{\today}

\begin{document}
\maketitle

\begin{abstract}

Empirical research is changing rapidly as AI coding agents are becoming increasingly proficient in every step of it, from writing code to manipulating data. We argue that while the potential of coding agents for science is vast, their use in hypothesis testing poses a substantial threat to classical inference, as these agents enable a supercharged form of selective inference. In particular, we argue that coding agents carrying out empirical research introduce a new type of post-selection inference that combines human-like, open-ended search over an implicit space of estimators with programmatic replayability of both the selection and the estimation process. Indeed, we find that on simulated data with zero effects, four frontier agents report significant $p$-values between 54\% and 98\% of the time when asked to produce a compelling finding. We propose a simple procedure for detecting and correcting these distortions. After an agent has completed its analysis, we reset and replay it on subsamples of the data to estimate the agent's relative probability of selecting that analysis when it looks statistically favorable. For agents with stable selection behavior, the resulting corrected test is asymptotically valid and consistent conditional on selection. Importantly, the correction is learned from the black-box agent's actual choices rather than a model of how it searches, so it remains valid even as AI technology continues to evolve rapidly. In simulations, the correction robustly lowers the high false rejection rates to near the nominal 5\%. We illustrate the method on applications from two published papers.


\end{abstract}

\newpage

\section{Introduction}

Coding agents, large language model (LLM)-based systems that can interact with a computing environment by reading files, writing and executing code, and invoking command-line tools, first emerged around 2023 \citep{lucas2023openinterpreter} before gaining widespread use in early 2025 \citep{anthropic2025claudecode} with the introduction of Claude Code. Since then, ``agentic'' AI has rapidly gained a foothold in the market, with estimated annualized spending on AI coding applications reaching \$34 billion by September 2026, up from approximately \$4.4 billion in 2025 \citep{menloventures2026enterprise}. While their initial user base was mostly concentrated among software engineers and so-called ``vibe coders'', programmers who delegate all their coding to AI agents, their applicability to tasks ranging from data analysis and scientific simulations to literature reviews and paper editing has contributed to their growing adoption in empirical science.

Indeed, empirical researchers are increasingly delegating large parts of data analysis and scientific modeling to coding agents \citep{codreanu2026ai, hermle2026AIScience, openaiecon2026codex}. In a survey of social scientists in Feb--Mar 2026, \citet{lyttelton2026agents} found that 20\% of quantitative social scientists, including 39\% of economists, regularly used coding agents. A researcher typically provides a coding agent with data and a description of the question they want to test or the design they want to implement. Like a human analyst, the agent can then inspect the data, choose variables, estimate a few models, and inspect their results before deciding what implementation is worth reporting. While this recent development has the potential to drastically speed up empirical science, it also poses a threat to valid statistical inference. Unless the precise hypothesis and method to test it are fully specified in a type of agentic pre-registration design, the fact that coding agents typically pull a rich information set into their context window before formulating their hypothesis means that classic statistical inferential guarantees, which assume the hypothesis is specified before looking at the data, will typically fail to hold.

The adoption of coding agents leads to a new ``agentic'' form of selective inference with several unique traits. Like human researchers, coding agents can search over a tremendously vast space of models. Unlike humans, however, coding agents are programs that can be controlled and replayed on counterfactual versions of the data. Moreover, the final output of a coding agent's workflow is a piece of executable code. In this paper, we exploit this unique combination of \textit{replayability of the open-ended, black-box selection process} and \textit{executability of the selected analysis} to restore valid inference.

\begin{figure}[hb!]
    \centering
    \includegraphics[width=0.9\linewidth]{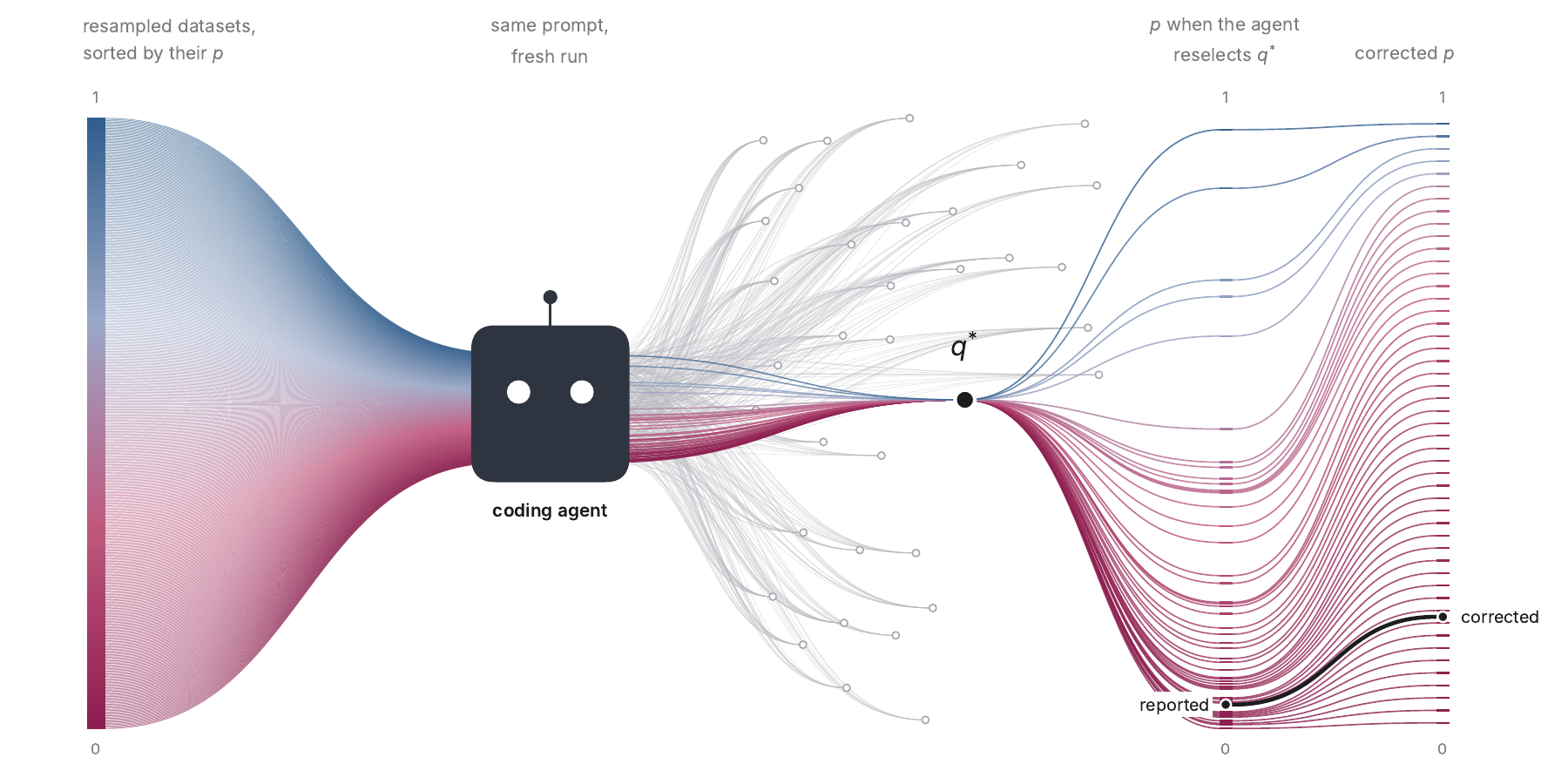}
    \caption{Correcting a coding agent's $p$-value for its own selection}
       \label{fig:selective_replay}
    \floatfoot{\textit{Note}: figure visually illustrates the idea behind our method. Under a fixed null hypothesis, exactly valid $p$-values follow a uniform distribution (left; color encodes $p$, with maroon for small values). When a coding agent (black robot figure) selectively reports results $q^*$ based on how favorable they look statistically, it distorts this distribution, piling the $p$-values near zero and causing the agent's reported $p$-values to overstate the evidence (``p when the agent reselects $q^*$''). Our method corrects these distortions (``corrected $p$'') by estimating the agent's relative probability of reselecting the same analysis $q^*$, illustrated by the many forking branches, some of which get mapped through $q^*$ by the agent.}
\end{figure}

We develop a method that detects when a coding agent chooses its analysis based on the results it sees and corrects the reported $p$-value for such data-driven selection. In particular, because coding agents are programs, we can reset and \textit{replay} them on resampled versions of the data. This offers a way around the black-box nature of agent-driven selection by empirically estimating how often agents reselect a given analysis when it looks statistically favorable on a dataset. We can estimate  this frequency by exploiting the fact that coding agents emit \textit{executable} analysis code. After an agent has selected and coded up its analysis, our replay method executes that code on subsamples of the data, sorting these subsamples by how favorable their $p$-values look compared to the agent's original $p$-value. If the agent is more likely to reselect its analysis on favorable subsamples, this offers evidence of data-driven selection by the agent. The ratio of the frequency with which the agent selects the analysis on favorable subsamples to its overall selection frequency, which we refer to as the selection multiplier $\lambda$, measures how strongly the agent's selection depends on the results. The idea is visualized in Figure \ref{fig:selective_replay}: the coding agent, by selecting an analysis $q^*$ from the space of potential analyses, distorts the uniformity characteristic of a calibrated $p$-value, leading to the $p$-values reported by the agent piling up near zero. Our method estimates this distortion by replaying the agent and corrects for it, restoring the uniformity and hence validity of the corrected $p$-values.

We show that $\lambda$ is all that is needed to 1) detect whether a given result was selected and 2) correct the agent's $p$-value and restore its statistical guarantees, building on \citet{sood2024selective}. More precisely, we establish that the resulting corrected $p$-value is asymptotically valid conditional on the agent's selection, that it loses no first-order power relative to the oracle correction where the agent's selection probabilities are known exactly, and that it inherits the consistency of the original test it corrects. We also show how to construct valid confidence bands for the corrected $p$-value, and provide conditions under which these results hold uniformly over a class of null distributions.

Our simulations and applications demonstrate the value of both detecting and correcting for coding agents' selection. When asked for ``one compelling, defensible finding'', the latest frontier coding agents, Claude Opus 5.5 and GPT-6.1-Sol (at high effort), report a significant ($p \leq 0.05$) result 54 and 88\% of the time, respectively, on simulated datasets where all true effects are zero. When given data from a seminal economics paper \citep{card1994minimum} and asked to find the most statistically compelling result, GPT-6.1-Sol is nearly 2 times more likely to report its chosen analysis when it looks statistically favorable, while Claude Sonnet 5.5 consistently commits to the same specification. In another application, to a replication analysis of data relating to immigration and welfare support \citep{brady2014does, breznau2022observing, borjas2026ideological}, GPT-6-Luna explicitly attempts to restrain its own search, yet it reselects its analysis 2.42 times as often when the analysis is significant. Across different settings and AI models, the agents exhibit inconsistent behavior: sometimes they strongly select results, while claiming they do not; sometimes they do not select.

The agents also attempt to self-correct, applying multiplicity corrections to their reported results. We argue and show empirically that these attempts do not restore coverage for a given analysis, as the family of analyses  they correct for is itself data-dependent. These findings highlight a major benefit of our proposed method: it does not rely on any model of how exactly coding agents behave and select analyses. Instead, it explicitly estimates the agents' selective behavior and uses this estimate to correct for the distortions it induces.

Formally, we treat the agents' selection rule as an arbitrary black-box selection mapping from the data to the reported results.
The only assumption our method requires on the agents is that their selection behavior stabilizes asymptotically, and that they are able to accurately report on the analyses they implement. Our method therefore applies to essentially any coding agent, regardless of the model family or generation. This is crucial because AI agents are a moving target, with specific findings about their behavior being short-lived \citep{queen2026shelflife}.

The proposed method also places no restrictions on the kinds of estimators and inference procedures the agents can implement, other than that their $p$-values are asymptotically exact. This is important because part of coding agents' appeal is that they can explore an analysis space as broad as humans can, holding the promise to accelerate empirical research. More broadly, our goal in developing this method was to intervene minimally on researchers' preferred agentic workflow, allowing researchers to simply layer it on top. Our method achieves this by estimating and correcting for any selective behavior \textit{after} the analysis is complete, at the cost of repeated agent runs.

\paragraph{Related literature.}
Our paper contributes to the literature on inference after data-dependent analysis, including sample splitting \citep{cox1975note}, simultaneous post-selection inference \citep{berk2013valid}, conditional selective inference \citep{fithian2014optimal, tibshirani2016exact, lee2016exact,tian2018selective,andrews2024inference}, and adaptive data analysis based on algorithmic stability \citep{dwork2015preserving,zrnic2023post}; see \citet{kuchibhotla2022post,neufeld2026inference} for useful reviews. We build most directly on two recent sub-strands of the literature dealing with black-box conditional selective inference \citep{markovic2019inference,liu2022black,liu2025flexible} and $p$-value-based selective inference \citep{sood2024selective}. Our contribution to this literature is to develop a computationally simple method to restore the validity of selected $p$-values that imposes minimal assumptions on the black-box selection rule and the data-generating process; see Section \ref{sec:comparison} for a detailed comparison. The paper also connects work on specification search and researcher discretion \citep{white2000reality,simmons2011false,mccloskey2026critical} with the growing literature on coding agents in empirical research \citep{korinek2025agents}. Recent studies empirically examine agents' analytical choices \citep{bertran2026many,gao2026nonstandard,mccully2026research, galiani2026ai, asher2026claude}, their sensitivity to prompting \citep{asher2026claude,miao2026agentic}, and their influence on human researchers' analytical decisions \citep{herdagdelens}, extending similar evidence of variation across human analysts \citep{silberzahn2018many,huntington2021influence,breznau2022observing}, though other research has cast doubt on the durability of empirical findings about LLMs \citep{queen2026shelflife}. Within this literature, \citet{miao2026agentic} propose an agentic bootstrap that evaluates agents' findings relative to the distribution of alternative findings they generate when their prompts are perturbed, while \citet{fishman2026editorial} study how the cheap specification search that coding agents enable changes editorial screening. We complement these contributions by developing a method for both detecting and correcting any distortions to classical statistical guarantees induced by coding-agent-driven empirical research. Moreover, because our correction is learned from the agent's actual choices rather than a model of how it searches, it does not require us to assume, understand, or anticipate its analytical behavior. This provides a durable framework for valid statistical inference even as AI technology continues to evolve rapidly.

\paragraph{Notation.} For a probability law $P$, $\Pr_P$ and $\E_P$ denote probability and expectation under $P$, and $P^n$ denotes the law of $n$ i.i.d.\ draws from $P$. We suppress the subscript $P$ when the relevant law is clear from the context. We write $\mathcal L(X)$ for the law of a random object $X$ and $\mathcal L(X\mid E)$ for its conditional law given an event $E$. Further, $[x]_+\coloneqq\max\{x,0\}$, $a\vee b\coloneqq \max\{a,b\}$, $\|\cdot\|_\infty$ is the supremum norm, and $\mathrm{KL}$ is the Kullback--Leibler divergence. Unless stated otherwise, limits are taken as $n\to\infty$, and $\stackrel{p}{\longrightarrow}$ and $\xrightarrow{d}$ denote convergence in probability and in distribution.

\section{Inference after coding agents}
\subsection{Setup} \label{sec:setup}

\paragraph{Coding agents.} We start by describing and formalizing, at a high level, how empirical research with coding agents takes place. We refer the readers to Appendix \ref{app:definitions} for useful background on how coding agents function. Consider the following empirical workflow. A researcher wants to investigate the effect of the minimum wage on employment using one or several datasets. They place those datasets in a folder on their computer and launch their coding agent of choice. Then they tell the coding agent their research question, point it to the data files, and outline the design that they want it to execute. The instructions for the coding agent can vary considerably in how precisely they describe the setting and the estimand the researcher wants to estimate. For example, the researcher could specify that they want the coding agent to estimate a staggered difference-in-differences design or a synthetic controls estimator, and provide additional files specifying the exact data structure and experimental design. Alternatively, the researcher might simply ask the coding agent to ``find me something interesting''. The coding agent then goes to work, inspects the data, writes code, executes preliminary analyses, revises its estimates, and finally settles on an estimate it reports back to the researcher.

To formalize this workflow, let $R=r$ denote the researcher's initial prompt, interpreted broadly to include the instructions and additional context available to the agent, including codebooks, project instructions, and descriptions of the research design. Let $D$ denote the data supplied to the agent, viewed as a random object that can comprise several files in different formats (e.g. .csv, .dta, .xlsx).

Importantly, throughout the paper, we treat the initial prompt and context $r$ as fixed and chosen independently of the data $D$. This rules out data-dependent revisions of the prompt by the human researcher, including feedback based on intermediate results. Such data-dependent revisions add another layer of human selection on top of the coding agent's own selection. This paper focuses on \textit{machine-driven} selective inference, that is, selection carried out by coding agents that, unlike the human researcher’s selection, can be replayed. Adding human decisions into the workflow would connect this setting to the well-established literature on $p$-hacking, which is not the focus of this article.

\paragraph{Formal setup.} We use $\mathcal Q$ to denote a collection of possible analyses produced by the agent. Following \citet{fithian2014optimal}, we think of a given $q\in\mathcal Q$ as specifying a model $\mathcal{M}$ and a corresponding null hypothesis $H_{0} \subset \mathcal M$,
\[
q = (\mathcal{M}, H_0) \in \mathcal{Q},
\]
and the alternative hypothesis is $\mathcal M \setminus H_0$. For each $q\in\mathcal Q$, we write $\mathcal M_q$ and $H_{0,q}$ for its model and null hypothesis, respectively. Then, the coding agent, after being given the initial prompt $r$ and the dataset $D$, can be modeled as a stochastic selection mapping,
\begin{equation} \label{eq:selection_mapping}
Q = \mathcal{A}_r(D, \Xi),
\end{equation}
where $\Xi$ captures randomness in the LLM's response and possibly in the outputs of its tool uses (e.g., when it executes a command whose output depends on a random seed). We hold the precise language model, tools, and runtime configurations used for the coding agent fixed as part of $\mathcal{A}_r$. The null hypothesis $H_{0,q}$ and statistical model $\mathcal M_q$ reflect any decisions the agent makes about sample restrictions, data filtering, outcome construction, etc., as those change the sampling population and specific outcome the null hypothesis is constructed around. Given this selection mapping, whenever we write $\Pr(Q=q)$ below, this is shorthand for $\Pr(\mathcal{A}_r(D, \Xi) = q)$. We assume that $\Xi$ is independent of $D$ and that $\mathcal A_r$ is jointly measurable so that $\Pr(Q=q\mid D=d)=\Pr(\mathcal A_r(d,\Xi)=q)$, the probability that the agent selects $q$ on a given dataset $d$, is defined for every $d$.

We assume that the agent also emits code to implement an inferential procedure that tests the null, using a p-value $p_q(D)$. In line with classical conditional selective inference, we assume each analysis $q$ has a fixed, valid p-value, that is, $q \mapsto p_q(D)$. That is, the agent can select $q$ adaptively based on the data, but given $q$, it does not ``p-hack'' further by choosing among alternative p-value procedures based on the data. Note that this does account for a broader definition of p-hacking where the agent commits to a different model or null $q'$ based on the corresponding p-value $p_{q'}$. Moreover, this assumption can be relaxed to allow for inferential selection as well, but we leave that for later work.

\begin{example}[Motivating example: choosing a population]\label{ex:motiv_example}
Let us return to the minimum-wage example. Suppose the researcher provides an initial prompt asking the agent to investigate the employment effects of a minimum wage increase. For illustration, suppose the agent considers two candidate populations: teenagers and adults. Each population $g$ defines an analysis $q_g$ consisting of a model $\mathcal M_g$ and a null hypothesis $H_{0,g}$ of no average employment effect, $q_g=(\mathcal M_g,H_{0,g})$. For each $q_g$, the agent emits a test whose p-value $p_g(D)$ would be valid if $q_g$ were pre-specified.

The agent will generally look at the data before deciding which analysis to report. For example, it could choose to focus on teenagers if their estimated employment response is larger than the effect for adults. Even if the agent accurately describes the selected analysis and correctly implements the prescribed procedure, the reported $p$-value will generally not be valid conditional on the agent's selection. This is because the agent uses the same data to select $q_g$ and test $H_{0,g}$.

Correcting for this selection requires answering the following question: ``How likely was the agent to select this analysis on this particular dataset?'' In practice, we propose to estimate this answer by rerunning the agent with the exact same prompt but a new version of the dataset. Does the agent still focus on teenagers when this new dataset shows smaller effects for them? For this, we only need to be able to \textit{replay} the agent and observe its selection, with no requirement to know or intervene on its analytical behavior, in particular the steps it takes to arrive at its final reported analysis.
\end{example}

\paragraph{Selective inference.}
Example \ref{ex:motiv_example} illustrates the distinction between validity for a fixed analysis and validity conditional on the agent selecting that analysis. Even if $p_q(D)$ is valid for a prespecified analysis $q=(\mathcal M_q,H_{0,q})$, $p_q(D)$ need not be valid if the agent selects analysis $q$ after inspecting the data $D$.

To see how the selective inference problem arises more formally, first consider a pre-specified analysis $q \in \mathcal{Q}$ that does not depend on the data. Let $\mathcal{P}_{0,q}$ denote the class of data-generating laws under which the fixed null hypothesis $H_{0,q}$ holds. Then, classical statistical validity requires,
\begin{equation} \label{eq:p_value_fixed}
\Pr_P\!\left(p_q(D)\leq\alpha\right)\leq\alpha,
\qquad
P\in\mathcal P_{0,q},\quad \alpha\in(0,1),
\end{equation}
that is, the p-value is valid for any given significance level $\alpha$, for repeated samples from $P$. Now, assume the coding agent \textit{selected} analysis $q$ after looking at the data. Then the guarantee
\[
\Pr_P\!\left(p_q(D)\leq\alpha\mid Q=q\right)
\leq\alpha
\]
need not hold, because the distribution of $p_q(D)$ conditional on $Q=q$ will in general not coincide with the fixed-analysis distribution above. For a visual representation of the difference between the two distributions, see Figure \ref{fig:selective_replay}. Note that this does not require the coding agent to search for favorable results deliberately. Simply inspecting the data, trying alternative specifications, or revising an analysis after inspecting the results makes the selected analysis $Q$ depend on the same data used to compute the p-value. %

Fully pre-specifying the analysis in the researcher's initial prompt would avoid this source of selection, but would also limit the flexibility that makes coding agents useful as empirical analysts and does not reflect their typical use \citep[see e.g.][]{hitzig2026agentic, codreanu2026ai}. We therefore allow the researcher's initial prompt $r$ to constrain the agent's analysis to varying degrees, ranging from complete preregistration to a generic open-ended prompt (``find me something interesting''). One could also potentially resolve the selection problem through data splitting, but this would similarly restrict the agent's flexibility. See \citet{fithian2014optimal} for a more detailed comparison between conditional selective inference and data splitting.

Our objective is to restore the validity of the p-value conditional on the selected analysis, following the conditional selective inference literature \citep{fithian2014optimal, tibshirani2016exact, lee2016exact}. Specifically, for any selected analysis $q$ with $\Pr_P\!\left(Q=q\right)>0$, we aim to construct a p-value  $p_q^{\mathrm{sel}}$
satisfying
\[
\Pr_P\!\left(p_q^{\mathrm{sel}}\leq\alpha\mid Q=q\right)
\leq\alpha,
\qquad
P\in\mathcal P_{0,q},\quad \alpha\in(0,1).
\]
In doing so, we aim to provide valid inferences while minimally intervening in researchers' empirical workflow, allowing them to exploit the versatility of coding agents as empirical analysts.

\begin{example}[Technical example: stochastic selection] \label{ex:stochastic_selection}
Consider two distinct analyses $q_1$ and $q_2$, each
with its own fixed p-value procedure. Suppose that, given the data, the agent selects $q_1$ with probability $1-p_{q_1}(D)$ and otherwise selects $q_2$, so that more favorable evidence for $q_1$ makes its selection more likely.

Fix $P\in\mathcal P_{0,q_1}$ and suppose that $p_{q_1}(D)$ is exactly calibrated, so that $U_1=p_{q_1}(D)$ is uniformly distributed on $[0,1]$.

Then, the probability of selecting $q_1$ is
\[
\Pr_P(Q=q_1)=\int_0^1(1-t)\,dt=\frac12
\]
and the cdf of $U_1$ conditional on selection is
\[
\Pr_P(U_1\le u\mid Q=q_1)
=
\frac{\int_0^u(1-t)\,dt}{1/2}
=2u-u^2.
\]
A nominal $5\%$ test therefore rejects with probability
$0.0975$ conditional on $Q=q_1$, even though its
fixed-analysis p-value is exactly calibrated.

The transformed p-value,
\[
p_{q_1}^{\mathrm{sel}}
=
2p_{q_1}(D)-p_{q_1}(D)^2,
\]
however, is uniformly distributed conditional on $Q=q_1$, and restores exact selective validity.
\end{example}

\paragraph{Correcting the p-value.}
In Example \ref{ex:stochastic_selection}, the correction evaluates the conditional cdf of the selected p-value at its observed value. We now develop this construction for a general analysis $q$. For this, fix a null law $P\in\mathcal P_{0,q}$ under which $\Pr_P(Q=q)>0$. For ease of exposition, assume first that the fixed-analysis p-value is exactly uniform under the null, that is,
\[
\Pr_P\!\left(p_q(D)\leq u\right)=u,
\quad u\in[0,1].
\]
Our theoretical results below replace this assumption with asymptotic exact calibration. We suppress the dependence on $P$ and $r$ in the objects defined below. Here and below, $p_q(D)$ is the p-value obtained by applying the fixed procedure associated with $q$ to $D$, whether or not the agent selects $q$ based on $D$.

Define
\[
H_q(u):=\Pr_P\!\left(p_q(D)\leq u\mid Q=q\right)
\]
as the conditional cdf of the fixed-analysis p-value among datasets on which the agent selects $q$. If $H_q$ were known, the canonical probability integral transform would give the oracle selection-corrected p-value
\[
p_q^{\mathrm{sel}}=H_q\!\left(p_q(D)\right),
\]
which satisfies
\begin{equation}\label{eq:p_select}
p_q^{\mathrm{sel}}\mid Q=q
\sim\operatorname{Unif}(0,1).
\end{equation}

By Bayes' rule, this correction has the following intuitive interpretation:
\begin{align}
H_q(u)
&=
\Pr_P\!\left(p_q(D)\leq u\right)
\frac{\Pr_P\!\left(Q=q\mid p_q(D)\leq u\right)}
     {\Pr_P(Q=q)} \label{eq:H_q_interpretation}
\\
&=
u\,\lambda_q(u),
\end{align}
where
\[
\lambda_q(u)
:=
\frac{\Pr_P\!\left(Q=q\mid p_q(D)\leq u\right)}
     {\Pr_P(Q=q)}
\]
is the selection multiplier, that is, the relative probability of selecting analysis $q$ given that the p-value under that analysis is at least as favorable as $u$. The second equality follows from the
assumption that the fixed-analysis p-value  is exactly valid under the null. Thus, $\lambda_q(u)$ compares the probability of selecting $q$ among datasets where that same analysis produces a p-value at most $u$ with its overall selection probability under the same null law $P$.
Writing $d$ for the observed dataset, this gives,
\[
p_q^{\mathrm{sel}}
=
p_q(d)\,
\lambda_q\!\left(p_q(d)\right).
\]
That is, the selective p-value equals the p-value actually reported by the coding agent, $p_q(d)$, adjusted by the relative probability of the agent selecting the same analysis on datasets that are at least as statistically favorable as $d$ (where the p-value is at least as small as the reported one). When $\lambda_q(u)>1$, analysis $q$ is more likely to be selected
on datasets where its evidence is favorable, and a p-value equal
to $u$ must be adjusted upward. Conversely, $\lambda_q(u)<1$ means that $q$
is less likely to be selected in this favorable tail, and the
correction adjusts the p-value downward. If the event $\{Q=q\}$ is independent of $p_q(D)$ under $P$, then $\lambda_q(u)=1$ for every $u\in(0,1]$ and no correction is required. In the stochastic selection Example \ref{ex:stochastic_selection} above, $\Pr_P(Q=q_1\mid U_1\le u)=1-u/2$, compared with $\Pr_P(Q=q_1)=1/2$,
so that $\lambda_{q_1}(u)=2-u$,
recovering $H_{q_1}(u)=u\lambda_{q_1}(u)=2u-u^2$.

For a single reported analysis, the multiplier depends on just
two selection probabilities. Define
\[
a_{1,q}(u)
:=
\Pr_P\!\left(Q=q\mid p_q(D)\leq u\right),
\qquad
a_{0,q}(u)
:=
\Pr_P\!\left(Q=q\mid p_q(D)>u\right).
\]
Since $p_q(D)$ is uniform under the null,
\[
\Pr_P(Q=q)
=
u\,a_{1,q}(u)+(1-u)\,a_{0,q}(u),
\]
and therefore
\begin{equation} \label{eq:H_q_population_decomp}
H_q(u)
=
\frac{u\,a_{1,q}(u)}
     {u\,a_{1,q}(u)+(1-u)\,a_{0,q}(u)}.
\end{equation}
Evaluating this expression at the p-value on the observed dataset $d$, $p_q(d)$, then gives the desired correction. Thus, to correct the reported p-value for selection, we only need to estimate two tractable objects: the probability of selecting analysis $q$ on alternative datasets with $p$-values smaller and larger than $p_q(d)$. Assuming we can sample such alternative datasets, which we will return to below, $H_q(u)$ can be estimated by simply replaying the agent on them.

We emphasize once more that the exact uniformity of the fixed-analysis $p$-value was a simplifying assumption to ease exposition in this section. The theoretical results below require only \textit{asymptotic} exact calibration (Assumption~\ref{asspt:base_calibration}). %

\subsection{Estimator}

\subsubsection{Selective p-value} \label{sec:selective_pvalue}

We estimate the selection correction by combining two operations on random subsamples of the original data: executing the agent's emitted program and replaying the agent. Section~\ref{sec:subsampling} describes how we draw subsamples of size $m$ from the original data $D_n$ of size $n>m$. We will index a dataset $D$ by its sample size $k$, e.g. $D_k$, where needed for clarity. The reported estimate and fixed-analysis p-value continue to use the full dataset. We use the subsampled datasets to estimate the selection multiplier at sample size $m$. The selection-stability assumption below then relates this multiplier to its full-sample counterpart.

We first execute the original emitted inferential program $p_q(\cdot)$, held fixed throughout the correction, on each subsample. This gives the p-values
\[
p_q^{(1)},p_q^{(2)},\ldots
\]
for the same fixed analysis $q$. For a threshold $u\in(0,1)$, we assign a subsample $D'$ to the tail stratum if $p_q(D')\leq u$ and to the body stratum if $p_q(D')>u$. For the final correction, we set $u=p_q(D_n)$, where $D_n$ is the observed full dataset.

Within each stratum, we replay the agent once on each subsample, in the order in which the subsamples were drawn. Each replay starts with the original prompt $r$ and independent agent randomness. We record whether the replay selects the same analysis $q=(\mathcal M_q,H_{0,q})$ as the original run.

Recall that
\[
a_{1,q}(u)
=
\Pr\!\left(
Q=q
\mid
p_q(D)\leq u
\right),
\qquad
a_{0,q}(u)
=
\Pr\!\left(
Q=q
\mid
p_q(D)>u
\right),
\]
and
\[
H_q(u)
=
\frac{
u a_{1,q}(u)
}{
u a_{1,q}(u)+(1-u)a_{0,q}(u)
}.
\]
Thus, to estimate $H_q(u)$, we only need to estimate the probability that the coding agent selects $q$ in the favorable tail and in the remaining body of the distribution of $p_q(D)$.

Let $B_1\geq1$ and $B_0\geq1$ denote the numbers of coding-agent replays in the tail and body strata, respectively, with total replay budget $B=B_1+B_0$. We draw subsamples until the tail contains $B_1$ and the body $B_0$ of them. Subsamples that fall into a stratum after it is full are discarded. We then estimate,
\begin{equation} \label{eq:a_1_a_0}
\widehat a_{1,q}(u)
=
\frac{1}{B_1}
\sum_{b=1}^{B_1}
\mathbf{1}\left\{Q_1^{(b)}=q\right\},
\qquad
\widehat a_{0,q}(u)
=
\frac{1}{B_0}
\sum_{b=1}^{B_0}
\mathbf{1}\left\{Q_0^{(b)}=q\right\},
\end{equation}
where $Q_1^{(b)}$ denotes a replay on a subsample with $p_q(D)\leq u$ and $Q_0^{(b)}$ a replay on a subsample with $p_q(D)>u$. We can then estimate
\begin{equation}
\label{eq:estimator_H}
\widehat H_q(u)
=
\frac{
u\widehat a_{1,q}(u)
}{
u\widehat a_{1,q}(u)
+
(1-u)\widehat a_{0,q}(u)
}.
\end{equation}

Evaluating this plug-in estimator at the observed p-value $u=p_q(D_n)$ gives the estimated selective p-value
\[
\widehat p_q^{\mathrm{sel}}
=
\widehat H_q(u).
\]
The corresponding estimate of the selection multiplier is
\begin{equation}\label{eq:estimator_lambda}
\widehat\lambda_q(u)
=
\frac{\widehat H_q(u)}{u},
\qquad u\in(0,1).
\end{equation}

These expressions require at least one replay in each stratum and a positive denominator in \eqref{eq:estimator_H}. If either stratum has no replay, or the denominator is zero, we report the correction as unavailable. For the mathematical definition of the test, we set $\widehat p_q^{\mathrm{sel}}=1$ on these outcomes.

The most expensive step of the procedure is agent replays. An agent can try many things when selecting an analysis, which can lead to highly skewed ratios of selection to execution compute time: in some of our experiments, the agent took up to $157{,}000$ times as long to select an analysis as it took to execute the selected analysis code. Also, coding agents typically incur monetary cost in the form of tokens. It therefore matters how a fixed budget $B$ is divided between the tail and body, i.e., how $B_1$ and $B_0$ are chosen. Because running the code is cheap, we can execute it on every subsample drawn and sort the subsamples into strata before spending agent calls.

To guide allocation, we approximate the replay indicators within each stratum as independent Bernoulli variables with success probability $a_{s,q}(u)$ in stratum $s\in\{0,1\}$, and treat the two strata as independent. At a fixed threshold $u$, the delta method gives the following approximation to the variance due to replay sampling:
\[
\operatorname{Var}\!\left\{\widehat H_q(u)\right\}
\approx
\left(
\frac{u(1-u)a_{0,q}(u)}{\rho_q(u)^2}
\right)^2
\frac{a_{1,q}(u)\{1-a_{1,q}(u)\}}{B_1}
+
\left(
\frac{u(1-u)a_{1,q}(u)}{\rho_q(u)^2}
\right)^2
\frac{a_{0,q}(u)\{1-a_{0,q}(u)\}}{B_0},
\]
where
\[
\rho_q(u)
=
u a_{1,q}(u)
+
(1-u)a_{0,q}(u).
\]
For a fixed total number of replays $B=B_1+B_0$ and $0<a_{s,q}(u)<1$, minimizing this variance approximation over positive continuous allocations gives the Neyman allocation \citep{neyman1992two},
\[
\frac{B_1}{B_0}
=
\sqrt{
\frac{
a_{0,q}(u)\{1-a_{1,q}(u)\}
}{
a_{1,q}(u)\{1-a_{0,q}(u)\}
}
}.
\]
Since the selection probabilities are unknown, we first estimate them using a pilot batch of replays in each stratum and substitute these estimates into the allocation rule. For allocation purposes only, we regularize the pilot proportions to lie strictly between zero and one. We impose minimum replay counts $B_s\geq b_{\min}(B)$, $s=0,1$, where $b_{\min}(B)\to\infty$ and $b_{\min}(B)/B\to0$. These constraints ensure that neither stratum receives only a fixed number of replays as the total budget grows. The pilot replays serve only to choose the allocation but not to compute the estimated correction or its confidence bands (see below). In our experiments, this simple rule reduced the number of agent replays required for an equal amount of precision by a factor of three to five.

\subsubsection{Resampling datasets} \label{sec:subsampling}
A crucial ingredient of the estimator is the resampled datasets on which we replay the agent. Since a coding agent can inspect any aspect of its inputs, its selection behavior could respond to practically any statistic of the data, beyond just the $p$-value. For example, sampling with replacement can repeat sampling units and thereby change the agent's behavior. We therefore use subsampling without replacement \citep{politis1994large}.

Subsampling has two useful properties for our setting. First, under i.i.d.\ sampling units, a uniformly selected subsample has exactly the same unconditional distribution as a fresh dataset of $m$ distinct units. Second, although conditioning on the original selection can distort the distribution of the parent dataset, we can bound the resulting distortion in the subsample distribution, and show that it vanishes as $m/n \to 0$, see Proposition \ref{prop:subsampling_selection} below.

\paragraph{The total variation distance as the worst-case change in selection behavior.}
To see the challenge more formally, let $D_n=(O_1,\ldots,O_n)\sim P^n$. Because the coding agent can inspect the entire dataset it receives, we need a notion of resampling fidelity that controls not just a particular statistic, but any arbitrary downstream behavior of the agent. The total variation distance---the ``testing metric''---provides exactly such a measure. Recall that for any two probability laws $\mathcal P_1$ and $\mathcal P_2$,
\[
d_{\mathrm{TV}}(\mathcal P_1,\mathcal P_2)
=
\sup_{0\leq f\leq 1}
\left|
\E_{\mathcal P_1}[f(D)]
-
\E_{\mathcal P_2}[f(D)]
\right|,
\]
where the supremum is taken over all measurable functions $f$ with values in $[0,1]$. For a fixed analysis $q$, define
\[
s_q(d)
=
\Pr(Q=q\mid D=d),
\]
the probability that the coding agent selects $q$ when presented with dataset $d$. Since $s_q(d)\in[0,1]$, it follows immediately that, for any two probability laws $\mathcal P_1$ and $\mathcal P_2$ over complete datasets supplied to the coding agent,
\[
\left|
\Pr_{\mathcal P_1}(Q=q)
-
\Pr_{\mathcal P_2}(Q=q)
\right|
=
\left|
\E_{\mathcal P_1}[s_q(D)]
-
\E_{\mathcal P_2}[s_q(D)]
\right|
\leq
d_{\mathrm{TV}}(\mathcal P_1,\mathcal P_2).
\]
Moreover, if the coding agent is allowed to implement an arbitrary randomized measurable selection rule, then any measurable function $f$ with values in $[0,1]$ can be realized as a selection probability for some analysis $q$. Hence,
\[
\sup_{\mathcal A,q}
\left|
\Pr_{\mathcal P_1}(Q=q)
-
\Pr_{\mathcal P_2}(Q=q)
\right|
=
d_{\mathrm{TV}}(\mathcal P_1,\mathcal P_2).
\]
Thus, the total variation distance is exactly the worst-case change in the coding agent's selection behavior induced by replacing one dataset law with another.

\paragraph{Bootstrap vs subsampling.}
With this in hand, we compare the marginal distribution of a bootstrap dataset with that of a fresh dataset. Let $D_m^*$ denote an $m$-out-of-$n$ bootstrap sample obtained by drawing $m\leq n$ observations with replacement from $D_n=(O_1,\ldots,O_n)\sim P^n$, and suppose that $P$ is nonatomic. Here, the unconditional distribution averages over both the original sample and the bootstrap indices. There are $n^m$ possible ordered sequences of bootstrap indices, of which $m!\binom{n}{m}$ contain only distinct indices. Hence,
\[
\Pr\left(\text{all bootstrap draws are distinct}\right) =  \frac{m!\binom{n}{m}}{n^m}.
\]
Whenever at least one bootstrap index is repeated, the bootstrap produces a type of dataset that would occur with probability zero under a truly fresh draw from $P^m$. Conditional on all bootstrapped indices being distinct, by contrast, the bootstrap dataset has distribution exactly $P^m$. This gives, by the definition of the TV distance,
\begin{equation} \label{eq:TV_bootstrap}
d_{\mathrm{TV}}
\left(
\mathcal L(D_m^*),
P^m
\right)
=
1-\frac{m!\binom{n}{m}}{n^m}.
\end{equation}
Thus, the total variation discrepancy is exactly the probability that sampling with replacement produces at least one duplicated sampling unit. In particular, for the ordinary $n$-out-of-$n$ bootstrap,
\[
d_{\mathrm{TV}}
\left(
\mathcal L(D_n^*),
P^n
\right)
=
1-\frac{n!}{n^n}
\longrightarrow 1,
\]
so the total variation distance converges to one. This establishes a worst-case discrepancy for unrestricted selection rules and of course does not imply that every coding agent reacts to bootstrap duplicates, though in our simulations we frequently noticed agents reacting to unusual features of the data when they appeared.

Subsampling without replacement avoids this duplicated-index effect. If $S_m$ is a uniformly random subset of $m$ distinct indices, chosen independently of $D_n$, then
\[
\mathcal L(D_{S_m})=P^m
\]
exactly. This identity is marginal over the original sample and the subsampling indices. It does not assert that resampling from a fixed observed dataset produces independent fresh draws from $P^m$.

Two more issues remain, however, driven by the fact that we resample from the single observed ``parent'' dataset $D_n$.

First, conditional on the analysis $q$ being selected, the law of the observed dataset $D_n$ is not necessarily equal to $P^n$, which represents the ex-ante distribution of the datasets. That is, we have,
\[
\mathcal L(D_n\mid Q=q) \neq P^n.
\]
Intuitively, observing the agent select analysis $q$ tells us something about the data $D_n$, in particular, that it is more likely to have features that made $q$ an attractive analysis to the agent. Subsamples of the original dataset may therefore inherit some of these features and, as a result, not behave like typical datasets drawn from $P^m$.

Second, even absent selection, all subsamples are drawn from the same observed parent dataset $D_n$. If the subsampling indices are drawn independently across subsamples, these datasets are independent conditional on $D_n$, but generally dependent unconditionally. Increasing the number of subsamples therefore approximates the resampling distribution associated with the observed $D_n$. However, it does not directly establish the convergence to the corresponding population quantities under $P^m$. That step needs a separate concentration argument as $n$ grows.

The following proposition bounds the distortion caused by conditioning on the original selection event. We impose the following sampling assumption.

\begin{assumption}[Subsampling]\label{asspt:sampling_units}
The observed dataset $D_n=(O_1,\dots,O_n)\sim P^n$ consists of $n$ i.i.d.\ sampling units. Each subsample $D_{S_m}$ is constructed by selecting a uniformly random subset of $m$ distinct indices from $\{1,\dots,n\}$ and placing the selected units in a uniformly random order. These index selections and orderings are independent across subsamples and independent of $D_n$ and the agent's randomness $\Xi$. Each agent replay uses fresh, independent randomness.
\end{assumption}

\begin{proposition}[Subsampling bounds the selection contamination] \label{prop:subsampling_selection}
  Let $\rho_{q,n} \coloneqq \Pr(Q=q) > 0$. Then, under Assumption \ref{asspt:sampling_units},
    \[
d_{\mathrm{TV}}
\left(
\mathcal L(D_{S_m}\mid Q=q),
P^m
\right)
\leq
\sqrt{
\frac{m}{2n}
\log\frac{1}{\rho_{q,n}}
}.
\]
\end{proposition}

Hence, whenever
\[
\frac{m}{n}\log\frac{1}{\rho_{q,n}}
\to 0,
\]
the conditional law of a random size-$m$ subsample from a dataset that induced selection of $q$ approaches the fresh-data law $P^m$ in total variation. Under the recurrence assumption imposed below (Assumption~\ref{asspt:recurrence}), which keeps $\rho_{q,n}$ bounded away from zero, this condition is implied by the standard subsampling condition $m/n\to 0$. Otherwise, if the selection probability vanishes, the condition requires its rate to be controlled relative to the subsample size.

This bound controls the distortion caused by conditioning on the original selection event. Convergence of averages computed
from subsamples sharing the same parent dataset
requires an additional, standard concentration argument, which we provide below.

In practice, subsampling requires specifying sampling units that satisfy Assumption~\ref{asspt:sampling_units}. We assume that the research design provides enough information for the researcher, or a helper coding agent, to identify these units. In an i.i.d.\ cross section, each sampling unit can simply be an observation. In clustered or panel settings, each sampling unit can instead comprise an entire cluster or a panel unit's full sequence of observations, provided these units are i.i.d. Dependence within a clustered sampling unit is allowed, since its observations are kept together when subsampling. Dependence across sampling units, as in time series or spatial data, is outside the scope of this paper. An extension using, for example, block subsampling would require additional assumptions and a separate validity argument.

\subsubsection{Measuring selection}
\label{sec:equivalence}

Estimating the selection probabilities requires identifying whether each replay agent selects the same analysis $q^*=(\mathcal M_{q^*},H_{0,q^*})$ as the original agent. In a typical agentic workflow, the agent can provide a textual summary, numerical results, and code without explicitly stating its statistical model and null hypothesis. Even when it states both, different runs may express the same analysis using different notation, variable names, or textual descriptions. Therefore, requiring an exact match based on the agent's full outputs is far too restrictive to be of practical use.

Our matching step therefore compares the \emph{statistical content} of the reported analyses, ignoring purely representational differences. Two runs match if they report the same statistical model, including the same target population and outcome, and the same null hypothesis. The code and sequence of exploratory steps can then be different between the runs.

\paragraph{The commitment prompt.}
To implement this matching step, we exploit the fact that coding agents are \textit{promptable}. After each run has completed its analysis, we append a fixed \textit{commitment prompt} asking the agent to state, in a self-contained manner, the statistical model and null hypothesis used in that analysis. The prompt asks the agent to describe the completed analysis without revising it. This reporting step is considered to be included in the selection mapping $\mathcal A_r$ and is applied in the same way to the original run and every replay.

Each replay agent first commits to its own analysis $q'$. After an agent commits, we provide the original agent's description of $q^*$ and append a fixed \textit{comparison prompt}. This prompt asks whether the two descriptions specify the same statistical model and null hypothesis, ignoring purely representational differences such as notation or variable relabeling. The comparison leaves the committed analysis $q'$ unchanged and produces a binary match decision.

This protocol allows agents to describe their analyses without imposing a common notation or a fixed catalogue of possible analyses. Our use of comparative prompting is motivated by work on LLMs as judges \citep[see, e.g.][]{zheng2023judging}. Under the reporting and matching assumptions stated below, the binary response identifies the event $Q'=q^*$, which is all that our method requires.

The below assumption formalizes what is required for this approach to accurately measure the target event $\{Q' = q^*\}$: truthful reporting by the agent.
\begin{assumption}[Trust your agent]
\label{asspt:trust}
Under the data-generating and replay laws considered in this paper, the following statements hold almost surely.
\begin{enumerate}[label=(\roman*)]
\item \textit{Correct implementation.}
Whenever the procedure evaluates an analysis $q$ on a dataset $D'$, the emitted code returns the fixed $p$-value $p_q(D')$. This applies to the original dataset and to execution of the original program on subsamples.

\item \textit{Faithful reporting.}
The commitment report accurately states the statistical model and null hypothesis selected and implemented by the agent.

\item \textit{Correct matching.}
For a replay selecting
$Q'=\mathcal A_r(D',\Xi')$, the binary comparison response $I'$ satisfies
\[
I'=\mathbf{1}\{Q'=q^*\},
\] where $q^*$ is the originally selected analysis.
\end{enumerate}
\end{assumption}

Our validity results below take these accuracy assumptions as given. The phrase \textit{trust your agent} reflects a basic premise of delegating empirical work: an agent trusted to carry out an analysis should also be able to accurately explain what it has done and to compare it against other analyses.

Under Assumption \ref{asspt:trust}, the replay procedure identifies selection through the reported model and null hypothesis, not through the code the agent emits. The selection correction focuses on addressing the effect of choosing $q$ adaptively from the data. Implementation errors, inaccurate reports, and errors in matching are outside the scope of the present guarantees. Assessing and relaxing these accuracy assumptions is complementary to work on agent evaluation and verification \citep[see, e.g.][]{gu2024blade, deng2025swe, majumder2025discoverybench, huang2025automated, candes2025probably, lee2026fuse}. In practice, if the researcher is using a lower-quality AI model as their analysis agent, we recommend having a higher-quality agent review its matching declarations.

\subsubsection{Summary} \label{sec:estimator_summary}

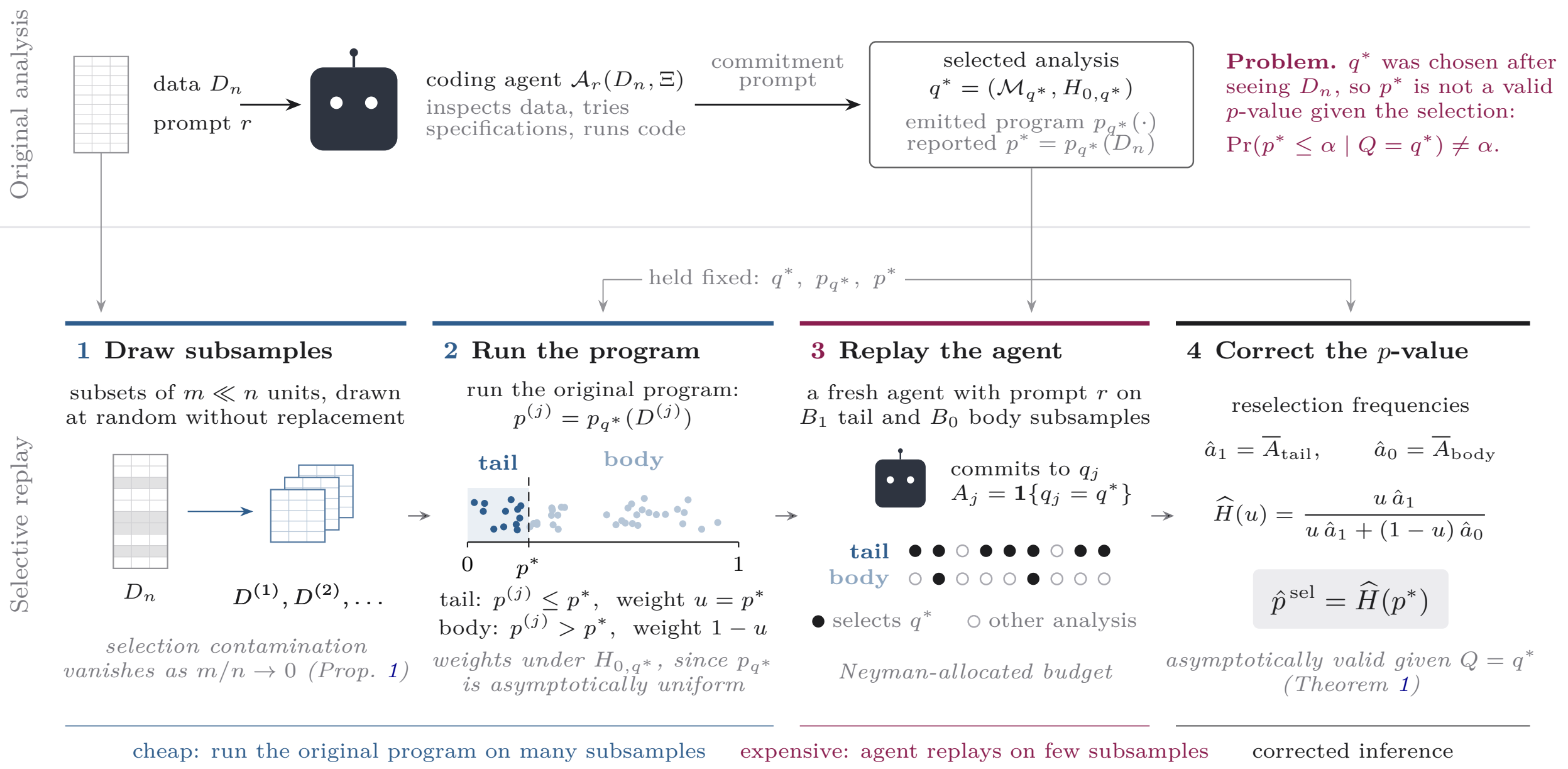
\begin{figure}[htb!]
\centering
\resizebox{\textwidth}{!}{%
\begin{tikzpicture}[x=1cm,y=1cm]

\node[rotate=90,font=\footnotesize,text=srwmuted] at (-0.3,0.95) {Original analysis};

\srwDataset{(0.3,0.4)}{0.62}{10}{0.11}{srwmuted}{}
\node[srw txt,anchor=west,align=left] at (1.08,1.18) {data $D_n$};
\node[srw txt,anchor=west,align=left] at (1.08,0.72) {prompt $r$};
\coordinate (inp) at (0.61,0.4);

\draw[srw flow] (2.2,0.95) -- (2.9,0.95);
\srwRobot{(3.5,0.9)}{0.95}
\node[srw txt,anchor=west,align=left] at (4.2,0.95)
  {coding agent $\mathcal A_r(D_n,\Xi)$\\[1pt]
   {\color{srwmuted}inspects data, tries}\\[-1pt]
   {\color{srwmuted}specifications, runs code}};

\draw[srw flow] (7.4,0.95) -- node[above=1pt,font=\scriptsize,text=srwmuted,align=center]
  {commitment\\[-2pt]prompt} (9.3,0.95);
\node[draw=srwink!70,line width=0.5pt,rounded corners=2pt,fill=white,
      minimum width=3.7cm,minimum height=1.35cm,align=center,font=\scriptsize,text=srwink]
  (qs) at (11.25,0.95)
  {selected analysis\\[1pt] $q^\ast=(\mathcal M_{q^\ast},H_{0,q^\ast})$\\[3pt]
   {\color{srwmuted}emitted program $p_{q^\ast}(\cdot)$}\\[-1pt]
   {\color{srwmuted}reported $p^\ast=p_{q^\ast}(D_n)$}};

\node[anchor=west,align=left,font=\scriptsize,text=srwagent] at (13.35,0.95)
  {\textbf{Problem.} $q^\ast$ was chosen after\\seeing $D_n$, so $p^\ast$ is not a valid\\$p$-value given the selection:\\[3pt]
   $\Pr(p^\ast\le\alpha\mid Q=q^\ast)\neq\alpha$.};

\draw[srwrule,line width=0.6pt] (-0.55,-0.45) -- (16.95,-0.45);
\node[rotate=90,font=\footnotesize,text=srwmuted] at (-0.3,-3.85) {Selective replay};

\def\srwTop{-1.55}
\foreach \xa/\xb/\c in {0.2/4.1/srwcodeblue,4.4/8.3/srwcodeblue,8.6/12.6/srwagent,12.9/16.95/srwink}
  \draw[\c,line width=1.4pt] (\xa,\srwTop) -- (\xb,\srwTop);

\foreach \x/\n/\c/\t in {0.2/1/srwcodeblue/Draw subsamples,
                         4.4/2/srwcodeblue/Run the program,
                         8.6/3/srwagent/Replay the agent,
                         12.9/4/srwink/Correct the $p$-value}
  \node[anchor=west,font=\footnotesize,text=srwink] at (\x,-1.88)
    {\textcolor{\c}{\textbf{\n}}\enspace\textbf{\t}};

\draw[srw feed] (inp) -- (0.61,\srwTop+0.12);
\draw[srw feed] (qs.south) -- (11.25,\srwTop+0.12);
\draw[srwmuted!80,line width=0.4pt] (6.35,-1.05) -- (14.9,-1.05);
\foreach \x in {6.35,14.9} \draw[srw feed] (\x,-1.05) -- (\x,\srwTop+0.12);
\node[font=\scriptsize,text=srwmuted,fill=white,inner sep=2pt] at (8.3,-1.05)
  {held fixed: $q^\ast,\ p_{q^\ast},\ p^\ast$};

\node[srw txt] at (2.15,-2.5)
  {subsets of $m\ll n$ units, drawn\\ at random without replacement};
\srwDataset{(0.75,-4.35)}{0.62}{10}{0.13}{srwmuted}{2,4,5,8}
\node[font=\scriptsize,text=srwink,anchor=north] at (1.06,-4.4) {$D_n$};
\draw[srw feed,draw=srwcodeblue] (1.6,-3.7) -- (2.35,-3.7);
\foreach \dx/\dy in {0.32/0.28,0.16/0.14,0/0}
  \srwDataset{($(2.55,-4.05)+(\dx,\dy)$)}{0.62}{5}{0.12}{srwcodeblue}{}
\node[font=\scriptsize,text=srwink,anchor=north] at (3.0,-4.4) {$D^{(1)},D^{(2)},\dots$};
\node[srw note] at (2.15,-5.4)
  { selection contamination\\ vanishes as $m/n\to0$ (Prop.~\ref{prop:subsampling_selection})};

\node[srw txt] at (6.35,-2.5)
  {run the original program:\\ $p^{(j)}=p_{q^\ast}(D^{(j)})$};
\def\axl{4.8}\def\axr{7.9}\def\axy{-4.05}\def\uu{5.5}
\fill[srwcodeblue!10] (\axl,\axy) rectangle (\uu,{\axy+0.62});
\pgfmathsetseed{11}
\foreach \i in {1,...,44}{%
  \pgfmathsetmacro{\px}{\axl+0.06+rnd*(\axr-\axl-0.12)}
  \pgfmathsetmacro{\py}{\axy+0.12+rnd*0.40}
  \pgfmathparse{\px<\uu ? 1 : 0}
  \ifnum\pgfmathresult=1 \fill[srwcodeblue] (\px,\py) circle (0.04);
  \else \fill[srwcodeblue!35] (\px,\py) circle (0.04); \fi}
\draw[srwink,line width=0.45pt] (\axl,\axy) -- (\axr,\axy);
\foreach \x in {\axl,\axr} \draw[srwink,line width=0.45pt] (\x,\axy) -- ++(0,-0.06);
\draw[srwink,densely dashed,line width=0.45pt] (\uu,{\axy-0.06}) -- (\uu,{\axy+0.74});
\node[font=\scriptsize,anchor=north] at (\axl,{\axy-0.05}) {$0$};
\node[font=\scriptsize,anchor=north] at (\uu,{\axy-0.05}) {$p^\ast$};
\node[font=\scriptsize,anchor=north] at (\axr,{\axy-0.05}) {$1$};
\node[srw lab=srwcodeblue,anchor=south] at ({(\axl+\uu)/2},{\axy+0.7}) {tail};
\node[srw lab=srwcodeblue!55,anchor=south] at ({(\uu+\axr)/2},{\axy+0.7}) {body};
\node[srw txt] at (6.35,-4.85)
  {tail: $p^{(j)}\le p^\ast$, \ weight $u=p^\ast$\\ body: $p^{(j)}>p^\ast$, \ weight $1-u$};
\node[srw note] at (6.35,-5.55)
  {weights under $H_{0,q^\ast}$, since $p_{q^\ast}$\\ is asymptotically uniform};

\node[srw txt] at (10.6,-2.5)
  {a fresh agent with prompt $r$ on\\ $B_1$ tail and $B_0$ body subsamples};
\srwRobot{(9.75,-3.38)}{0.55}
\node[srw txt,anchor=west,align=left] at (10.2,-3.38)
  {commits to $q_j$\\ $A_j=\mathbf 1\{q_j=q^\ast\}$};
\def\oy{-4.15}
\node[srw lab=srwcodeblue,anchor=east] at (9.75,\oy) {tail};
\node[srw lab=srwcodeblue!55,anchor=east] at (9.75,{\oy-0.32}) {body};
\foreach \k [count=\i] in {1,1,0,1,1,1,0,1,1}{%
  \ifnum\k=1 \srwHit{({9.65+0.27*\i},\oy)} \else \srwMiss{({9.65+0.27*\i},\oy)} \fi}
\foreach \k [count=\i] in {0,1,0,0,0,1,0,0,0}{%
  \ifnum\k=1 \srwHit{({9.65+0.27*\i},{\oy-0.32})} \else \srwMiss{({9.65+0.27*\i},{\oy-0.32})} \fi}
\node[srw txt,text=srwmuted] at (10.6,-4.95)
  {\tikz[baseline=-0.6ex]{\srwHit{(0,0)}}\ selects $q^\ast$ \quad
   \tikz[baseline=-0.6ex]{\srwMiss{(0,0)}}\ other analysis};
\node[srw note] at (10.6,-5.55) {Neyman-allocated budget};

\node[srw txt] at (14.9,-2.5) {reselection frequencies};
\node[font=\scriptsize,text=srwink] at (14.9,-2.98)
  {$\hat a_1=\overline A_{\mathrm{tail}}, \qquad \hat a_0=\overline A_{\mathrm{body}}$};
\node[font=\scriptsize,text=srwink] at (14.9,-3.75)
  {$\displaystyle \widehat H(u)=\frac{u\,\hat a_1}{u\,\hat a_1+(1-u)\,\hat a_0}$};
\node[fill=srwrule!70,rounded corners=2pt,inner xsep=6pt,inner ysep=4pt,
      font=\small,text=srwink] at (14.9,-4.7)
  {$\hat p^{\,\mathrm{sel}}=\widehat H(p^\ast)$};
\node[srw note] at (14.9,-5.55)
  {asymptotically valid given $Q=q^\ast$\\ (Theorem~\ref{thm:asymptotic_validity})};

\foreach \a/\b in {4.12/4.38,8.32/8.58,12.62/12.88}
  \draw[-{Stealth[length=1.3mm,width=1.1mm]},line width=0.5pt,draw=srwmuted] (\a,-3.75) -- (\b,-3.75);

\foreach \xa/\xb/\c/\t in {0.2/8.3/srwcodeblue/{cheap: run the original program on many subsamples},
                           8.6/12.6/srwagent/{expensive: agent replays on few subsamples},
                           12.9/16.95/srwink/{corrected inference}}{%
  \draw[\c!60,line width=0.5pt] (\xa,-6.15) -- (\xb,-6.15);
  \node[font=\scriptsize,text=\c,anchor=north] at ({(\xa+\xb)/2},-6.22) {\t};}

\end{tikzpicture}%
}
\centering
\caption{Agentic Selection Correction: Workflow Overview}
\label{fig:workflow}
\floatfoot{\textit{Note:} figure depicts the workflow for the researcher-run ``Original Analysis'' and the ``Selective Replay'' that replays fresh coding agents (3) on subsampled datasets (1) which are split into tail and body based on how favorable the original p-value is on them (2) to estimate the relative probability of selecting the same null hypothesis $H_{0,q^*}$ and statistical model $\mathcal M_{q^*}$ as the ``Original Analysis''. These estimates are then used to correct the p-value for selection (4). See Section \ref{sec:estimator_summary} and the preceding sections for details. }
\end{figure}

Figure~\ref{fig:workflow} summarizes the proposed procedure. In the ``Original Analysis'' pane, the researcher provides the agent with a prompt $r$, which is chosen independently of the data, and the dataset $D_n$. The agent then selects an analysis $q^*=(\mathcal M_{q^*},H_{0,q^*})$ and reports $p^*=p_{q^*}(D_n)$. Because the agent chooses $q^*$ based on $D_n$, the fixed-analysis calibration of this procedure need not hold conditional on $Q=q^*$. The ``Selective Replay'' pane thus estimates how much more likely the agent is to select $q^*$ on datasets where $q^*$ looks at least as favorable as on $D_n$.

In the ``Selective Replay'' pane, we draw subsamples $D^{(1)},D^{(2)},\ldots$ of size $m<n$ using the subsampling scheme from Assumption~\ref{asspt:sampling_units}. We execute the original program on each subsample and assign it to a tail stratum, where $p_{q^*}(D^{(j)})\leq p^*$, or a body stratum, where $p_{q^*}(D^{(j)})>p^*$. We keep drawing until the tail contains $B_1$ subsamples and the body $B_0$, and replay the agent once on each, using the original prompt $r$ and independent agent randomness $\Xi'$. The allocation of the $B_1$ tail and $B_0$ body replays can be chosen using pilot estimates and the Neyman allocation rule described above. After completing its initial response, each replay agent first commits to its own analysis, after which it is asked to compare its report with the original agent's report for $q^*$ and decide if there is a match.

Under Assumption~\ref{asspt:trust}, the recorded matches identify whether each replay selects $q^*$. The estimated tail and body reselection frequencies then yield the correction
\[
\widehat p_{q^*}^{\mathrm{sel}}
=
\widehat H_{q^*}(p^*).
\]
When the correction is unavailable, we follow the non-rejection convention described above.
Under the conditions of Theorem~\ref{thm:asymptotic_validity}, the resulting $p$-value is asymptotically valid conditional on $Q=q^*$.

\subsection{Comparison to existing approaches} \label{sec:comparison}

\paragraph{Black-box selective inference.} As discussed in the Introduction, our paper builds on a rich literature in conditional selective inference, which aims to restore the statistical guarantees of uncertainty statements about an analysis \textit{conditional} on it being chosen based on the data. In contrast to most of that literature, the selection rule we study here cannot be characterized, seeing as it is an unknown stochastic mapping from the data to the reported analysis that is implicitly defined by the billions to trillions of parameters of the AI model conducting the analysis. That is, coding agents are black-box selectors.

\citet{markovic2019inference} and \citet{liu2022black} also rerun a black-box selector to correct inference, targeting the law of an estimator $\widehat\theta$ conditional on selection. Since selection generally also depends on other features of the data, whose law involves unknown nuisance parameters, they assume that the algorithm depends on the data only through a summary statistic specified by the analyst, such as $X^\top y$ for the lasso, that is asymptotically jointly Gaussian with $\widehat\theta$. This statistic can then be written as a linear function of $\widehat\theta$ plus an independent residual. Conditioning on the residual leaves a selection probability that depends only on $\widehat\theta$, and a Gaussian law for $\widehat\theta$ that depends only on $\theta$. They then learn this selection probability, either by rerunning the algorithm on synthetic values of the summary statistic \citep{markovic2019inference} or by fitting a classifier on bootstrapped data \citep{liu2022black}, and reweight the Gaussian law by it. Both methods thus treat the map from a specified statistic to the selected model as a black box. For a coding agent, however, which can respond to any feature of the files it reads, the assumption that selection depends on a low-dimensional summary statistic that is Gaussian is unrealistic, as the only statistic guaranteed to capture its selection is the dataset itself (Section~\ref{sec:subsampling}).

More recently, \citet{liu2025flexible} proposed a method that allows for arbitrary black-box selection mappings by learning the selection-conditional distribution of an estimator through flow-based transport maps. Their approach relaxes the assumption that the selection mapping has a low-dimensional Gaussian summary, but instead requires the ability to simulate datasets under a suitable null model. For coding agents that can inspect arbitrary features of the data, the null hypothesis attached to the selected analysis generally does not specify the joint distribution of all the data features that may influence selection, making such simulation difficult.

Instead, we take a different approach and target the $p$-value of the selected null hypothesis. Since the null law of a calibrated, fixed-analysis $p$-value is asymptotically uniform regardless of the nuisance parameters, we can simply condition only on the selection event and average out all the other information in the data. In particular, we leverage the fact that correcting the reported $p$-value requires estimating just two selection probabilities, which condition only on the realization of the original analysis's $p$-value. This averaging in turn only requires being able to sample from the dataset's generating law, but not from the estimator's null law. We  show that this can be achieved by subsampling from the original dataset, with any remaining influence of the selection event washing out under standard subsampling conditions and asymptotically stable selection. This avoids both the need to summarize the selection mapping through a low-dimensional Gaussian statistic in \citet{markovic2019inference, liu2022black} and the need to specify a full generative null model in \citet{liu2025flexible}. In exchange, we  obtain a test for only a single selected null hypothesis instead of the full selection-conditional distribution of the parameter, though in principle we could also construct confidence sets for the parameter through test inversion, at higher computational cost \citep{sood2024selective}.

\paragraph{Agentic empirical science.} The paper closest to ours in the emerging statistical literature on coding agents is \citet{miao2026agentic}. That paper studies how different prompts affect the distribution of analyses that coding agents consider, and then proposes an ``m-value'' that reflects how extreme a given analysis produced by a coding agent is relative to this distribution of analyses, on the same dataset. In our notation, let $d$ denote the observed data, $T_q(d)$ a scalar statistic reported by analysis $q$, and $\Pi$ a reference distribution over analyses. Then, for the original analysis $q^*$, the (two-sided) m-value they propose is,
\[
m_{\Pi}\left(q^* ; d\right)=2 \min \left\{\operatorname{Pr}_{q^{\prime} \sim \Pi}\left(T_{q^{\prime}}(d) \leq T_{q^*}(d)\right), \operatorname{Pr}_{q^{\prime} \sim \Pi}\left(T_{q^{\prime}}(d) \geq T_{q^*}(d)\right)\right\}.
\]
In practice, the authors estimate this quantity by an ``agentic bootstrap'', which replays coding agents with various prompts $r'$ (e.g. pro, anti, and neutral immigration personas), logs all the analyses considered by the replayed agents, and constructs an estimate $\hat{\Pi}$ of the reference distribution based on that. Their m-value then measures the atypicality of $q^*$ relative to this constructed distribution. An important pitfall of this approach, as noted by the authors as well, is that the $m$-value is \textit{endogenous} to the reference distribution. For example, adding an ``extremely anti-immigration'' persona to the set of prompts could shift the reference distribution toward more negative estimates and assign previously atypical anti-immigration results non-significant $m$-values, even when neither the reported analyses nor the underlying data have changed. Additionally, a significant $m$-value indicates that a result is atypical relative to the reference distribution, but does not establish whether that result provides statistically compelling evidence against its own null hypothesis.

By contrast, we hold the original prompt $r$ and the agent configuration fixed but vary the \textit{data} to ask instead how the agent's selection of $q^*$ distorts the distribution of its reported $p$-value under the null. This yields a method to \textit{correct} this distortion and restore the validity of statistical inference conditional on having selected $q^*$, following the selective inference literature. Importantly, our method does not require constructing a reference distribution but rather treats the distribution under the original null as the reference, approximating it by resampling the data. The distribution of alternative analyses is left unspecified, as we only need to know \textit{whether} the agent selected an alternative analysis, but do not need to know or explicitly model this alternative analysis space. Also, as we point out below, our method only depends on minimal assumptions about how agents behave, whereas the $m$-value assumes that one can trace out the distribution of possible analyses through injecting contrasting personas in prompts. As we show below, it appears that the latest generation of models has become robust to such persona-based steering.

\section{Statistical Results} \label{sec:statistical_results}
In this section, we establish the statistical properties of our agentic selection-correction estimator.
Throughout this section, we will make the sample-size index explicit in our notation. The subsamples have size $m=m_n$, where $m_n\rightarrow \infty$ and $m_n/n\rightarrow 0$ as $n\rightarrow \infty$.  For a generic sample size $k$ and a fresh dataset $D_k\sim P^k$, let
\[
U_{q,k}:=p_q(D_k),\qquad
Q_k:=\mathcal A_r(D_k,\Xi),\qquad
\rho_{q,k}:=\Pr_P(Q_k=q),
\]
and define the fixed-analysis cdf
\[
F_{q,k}(u):=\Pr_P(U_{q,k}\le u).
\]
Whenever $\rho_{q,k}>0$, define the conditional cdf
\[
H_{q,k}(u)
:=
\Pr_P(U_{q,k}\le u\mid Q_k=q).
\]
Whenever both $\rho_{q,k}>0$ and $F_{q,k}(u)>0$, define the selection multiplier
\begin{equation}
\label{eq:lambda_k}
\lambda_{q,k}(u)
:=
\frac{\Pr_P(Q_k=q\mid U_{q,k}\le u)}
     {\rho_{q,k}}.
\end{equation}
Below, we evaluate these objects at $k=m_n$ and $k=n$.

\subsection{Assumptions}

We impose the following assumptions, in addition to Assumptions~\ref{asspt:sampling_units} and~\ref{asspt:trust}.
They restrict the calibration of the fixed inferential procedure and the behavior of the selection process across datasets and sample sizes.

\begin{assumption}[Asymptotic exact calibration]
\label{asspt:base_calibration}
For the fixed analysis $q$ under consideration and every $P\in\mathcal P_{0,q}$,
\[
\sup_{u\in[0,1]}
\left|F_{q,k}(u)-u\right|
\longrightarrow 0
\qquad\text{as } k\to\infty.
\]
\end{assumption}

This assumption requires the fixed-analysis $p$-value $p_q(D_k)$ to converge in distribution to $\operatorname{Unif}(0,1)$ under each null law $P\in\mathcal P_{0,q}$. Note that it does permit departures from exact calibration in finite samples. Two-sided t-tests of regression coefficients, Wald tests, tests based on GMM, and bootstrap tests all satisfy Assumption \ref{asspt:base_calibration} under standard regularity conditions. The assumption concerns only the procedure associated with the fixed analysis $q$ and places no restriction on the agent's selection behavior.

\begin{assumption}[Selection stability]
\label{asspt:lambda_stability}
The selection multiplier $\lambda_{q,k}(u)$ satisfies one of the following conditions:
\begin{enumerate}[label=(\alph*)]
\item \textit{One-sided stability.}
For the subsample sequence $m_n\to\infty$
with $m_n/n\to0$, for every $u\in(0,1]$,
\[
\left[\lambda_{q,n}(u)
-\lambda_{q,m_n}(u)\right]_+
\longrightarrow0.
\]

\item \textit{Two-sided stability.}
Under the same subsample sequence, for every
$u\in(0,1]$,
\[
\left|\lambda_{q,n}(u)
-\lambda_{q,m_n}(u)\right|
\longrightarrow0.
\]
\end{enumerate}
\end{assumption}

The multiplier compares the probability of selecting $q$ on datasets where its fixed-analysis $p$-value is at most $u$ with the unconditional probability of selecting $q$. This assumption restricts how that relative probability can change between sample sizes $m$ and $n$, while allowing $\rho_{q,k}$ to vary with $k$.

Condition (a) allows the subsample multiplier to exceed the full-sample multiplier, but requires any excess in the opposite direction to vanish asymptotically.
Under the remaining conditions of Theorem~\ref{thm:asymptotic_validity}, this one-sided condition is sufficient for asymptotic validity. Replacing the positive part in (a) with $|\lambda_{q,n}(u)-\lambda_{q,m_n}(u)|$ gives the two-sided stability in (b) and, together with those conditions, asymptotic exactness. Both conditions are imposed pointwise for each null law $P\in \mathcal{P}_{0,q}$. See Section \ref{sec:uniformity} for a discussion of uniform validity.

\begin{assumption}[Recurrence]
\label{asspt:recurrence}
For the fixed analysis $q$ under consideration, there exists $\rho_\star>0$ such that, for all $n$ along the sequence under consideration,
\[
\min\{\rho_{q,n},\rho_{q,m_n}\}\ge\rho_\star.
\]
\end{assumption}

This assumption keeps the probability of selecting $q$ bounded away from zero at both the full-sample size $n$ and the subsample size $m_n$, which is needed so that we can use simple subsampling. This assumption is a restriction on the selected analysis and the agent, not a universal property of coding agents. A well-defined and circumscribed prompt can make this condition more plausible by limiting the range of analyses the agent considers, but does not guarantee it.

By contrast, an open-ended prompt such as ``find me something interesting'' could lead the agent to select among many different analyses, making any particular $q$ difficult to reproduce. When selection is very rare, obtaining informative replay estimates may require many agent executions. Cases where the selection probabilities vanish with sample size are excluded by Assumption \ref{asspt:recurrence}.

\subsection{Validity of the replay correction}
\label{sec:asymptotic_validity}

We first establish the ``oracle'' validity of the selective $p$-value, assuming that the conditional cdf $H_{q,n}$ is known under the null law $P$. The finite-sample statement follows from applying the cdf transform under the conditional law given $Q_n=q$; see also \citet{sood2024selective} for related $p$-value-based selective corrections.

\begin{proposition}[Corrected p-value: oracle validity]
\label{prop:selected_rank_general}
Fix an analysis $q$ and a law $P\in\mathcal P_{0,q}$. For every $n$ for which $\rho_{q,n} > 0$,
\[
\Pr_P\!\left(
H_{q,n}(U_{q,n})\le\alpha
\mid Q_n=q
\right)\le\alpha,
\qquad \alpha\in[0,1].
\]
Moreover, under Assumptions~\ref{asspt:base_calibration} and \ref{asspt:recurrence},
\[
H_{q,n}(U_{q,n})\mid Q_n=q
\xrightarrow{d}\operatorname{Unif}(0,1)
\qquad\text{as }n\to\infty.
\]
\end{proposition}

The oracle correction is the cdf transform applied to the distribution conditional on selection. The finite-sample inequality holds even when this distribution is discrete. When $H_{q,n}$ is continuous, the corrected $p$-value is exactly uniform conditional on the selection. More generally, asymptotic exact calibration of the original $p$-value, together with the selection probability being bounded away from zero, gives the stated asymptotic exactness. Note that apart from recurrence, this oracle result places no restriction on the agent's selection behavior.

Next, we study the asymptotic validity of the estimated correction. Write $u_n=U_{q,n}$. Recall that $\widehat a_{1,q}(u_n)$ and $\widehat a_{0,q}(u_n)$ estimate the probabilities of re-selecting $q$ in the tail and body strata, respectively.
 Whenever both replay frequencies are defined and the denominator is positive, the estimator of the selective $p$-value is
\begin{equation}
\label{eq:plugin_selective_p_theory}
\widehat p_{q,n}^{\mathrm{sel}}
:=
\frac{u_n\widehat a_{1,q}(u_n)}
{u_n\widehat a_{1,q}(u_n)
 +(1-u_n)\widehat a_{0,q}(u_n)}.
\end{equation}
Otherwise, we set $\widehat p_{q,n}^{\mathrm{sel}}=1$, following the convention of Section~\ref{sec:selective_pvalue}.

\begin{theorem}[Estimated selective p-value: asymptotic validity]
\label{thm:asymptotic_validity}
Fix an analysis $q$ and a law $P\in\mathcal P_{0,q}$ and
suppose Assumptions~\ref{asspt:sampling_units},
\ref{asspt:trust}, \ref{asspt:base_calibration},
\ref{asspt:lambda_stability}(a), and~\ref{asspt:recurrence} hold.
Let $m_n\to\infty$ with $m_n/n\to0$. Draw subsamples as in Assumption~\ref{asspt:sampling_units} until the tail and body strata contain $B_{1,n}$ and $B_{0,n}$ subsamples, and replay the agent once on each. Use fresh agent randomness $\Xi'$ that is i.i.d.\ across replays and independent of $D_n$, the original randomness $\Xi$, and the subsample draws.
The replay counts can be chosen adaptively, but must satisfy
\[
B_{s,n}\stackrel{p}{\longrightarrow}\infty,
\quad s\in\{0,1\},
\]
under the conditional law given $Q_n=q$.
Let $\widehat p_{q,n}^{\mathrm{sel}}$ be the estimator in \eqref{eq:plugin_selective_p_theory}.

Then, for every $\alpha\in(0,1)$,
\begin{equation} \label{eq:asympt_validity}
\limsup_{n\to\infty}
\Pr_P\!\left(
\widehat p_{q,n}^{\mathrm{sel}}\le\alpha
\mid Q_n=q
\right)
\le\alpha.
\end{equation}
If additionally Assumption \ref{asspt:lambda_stability}(b) holds, then, writing $u_n=U_{q,n}$,
\begin{equation}\label{eq:oracle_equivalence}
\widehat p_{q,n}^{\mathrm{sel}}-H_{q,n}(u_n)
\stackrel{p}{\longrightarrow}0,
\qquad
\widehat\lambda_q(u_n)-\lambda_{q,n}(u_n)
\stackrel{p}{\longrightarrow}0,
\end{equation}
conditional on $Q_n=q$, where $\widehat \lambda_q(u_n)=\widehat p_{q,n}^{\mathrm{sel}}/u_n$. In particular,
\[
\widehat p_{q,n}^{\mathrm{sel}}\mid Q_n=q
\xrightarrow{d}\operatorname{Unif}(0,1).
\]
\end{theorem}

Theorem~\ref{thm:asymptotic_validity} establishes that selection probabilities estimated from subsampled replay datasets can provide asymptotically valid inference for the full-sample analysis, conditional on the original agent selecting $q$ and under the stated assumptions.

Assumption \ref{asspt:lambda_stability}(a), which requires the subsample multiplier $\lambda_{q,m_n}$ to be at least as large in the limit as the full-sample one, $\lambda_{q,n}$, gives the asymptotic validity in Eq. \eqref{eq:asympt_validity} because it makes the estimated correction at least as large as the oracle correction in the limit, which is valid by Proposition \ref{prop:selected_rank_general}. Under the stronger assumption that the difference between the two multipliers goes to zero (Assumption \ref{asspt:lambda_stability}(b)), we get the stronger result in Eq. \eqref{eq:oracle_equivalence}, which implies that the corrected p-value is exactly uniformly distributed in the limit, conditional on $Q_n = q$.

\subsection{Uniformity over null distributions}
\label{sec:uniformity}
Theorem~\ref{thm:asymptotic_validity} establishes pointwise asymptotic validity for each fixed null distribution $P$. To extend this pointwise result to uniform validity, we introduce strengthened assumptions, which ensure that the worst-case conditional size distortion over a prescribed class vanishes as the sample size increases.

Throughout this subsection, we fix an analysis $q$ and define a nonempty class $\mathfrak P_{0,q}\subseteq\mathcal P_{0,q}$ of null distributions. The agent, prompt, and fixed-analysis inferential procedure stay unchanged as $P$ varies. We use superscripts $P$ to make the dependence of $F^P_{q,k}$, $H^P_{q,k}$, $\rho^P_{q,k}$, and $\lambda^P_{q,k}$ on the data-generating law explicit.

In order to state the uniform results, we maintain the sampling, implementation, and replay-design conditions of Theorem~\ref{thm:asymptotic_validity} for every $P\in\mathfrak P_{0,q}$. We define $m_n$ to be a deterministic sequence, which is common to all distributions in this class.

Based on this, we strengthen the calibration, recurrence, selection-stability, and replay-growth conditions as follows.

\begin{assumption}[Uniform calibration]
\label{asspt:uniform_calibration}
The fixed-analysis p-value satisfies
\[
\kappa_n
:=
\sup_{P\in\mathfrak P_{0,q}}
\max_{k\in\{n,m_n\}}
\sup_{u\in[0,1]}
\left|F^P_{q,k}(u)-u\right|
\to0
\quad\text{as }n\to\infty.
\]
\end{assumption}
\begin{assumption}[Uniform recurrence]
\label{asspt:uniform_recurrence}
There exist $\rho_\star>0$ and an integer $n_0\ge1$ such that, for every $n\ge n_0$,
\[
\inf_{P\in\mathfrak P_{0,q}}
\min_{k\in\{n,m_n\}}
\rho^P_{q,k}
\ge \rho_\star.
\]
\end{assumption}

\begin{assumption}[Uniform selection stability]
\label{asspt:uniform_lambda_stability}
The selection multipliers satisfy one of the following conditions:
\begin{enumerate}[label=(\alph*)]
\item \textit{One-sided stability.} For every fixed $u\in(0,1]$,
\[
\sup_{P\in\mathfrak P_{0,q}}
\left[\lambda^P_{q,n}(u)-\lambda^P_{q,m_n}(u)\right]_+
\to0.
\]

\item \textit{Two-sided stability.} For every fixed $u\in(0,1]$,
\[
\sup_{P\in\mathfrak P_{0,q}}
\left|\lambda^P_{q,n}(u)-\lambda^P_{q,m_n}(u)\right|
\to0.
\]
\end{enumerate}
\end{assumption}

\begin{assumption}[Uniform replay growth]
\label{asspt:uniform_replay_growth}
Let $B_{s,n}$ denote the number of completed replays in stratum $s\in\{0,1\}$, with $B_{s,n}=0$ when no replay is completed. For every fixed integer $M\ge1$,
\[
\sup_{P\in\mathfrak P_{0,q}}
\Pr_P\!\left(
\min_{s\in\{0,1\}}B_{s,n}<M
\,\middle|\,Q_n=q
\right)
\to0.
\]
\end{assumption}

These conditions are a direct strengthening of the assumptions required for Theorem~\ref{thm:asymptotic_validity} in that they hold uniformly over the chosen class of null distributions. As is standard for uniformity results, these assumptions are most plausible on a restricted class of null laws for which the inferential procedure satisfies uniform regularity conditions, and under which the features of $P$ that affect the agent's selection behavior are well separated relative to the sampling error, as discussed below. Under these assumptions, we have the following uniformity result.

\begin{corollary}[Uniform selective validity and exactness]
\label{cor:uniform_validity}
Suppose Assumptions~\ref{asspt:sampling_units} and~\ref{asspt:trust} hold for every $P\in\mathfrak P_{0,q}$, and use the subsampling and within-stratum replay design of Theorem~\ref{thm:asymptotic_validity}. Let $m_n$ be a deterministic sequence, common to all $P\in\mathfrak P_{0,q}$, satisfying
\[
m_n\to\infty,\quad m_n/n\to0.
\]
Let $\widehat p^{\mathrm{sel}}_{q,n}$ be the estimator of Theorem~\ref{thm:asymptotic_validity}, with the same failure convention.

Then, under Assumptions~\ref{asspt:uniform_calibration}, \ref{asspt:uniform_recurrence}, \ref{asspt:uniform_lambda_stability}(a), and \ref{asspt:uniform_replay_growth},
\[
\sup_{P\in\mathfrak P_{0,q}}
\sup_{\alpha\in[0,1]}
\left[
\Pr_P\!\left(
\widehat p^{\mathrm{sel}}_{q,n}\le\alpha
\,\middle|\,Q_n=q
\right)-\alpha
\right]_+
\to0.
\]
If, in addition, Assumption~\ref{asspt:uniform_lambda_stability}(b) holds, then
\[
\sup_{P\in\mathfrak P_{0,q}}
\sup_{\alpha\in[0,1]}
\left|
\Pr_P\!\left(
\widehat p^{\mathrm{sel}}_{q,n}\le\alpha
\,\middle|\,Q_n=q
\right)-\alpha
\right|
\to0.
\]
\end{corollary}

Assumption \ref{asspt:uniform_calibration} on uniform calibration requires the fixed-analysis p-value to be asymptotically uniform at both sample sizes, with an approximation error that vanishes \emph{uniformly} over the class of null distributions. Assumption \ref{asspt:uniform_recurrence} on uniform recurrence requires a common positive lower bound on all selection probabilities within this class. Note in particular that positive selection probabilities for each fixed law do not guarantee such a bound, since the selection probability could tend to zero along a sequence of laws in the class. Finally, Assumption \ref{asspt:uniform_replay_growth} on uniform replay growth requires enough completed replays in both strata, also uniformly over the class. These assumptions are natural uniform analogues of their pointwise counterparts and impose relatively mild requirements on the data-generating process and the replay design.

Assumption \ref{asspt:uniform_lambda_stability} is a substantive restriction. It addresses a different problem: whether the correction learned from subsamples stays appropriate for the full sample. The challenge is that the null for $q$ can leave the effects of competing analyses and nuisance coefficients that influence the agent's choice completely unrestricted. In regular settings, an effect of order $n^{-1/2}$ can influence selection at size $n$, while its standardized signal at size $m_n$ vanishes as $\sqrt{m_n/n}\to0$. In other words, the same dilution that removes the trace of the original selection on subsamples (Proposition \ref{prop:subsampling_selection}) can also erase signals that are relevant to selection. This means that validity for every \emph{fixed} distribution can coexist with persistent size distortions along sequences of null distributions. This is a known limitation of subsampling \citep[Theorem~2.1]{andrews2010asymptotic,romano2012uniform}.

Intuitively, Assumption~\ref{asspt:uniform_lambda_stability} requires the subsample correction to be large enough to account for selection in the full sample. More precisely, at every fixed p-value threshold, the subsample multiplier must be at least as large as the full-sample multiplier, up to an error that vanishes uniformly over the null-law class. A larger correction can make the test conservative, which is allowed by the validity conclusion of Corollary~\ref{cor:uniform_validity}. If the two multipliers converge to each other uniformly, the corollary instead gives rejection probabilities that converge uniformly to the nominal level.

This still allows the agent to use the data to choose an analysis, but the correction obtained at the smaller sample size has to adequately account for that choice at the full sample size. This does \emph{not} follow simply because the individual tests are valid, or because the agent considers only true null hypotheses. Increasing the number of replays estimates the subsample correction more accurately, but cannot fix a mismatch between the corrections needed at the two sample sizes.

Assumption~\ref{asspt:uniform_lambda_stability} is plausible when the features of $P$ that affect the agent's selection behavior are well separated relative to their sampling error. It also holds, for example, if the agent responds only to $t$-statistics or $p$-values whose joint distribution is asymptotically invariant to the sample size.

\begin{remark}[Uniformly valid inference procedures]
Failure of Assumption~\ref{asspt:uniform_lambda_stability} does not rule out uniformly valid inference using other approaches. The main challenge with adopting existing approaches \citep[e.g.,][]{andrews2009hybrid,mccloskey2017bonferroni} is that it is unknown which nuisance parameters matter to the agent.
\end{remark}

\subsection{Power}\label{sec:replay_power}

Next, we characterize the asymptotic power of the corrected test implied by the corrected p-value $\widehat p_{q,n}^{\mathrm{sel}}$. For $\alpha\in(0,1)$, define the corrected test and its oracle counterpart as $\phi_{n,\alpha}:=\mathbf 1\{\widehat p_{q,n}^{\mathrm{sel}}\le\alpha\}$ and $\phi^{\mathrm{or}}_{n,\alpha}:=\mathbf 1\{H_{q,n}^{P_0}(U_{q,n})\le\alpha\}$, respectively, where $H_{q,n}^{P_0}$ is the oracle cdf under a null law $P_0\in\mathcal P_{0,q}$ and, by convention, $\phi_{n,\alpha}=0$ whenever the correction is unavailable. We are interested in the power of this test, conditional on the agent selecting $q$, that is, $\Pr_{P}(\phi_{n,\alpha}=1\mid Q_n=q)$.

We need one additional assumption for consistency, which ensures that both strata contain at least one replay. If this assumption does not hold, the proposed correction is unavailable and the test does not reject.

\begin{assumption}[Availability of replay observations]
\label{asspt:replay_availability}
Along the sequence of alternative laws $P_n$ under consideration, let $B_{s,n}$ denote the number of completed agent replays in stratum $s\in\{0,1\}$, with $B_{s,n}=0$ when no replay is available in that stratum. We assume that
\[
\Pr_{P_n}\!\left(
B_{1,n}\geq 1,\; B_{0,n}\geq 1
\mid Q_n=q
\right)\longrightarrow 1.
\]
\end{assumption}

We can now state the following power and consistency result.

\begin{proposition}[Oracle power and consistency]\mbox{}
\label{prop:replay_power}
\begin{enumerate}[label=(\alph*)]
\item \textit{Local alternatives.}
Fix a null distribution $P_0\in\mathcal P_{0,q}$. Suppose all conditions of Theorem~\ref{thm:asymptotic_validity} hold under $P_0$, with the two-sided selection-stability condition in Assumption~\ref{asspt:lambda_stability}(b) replacing the one-sided condition in Assumption~\ref{asspt:lambda_stability}(a).
Let $P_n^n$ be contiguous to $P_0^n$ and suppose $\liminf_n\Pr_{P_n}(Q_n=q)>0$.
Let $H_{q,n}^{P_0}$ denote the oracle cdf under $P_0$. Then, for every $\alpha\in(0,1)$,
\[
\Pr_{P_n}\!\left(\phi_{n,\alpha}\neq\phi^{\mathrm{or}}_{n,\alpha}\mid Q_n=q\right)\longrightarrow0.
\]

\item \textit{Consistency.}
Under any sequence of alternatives satisfying Assumption \ref{asspt:replay_availability}, suppose that,
conditional on $Q_n=q$,
\[
\frac{U_{q,n}}
{\widehat a_{0,q}(U_{q,n})}
\stackrel{p}{\longrightarrow}0,
\]
where the ratio is defined as $\infty$ whenever
its denominator is zero. Then
\[
\widehat p_{q,n}^{\mathrm{sel}}
\stackrel{p}{\longrightarrow}0
\]
conditional on $Q_n=q$, or equivalently,
\[
\Pr_{P_n}\!\left(\phi_{n,\alpha}=1\mid Q_n=q\right)\longrightarrow1
\qquad\text{for every }\alpha\in(0,1).
\]
\end{enumerate}
\end{proposition}

Part (a) shows that under contiguous alternatives, the test $\phi_{n,\alpha}$ based on the estimated selective p-value $\widehat p_{q,n}^{\mathrm{sel}}$ has the same asymptotic power as the corresponding ``oracle'' test $\phi^{\mathrm{or}}_{n,\alpha}$. Part (b) shows that $\phi_{n,\alpha}$ inherits consistency from the original p-value as long as both strata contain replays and the agent does not exclusively select the original analysis on subsamples in the favorable tail as the data grows. The latter condition is a version of a recurrence condition, imposed under the alternative on the less favorable subsamples in the body stratum.

\subsection{Confidence intervals under finite replay budget}
\label{sec:replay_uncertainty}
Theorem~\ref{thm:asymptotic_validity} establishes asymptotic validity of the corrected p-value as the number of replays diverges. Since, in practice, the number of replays is determined by a finite computational budget, we now construct confidence intervals for $H_{q,n}(u_n)$ and $\lambda_{q,n}(u_n)$ that are valid for such a fixed replay budget. The idea is that, conditional on the data, the number of replays that reselect $q$ in each stratum is distributed as a binomial, so we can construct an exact finite-sample confidence interval for the correction at the subsample level by leveraging classical results on confidence intervals for binomial proportions \citep{clopper1934use, berger1994p}. We then show that the stability Assumption \ref{asspt:lambda_stability} transfers these confidence intervals to the full-sample targets as well.

Write $u_n=U_{q,n}\in(0,1)$ and $m=m_n$. Define the stratum selection probabilities under the subsampling law given $D_n$,
\[
b_{s,n}:=\Pr\!\left(\mathcal A_r(D_{S_m},\Xi')=q\ \middle|\ D_n,\ \mathbf 1\{p_q(D_{S_m})\le u_n\}=s\right),
\qquad s\in\{0,1\},
\]
and, for $a=(a_1,a_0)\in[0,1]^2$ with $ua_1+(1-u)a_0>0$,
\[
g_u(a_1,a_0)=\frac{u\,a_1}{u\,a_1+(1-u)\,a_0},
\qquad
w_n:=g_{u_n}(b_{1,n},b_{0,n}).
\]
Then, $\widehat p^{\mathrm{sel}}_{q,n}=g_{u_n}(\widehat a_{1,q}(u_n),\widehat a_{0,q}(u_n))$ estimates $w_n$, which approximates $H_{q,n}(u_n)$ under Assumption~\ref{asspt:lambda_stability}. Let $X_s=B_s\widehat a_{s,q}(u_n)$ denote the number of replays that select $q$ in stratum $s$. Note that $g_u$ depends on $a$ only through the ratio $a_1/a_0$ and is increasing in it.

Conditional on $D_n$, and for a fixed number of replay counts, the subsamples replayed within a stratum are i.i.d.\ draws from the stratum's subsampling law and each replay uses fresh agent randomness $\Xi'$, so $X_1$ and $X_0$ are independent with $X_s\sim\operatorname{Bin}(B_s,b_{s,n})$ under Assumption \ref{asspt:trust}. The problem is therefore to bound $g_{u_n}(a_1,a_0)$, an increasing function of the ratio $a_1/a_0$, from two independent binomial counts. The classical formulations of this problem \citep{chan1999test, agresti2001small, fay2021practical} take the binomial counts as given. Lemma~\ref{lem:exact_ci} shows that, conditional on the data, our replays on subsamples from that data reduce to the classical problem, so that any valid interval for $a_1/a_0$ covers the correction $w_n$. Theorem~\ref{thm:replay_confidence} then transfers this coverage to the oracle selective $p$-value for the full sample, controlling the extrapolation from $m_n$ to $n$ and the conditioning on the selected dataset.

For $t\in[0,1]$ let $\Theta_-(t)=\{a:g_{u_n}(a)\le t\}$ and $\Theta_+(t)=\{a:g_{u_n}(a)\ge t\}$, let $S_t(X)$ be the signed root of the likelihood-ratio statistic for the constraint $g_{u_n}(a)=t$ (Appendix~\ref{app:ci_definition}), so that large values are evidence against $\Theta_-(t)$ and small values against $\Theta_+(t)$, and let $C(X)$ be the product of Clopper--Pearson intervals for $a_1$ and $a_0$ at level $1-\beta/2$, where $1-\gamma$ is the nominal coverage of the interval and $\beta \in (0,\gamma/2)$ is a small error margin. In practice, we use $\beta=\gamma/20$. Then, define the one-sided p-values,
\begin{align*}
p_-(t;X) & =\min\Big\{1,\ \beta+\sup_{a\in C(X)\cap\Theta_-(t)}\Pr_a(S_t(X')\ge S_t(X))\Big\},
\\
p_+(t;X)& =\min\Big\{1,\ \beta+\sup_{a\in C(X)\cap\Theta_+(t)}\Pr_a(S_t(X')\le S_t(X))\Big\},
\end{align*}
where $\Pr_a$ is the product law of independent $\operatorname{Bin}(B_1,a_1)$ and $\operatorname{Bin}(B_0,a_0)$ counts. $X=(X_1,X_0)$ denotes the observed counts and $X'=(X_1',X_0')$ a hypothetical replicate, so that $\Pr_a(S_t(X')\ge S_t(X))$ is the probability, under the null that the stratum selection probabilities were $a$, of a statistic at least as large as the one observed. Then, the confidence bounds are the endpoints of the hull of the acceptance sets,
\begin{equation}\label{eq:ci_bounds}
\underline p_n=\inf\bigl(\{1\}\cup\{t:p_-(t;X)\ge\gamma/2\}\bigr),
\qquad
\overline p_n=\sup\bigl(\{0\}\cup\{t:p_+(t;X)\ge\gamma/2\}\bigr).
\end{equation}
The null $g_u\left(a_1, a_0\right) \leq t$ constrains the ratio $a_1 / a_0$, but does not determine the individual selection probabilities $a_1$ and $a_0$. Since the distribution of the replay counts depends on both probabilities, we obtain a valid $p$-value by maximizing the tail probability over all pairs of parameters that are consistent with the null. However, some of the pairs $a_1, a_0$ in the null set could not plausibly have produced the observed counts.  \citet{berger1994p} propose to address this by maximizing only over a $1-\beta$ confidence rectangle $C(X)$ around the observed frequencies and correct for the induced error by adding $\beta$ to the resulting p-value. Then, the one-sided error is at most $\beta+(\gamma/2-\beta)=\gamma/2$.

 We write
\[
\mathcal I^p_{n,B}=[\underline p_n,\overline p_n],
\qquad
\mathcal I^\lambda_{n,B}=\mathcal I^p_{n,B}/u_n,
\]
with the convention $\mathcal I^p_{n,B}=[0,1]$ and $\mathcal I^\lambda_{n,B}=[0,1/u_n]$ when a stratum has no replay, or when $X_1=X_0=0$.

\begin{lemma}[Finite-sample validity]\label{lem:exact_ci}
Suppose Assumptions~\ref{asspt:sampling_units} and~\ref{asspt:trust} hold, let $B_1,B_0\ge1$ be fixed given $D_n$, and suppose that both strata have positive probability under the subsampling law given $D_n$. Then, for every $a\in[0,1]^2$ with $u_na_1+(1-u_n)a_0>0$,
\begin{equation} \label{eq:finite-sample-validity}
\Pr_a(\underline p_n>g_{u_n}(a))\le\gamma/2,
\qquad
\Pr_a(\overline p_n<g_{u_n}(a))\le\gamma/2,
\end{equation}
and consequently, on the event $u_nb_{1,n}+(1-u_n)b_{0,n}>0$,
\begin{equation} \label{eq:finite-sample-coverage}
\Pr(w_n\in\mathcal I^p_{n,B}\mid D_n)\ge1-\gamma .
\end{equation}
\end{lemma}

The bounds are defined for any counts, including $X_0=0$ or $X_1=0$, and are computed only at the observed counts and so remain computationally tractable at large replay budgets. Moreover, for $\gamma\le1/2$, the interval always contains the point estimate $\widehat p^{\mathrm{sel}}_{q,n}$ (see the proof of Theorem~\ref{thm:replay_confidence}). Shorter admissible finite-sample valid intervals for this ratio exist \citep{wang2015exact}, but their computational cost scales with the full sample space instead of the observed counts and only led to modest reductions in interval width in our experiments, so we do not use them.
We obtain the following result.

\begin{theorem}[Full-sample confidence intervals]
\label{thm:replay_confidence}
Fix $q$ and $P\in\mathcal P_{0,q}$. Suppose Assumptions~\ref{asspt:sampling_units}, \ref{asspt:trust}, \ref{asspt:base_calibration}, and~\ref{asspt:recurrence} hold, $m_n\to\infty$ with $m_n/n\to0$, and $B_1,B_0\ge1$ are fixed integers.
\begin{enumerate}[label=(\alph*)]
\item Under Assumption~\ref{asspt:lambda_stability}(a),
\[
\liminf_{n\to\infty}\Pr_P\!\left(
H_{q,n}(u_n)\le\overline p_n,\
\lambda_{q,n}(u_n)\le\overline p_n/u_n
\mid Q_n=q\right)\ge1-\gamma/2 .
\]
\item Under Assumption~\ref{asspt:lambda_stability}(b),
\[
\liminf_{n\to\infty}\Pr_P\!\left(
H_{q,n}(u_n)\in\mathcal I^{p}_{n,B},\
\lambda_{q,n}(u_n)\in\mathcal I^{\lambda}_{n,B}
\mid Q_n=q\right)\ge1-\gamma .
\]
\item Under Assumption~\ref{asspt:lambda_stability}(b), for every $\varepsilon>0$,
\begin{equation}
\label{eq:replay_ci_precision}
\lim_{\min\{B_0,B_1\}\to\infty}\
\limsup_{n\to\infty}\
\Pr_P\!\left(
\max\big\{
\operatorname{diam}\mathcal I^{p}_{n,B},
\operatorname{diam}\mathcal I^{\lambda}_{n,B}
\big\}>\varepsilon
\mid Q_n=q
\right)=0 .
\end{equation}
\end{enumerate}
\end{theorem}

Part (a) gives an upper bound on the selection-corrected $p$-value and the multiplier. As such, it can be used to conservatively reject the null at level $\alpha$ whenever $\overline p_n\le\alpha-\gamma/2$, for $\gamma<2\alpha$, which accounts for both selection and the finite number of replays.\footnote{\label{fn:replay_test_level}Note that rejecting when $\overline p_n\le\alpha$ yields a test with asymptotic level at most $\alpha+\gamma/2$ for a fixed number of replays; by part (c) and Theorem~\ref{thm:asymptotic_validity}, its asymptotic level approaches $\alpha$ as the number of replays grows.} Part (b) gives two-sided intervals. The width of these intervals only reflects uncertainty due to the finite number of replays. Approximation errors from subsampling at size $m_n$, from conditioning on the selected dataset, and from finite-sample calibration of the fixed-analysis $p$-value are shown to vanish as $n\to\infty$ in the proof, but are not reflected in the width, which is why the results are asymptotic in $n$. Part (c) shows that these intervals shrink to zero width as the replay budget grows, which we corroborate in our simulations below.

\begin{remark}[Data-dependent replay budgets]\label{rem:adaptive_ci}
Lemma~\ref{lem:exact_ci} treats the replay budget as being fixed in advance. In practice, one may want to allocate replays to the two strata based on a pilot, as in the Neyman allocation described in  Section~\ref{sec:selective_pvalue} above. Similarly, one may want to keep adding replays until $\overline p_n$ falls below some significance level $\alpha$. The first practice is a data-dependent allocation, whereas the second is a form of optional stopping. Both can invalidate the fixed-budget guarantee. For the first, one can restore validity by simply discarding the initial pilot draws and only using the post-pilot draws to compute the confidence bands, which we do in practice. For the second approach, one could instead construct time-uniform confidence sequences for the two selection probabilities \citep{howard2021time, waudbysmith2024estimating}.
\end{remark}

\section{Simulations}\label{sec:simulation_results}

We now implement four simulation designs to study whether our selection correction indeed restores correct coverage in practice. In the first design, the selection mapping is known --- and hence not driven by coding agents' selection --- to allow us to study the true coverage of our selection correction; in the other three, the selection is driven by open-ended coding agents, where we can compare the frequency of the original and corrected p-values but do not know the oracle correction or selection mapping.

\subsection{Implementation} \label{sec:implementation}

Throughout the simulation and application sections, we use GPT-6-Luna as the base model through the Codex CLI at high reasoning effort, unless stated otherwise.

All simulation designs and applications follow the procedure of Section~\ref{sec:estimator_summary}.  Each replay starts from an identical copy of the original agent's workspace. The workspace is a folder inside a Linux machine, with Python and R environments, access to the command line, file creation and editing abilities through the Codex CLI, and limited internet access --- no web-search tool and explicit instructions to only use the internet to read technical documentation. Each agent is provided with an initial prompt that varies with the specific simulation design and application, see Section \ref{app:prompts} for the full collection of prompts. The only thing that is modified across replays is the dataset, which is replaced by a randomly drawn subsample, and any corresponding statistics. In particular, we also modify any numerical information in the codebook about the data, like descriptions about the sample size. Indeed, in initial experiments, we found that agents noticed discrepancies between the codebook and the subsampled data and spent some of their analysis trying to reconcile them. This aligns with our arguments above that coding agents need to be replayed on ``genuine'' resampled datasets.
The agents only have access to their folder on the machine they run on and cannot see other agents or other folders, so that nothing in the agents' environment reveals that they are being replayed. Replays that were refused by the model provider were discarded unread and rerun on the same subsample.

\subsection{Simulation Designs} \label{sec:simulation_designs}

\paragraph{Fixed design.} For the known selection mapping, we simulate a data-generating process similar to Example \ref{ex:stochastic_selection}. In particular, we draw $400$ datasets with $n=1{,}000$ i.i.d. observations of six independent outcomes drawn from a standard normal distribution. From each outcome's observations, we then compute the sample mean and construct a z-score using the known standard error to test whether the outcome $j$ has a mean equal to zero. The corresponding $p$-values are independent and uniformly distributed under the null. We model the selection mapping as in Example \ref{ex:stochastic_selection} as picking the smallest $p$-value with probability 0.8 and picking a random one with probability 0.2. The exact selected cdf for a nominal p-value of $u$ then is,
\begin{equation} \label{eq:fixed_exact_H}
H_q(u)=0.2u+0.8\{1-(1-u)^6\},
\end{equation}
which allows us to compute the true selection-corrected p-value.

Based on this DGP, we apply the correction procedure of Section \ref{sec:estimator_summary} in the same way as we would to a coding agent. We draw subsets of $m=31$ observations, evaluate the selected test on each, split them into tail and body using the selected test's $p$-value calculated on the full data, and replay the selector on $B$ subsets per stratum, for $B \in \{25,100, 400, 1000\}$. A match is then declared if the replayed selector picks the same analysis $j$ as the ``original'' selector on the full data.

\paragraph{Open-ended agent design.} To get a better sense of how well our method performs with real coding agents, we simulate 100 datasets from an imaginary randomized control trial for an employment support program. The coding agent is given a codebook with details on the data and experimental design, which describes the following. The program has $2{,}000$ participants at four sites, with half of each site's participants assigned to the program at random. We simulate the following baseline characteristics: baseline age, sex, parenthood, education, household size, employment, earnings, and months unemployed. Additionally, we simulate the following outcome variables, each ``measured'' 6 and 12 months after the program: employment, weekly hours worked, monthly earnings, a financial-security index, food insecurity, debt in collections, psychological distress, self-rated health, and training enrollment. The full system of equations that we used to generate this data is given in Appendix \ref{app:simulation_details}. Treatment does not affect any of the outcome variables, so all program effects are exactly equal to zero. This allows us to calculate rejection rates under the null hypothesis, even if we cannot characterize the (agents') selection mapping as we can in the fixed design above.

The agents receive an open-ended prompt (Section \ref{app:prompts}) asking them to provide a team with a ``compelling, defensible finding to motivate its next study'' and encouraging the agent to ``explore the data as you see fit''. The codebook provided to the agent broadly specifies the design (least squares of an outcome on the treatment with robust standard errors) but leaves the choice of outcome, follow-up, sample restrictions, and precise regression specification open to the agent.

For each of the 100 datasets, we run the Luna agent once on the full dataset to get the original analysis. Then, we run replay agents on tail and body subsamples of the dataset of size $m=200$, giving $m/n=0.1$. We draw $40+200$ and $240+400$ tail and body subsamples respectively, to see the impact of increasing the replay size in this setting.

\paragraph{Real data designs.} One issue with simulated data is that its stylized nature might cause particular types of selection behavior that real data would not. To assess this, we also test the agents on data from real papers published in a top economics journal. The papers study: 1) a randomized evaluation of a workplace wellness program's effects on employee health and productivity \citep[][]{JonesMolitorReif2019}; 2) a randomized evaluation of the effect of a financial information intervention on mothers' labor supply \citep{CostaRamonSchaedeSlotwinski2026}; and 3) an online survey experiment testing how consumers make ethical purchasing decisions \citep[][]{KaufmannAndreKoszegi2024}. For the first two papers, we permute the randomly assigned treatment, similar to a randomization test, to make the null hypothesis of no effect hold true, while holding the other aspects of the data fixed so that it closely resembles the real data. For the third paper, we create a new column in the data given to the coding agent with an imaginary randomized intervention that was not actually implemented, which should similarly have null effects. We chose these papers based on the criteria that they had to be recently published in a top-5 economics journal, have at least $2{,}000$ participants with sampling at the individual level (to simplify subsampling), have publicly available replication data, and evaluate a non-canonical experiment (e.g. not the Rand Health study) to mitigate the AI model having memorized some of the results (see also the discussion in Section \ref{sec:card_krueger} below). The prompt given to the coding agent is nearly identical to the open-ended simulation design above, just dropping the name of the program, so that one prompt serves all three studies (see Section \ref{app:prompts}). The codebook contains details on the experimental design of the paper, the variables in the dataset including design variables and outcomes, and a broad description of the intended methods: OLS regressions of an outcome on the treatment indicator and group dummies with robust standard errors. We removed anything that could identify the studies, including proper nouns and open-text answers from study participants, renamed all variables, and changed the study years. Since memorized study results might influence selection, we also checked the logs of all 24,542 original and replay sessions for any mention of the study authors, title, or title paraphrases, but did not find any. Finally, while all three studies had a pre-registration plan, we do not include those in the codebook since it would obstruct the selective behavior that we are trying to model.

For each of the 3 studies, we draw 34 fresh datasets with the treatment permuted. Then, we run the Luna agent on each of the 102 full datasets at their original size, which we call the ``original'' analyses. For each original analysis, we then draw 40 tail and 200 body subsample datasets from the permuted fresh dataset with a relative subsample size of $m/n = 0.1$, replay the Luna agent on them, and estimate the corrected $p$-value and $\widehat \lambda$ together with their confidence bands.

\paragraph{Power design.} Finally, we test the replay test's power to reject a false null by injecting ``true'' treatment effects into 200 new simulated randomized control trials from the open-ended agent design above, 100 for each effect size. In particular, we modify the design to let the treatment (the employment support program) increase employment 12 months after the program by 6.2 and 9.6 percentage points in two sets of simulations, to see how our test's power increases with effect size. The effect sizes were chosen to target 80\% and 99\% power for a pre-specified test on a regression coefficient of the treatment with employment as the outcome. They correspond to a Cohen's d (effect over outcome standard deviation) of 0.127 and 0.190, respectively. The employment effect spills over into three other outcomes at 12 months: weekly hours, monthly earnings, and training enrollment. All other outcomes continue to be unaffected by treatment. We compare the rejection frequency of the replay test under this design against the rejection frequency of the pre-specified test, as well as the average rejection frequency of each ``original'' agent's program on 4,000 new datasets drawn with the same effect.

\subsection{Results}

\begin{figure}[!ht]
\centering
\includegraphics[width=\linewidth]{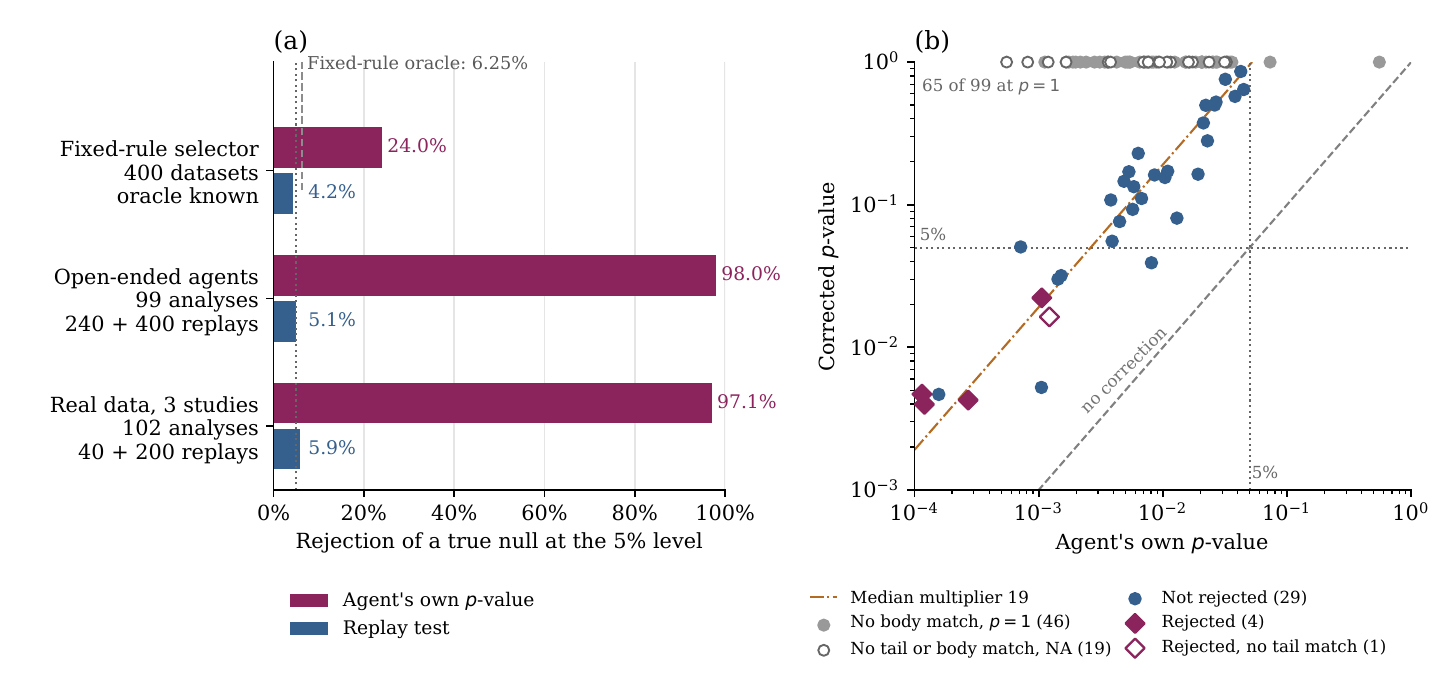}
\caption{Replay correction under the null}
\floatfoot{\textit{Notes:} Panel (a) plots the share of true nulls rejected at the 5\% level by the $p$-value reported by the agent/selector against those rejected by our replay test, which rejects when the one-sided 95\% upper bound for the corrected $p$-value lies below 0.05, in 1) the fixed design (400 datasets, 100 replays per stratum, the dashed line indicates the exact oracle correction); 2) the open-ended design (99 datasets on which a coding agent chooses its own model and null, on subsamples with $m/n=0.10$) with 240 tail and 400 body replays; and 3) the real data design (102 datasets from three studies with a placebo treatment, $m/n=0.10$, 40 tail and 200 body replays). Table~\ref{tab:simulation_interval_performance} gives the full simulation results. Panel (b) plots the corrected $p$-value of the open-ended agent analysis (simulated data) against the agent's own reported $p$-value. The dashed diagonal corresponds to no correction (reported = corrected $p$-value), and the dash-dotted line marks the median multiplier of 19 over the analyses re-selected in both strata. Analyses never re-selected on either body or tail have no point estimate and are shown at $p=1$ by convention with hollow circles. Those re-selected on tail but on no body subsample have a point estimate of $p=1$ and are indicated by filled gray circles. Diamonds mark the analyses the replay test rejects; the hollow diamond is an analysis with a body match but no tail match, which can still be rejected based on the upper endpoint of the confidence band.}
\label{fig:simulation_overview}
\end{figure}

Table \ref{tab:simulation_interval_performance} collects the results of all four simulation designs, and Figure \ref{fig:simulation_overview} summarizes the rejection rates of all designs (panel a) and plots the corrected against the reported p-values for the open-ended design (panel b). Without correction, the agents strongly inflate the false rejection rates: reporting a $p$-value of less than $0.05$ for nearly every original analysis. These strikingly high rejection rates occur under relatively mild selection pressure (``the team needs one compelling, defensible finding'') even as the original prompts specifically encourage the agent to ``explain your choice'' and ``discuss the exploratory nature of the finding''.

Panel a in the figure shows that the replay tests correct these rates to close to the nominal level in each design: 4.2\% for the fixed design, 5.1\% for the open-ended agents, and 5.9\% for the real data design pooled across the three papers. The replay test rejects the null when the upper endpoint of the one-sided 95\% confidence band around the corrected $p$-value lies below 0.05. Since we use many replays per analysis, we compare its rejection rates with the nominal 5\% level, which its level approaches as the number of replays grows (footnote~\ref{fn:replay_test_level}).

In Figure \ref{fig:simulation_overview}, panel (b), we plot the agent's reported $p$-values against our corrected ones for the open-ended design. The gray dashed diagonal is the 45-degree line and indicates no correction. The dash-dotted diagonal indicates the correction corresponding to a multiplier $\widehat \lambda$ of 19, our median estimated multiplier. Red symbols indicate corrected $p$-values that still reject the null, whereas blue ones do not reject. The mass of white and gray circles at the top of the panel indicates simulations in which there were either no matches to the original analysis in the tail replays, in the body replays, or in either. For simulations where there are only tail matches, the corrected $p$ is $1$ indicating maximal estimated selection. For simulations with only body matches, the point estimate for the corrected $p$ is zero but the upper bound of the confidence band can, and so a statistical decision is still possible, as in the hollow red diamond on the plot. In 19 out of 99 analyses, there is neither a tail nor a body match, in which case there is no point estimate and the analysis is shown at $p=1$.

Looking more closely at the fixed design in Panel A in Table \ref{tab:simulation_interval_performance}, the estimator (computed by simply replaying the selection rule on the subsamples) reaches the same rejection decision as the oracle correction in 96.2\% of simulations with $25$ replays, increasing to 99.8\% with $1{,}000$ replays. The root mean squared error (RMSE) of the corrected $p$-value relative to the known truth decreases at roughly $\sqrt{B}$ rate with $B$ the number of replays, falling to $0.019$ at $1{,}000$ replays, and the mean width of the 95\% confidence intervals decreases to $0.069$. The confidence bands' coverage of the known oracle correction also decreases with the number of replays, with some slight undercoverage at the larger replay budgets, which likely reflects asymptotic approximation error. The median multiplier $\widehat \lambda$ remains stable around $3.5$ with shrinking confidence bands. Overall, the theory of Section \ref{sec:replay_uncertainty} appears to be supported by these results.

Panels B--D in Table \ref{tab:simulation_interval_performance} demonstrate the performance of the estimator when run with a real, frontier coding agent on realistic, open-ended tasks. We ran the replay on 100 datasets in Panel B and on 100 datasets per effect size in Panel D; in each set, the original agent's code produced an error on one dataset, so we dropped it and report the results from the remaining 99. Panel B, row 2, shows that the correction achieves a near-nominal false rejection rate with enough replays (5.1\%) even when the agent engages in near-total selection of significant results, rejecting in 97/99 (98.0\% rounded) of simulations. The ``Width'' column gives the average width of the two-sided 95\% confidence bands around the $p$-value across simulations. The fairly large width of $0.71$ partly reflects the fact that we include simulations with only tail matches, whose confidence bands include $1$ as their upper endpoint. When dropping those simulations, the mean width drops to $0.44$ at 640 replays. Even so, the confidence bands are much wider in Panel B than in Panel A, reflecting two things. One, the higher variance of the selection choices made by actual coding agents on an open-ended task. Two, the extreme selection the agents engage in: they rarely reselect the original analysis when it does not deliver a significant $p$-value (only 0.29\% of body replays). Nonetheless, these wide bands do not prevent our estimator from detecting statistically significant evidence of strong selection, with median multipliers of $17$--$19$ (lower confidence bounds $6.9$--$7.2$).

\begin{table}[hb!]
\centering
\caption{Simulation results: replay correction}
\floatfoot{\textit{Notes:} table shows rejection frequencies of the agent's reported $p$-values against those of the replay test in the four simulation designs. The replay test rejects when the one-sided 95\% upper bound for the corrected $p$-value (Section~\ref{sec:replay_uncertainty}) is below 0.05. \textit{Re-selected}: share of tail and body replays choosing the original analysis, pooled across different analyses. \textit{Mean width}: width of two-sided 95\% interval for the corrected $p$-value, averaged over analyses with a tail match. \textit{$\widehat\lambda$ columns}: computed on simulations with analyses re-selected in both strata. ``95\% interval'' gives median confidence upper and lower endpoints. See Section \ref{sec:simulation_designs} for a description of the four designs. \textit{Panel A}: ``fixed design''. Decision, coverage, and RMSE are relative to the exact selective $p$-value from Eq. \eqref{eq:fixed_exact_H}. \textit{Panel B}: ``open-ended agent design''. \textit{Panel C}: ``real data designs''. The design of Panel B's first row on three real studies with a placebo treatment, 34 datasets each (24,440 replays). \textit{Panel D}: design from panel B but with a true effect injected on employment at 12 months, in percentage points. \textit{Pre-specified test}: power of a test known in advance that regresses employment at 12 months on the treatment indicator. \textit{Agent's analysis, new data}: share of 4,000 new simulated datasets, drawn with the same effect, on which the agent's program rejects, averaged over analyses; with no effect the same programs reject 4.1 to 5.9\% of new datasets. All numbers rounded to first decimal.}
\label{tab:simulation_interval_performance}
\footnotesize
\setlength{\tabcolsep}{3pt}
\begin{tabular*}{\linewidth}{@{\extracolsep{\fill}}lccccccc@{}}
\toprule
\multicolumn{8}{@{}l}{\rlap{\textit{Panel A. Fixed-rule selector: 400 datasets, $n = 1{,}000$, $m = 31$, replays per stratum}}} \\
\addlinespace[2pt]
 & \multicolumn{5}{c}{Replay test and corrected $p$-value} & \multicolumn{2}{c}{Multiplier $\widehat\lambda$} \\
\cmidrule(lr){2-6}\cmidrule(l){7-8}
Replays & Reject 5\% & Same as oracle & Coverage & RMSE & Width & Median & 95\% interval \\
\midrule
25 & 2.5\% & 96.2\% & 97.2\% & 0.153 & 0.439 & 3.4 & [1.8, 5.5] \\
100 & 4.2\% & 98.0\% & 94.0\% & 0.065 & 0.218 & 3.5 & [2.5, 4.6] \\
400 & 5.8\% & 99.5\% & 93.5\% & 0.031 & 0.109 & 3.5 & [3.0, 4.1] \\
1,000 & 6.0\% & 99.8\% & 92.0\% & 0.019 & 0.069 & 3.6 & [3.2, 4.0] \\
\end{tabular*}

\medskip

\begin{tabular*}{\linewidth}{@{\extracolsep{\fill}}lccccccc@{}}
\multicolumn{8}{@{}l}{\rlap{\textit{Panel B. Open-ended agents, simulated data: 99 datasets, $n = 2{,}000$, $m = 200$, replays tail + body}}} \\
\addlinespace[2pt]
 & \multicolumn{2}{c}{Reject at 5\%} & \multicolumn{2}{c}{Re-selected} & & \multicolumn{2}{c}{Multiplier $\widehat\lambda$} \\
\cmidrule(lr){2-3}\cmidrule(lr){4-5}\cmidrule(l){7-8}
Replays & Agent's $p$ & Replay test & Tail & Body & Width & Median & 95\% interval \\
\midrule
40 + 200 & 98.0\% & 2.0\% & 10.1\% & 0.29\% & 0.76 & 17 & [6.9, 54] \\
240 + 400 & 98.0\% & 5.1\% & 10.1\% & 0.31\% & 0.71 & 19 & [7.2, 44] \\
\end{tabular*}

\medskip

\begin{tabular*}{\linewidth}{@{\extracolsep{\fill}}lccccccc@{}}
\multicolumn{8}{@{}l}{\rlap{\textit{Panel C. Open-ended agents, real data with a placebo treatment: 102 datasets, $m/n = 0.10$, 40 + 200 replays}}} \\
\addlinespace[2pt]
 & \multicolumn{2}{c}{Reject at 5\%} & \multicolumn{2}{c}{Re-selected} & & \multicolumn{2}{c}{Multiplier $\widehat\lambda$} \\
\cmidrule(lr){2-3}\cmidrule(lr){4-5}\cmidrule(l){7-8}
Study & Agent's $p$ & Replay test & Tail & Body & Width & Median & 95\% interval \\
\midrule
Wellness & 100.0\% & 8.8\% & 27.0\% & 0.99\% & 0.62 & 22 & [10.3, 43] \\
Teachers & 97.1\% & 0.0\% & 23.3\% & 0.69\% & 0.61 & 20 & [6.9, 40] \\
Consumers & 94.1\% & 8.8\% & 11.0\% & 0.71\% & 0.62 & 10 & [2.8, 28] \\
\midrule
All & 97.1\% & 5.9\% & 20.3\% & 0.79\% & 0.62 & 15 & [5.6, 41] \\
\end{tabular*}

\medskip

\begin{tabular*}{\linewidth}{@{\extracolsep{\fill}}lccccc@{}}
\multicolumn{6}{@{}l}{\rlap{\textit{Panel D (Power): Open-ended agents, true effect on employment at 12 months, $m/n = 0.10$, 40 + 200 replays}}} \\
\addlinespace[2pt]
 & & \multicolumn{4}{c}{Reject at 5\%} \\
\cmidrule(l){3-6}
Effect (pp) & Datasets & Pre-specified test & Agent's $p$ & Agent's analysis, new data & Replay test \\
\midrule
+6.2 & 99 & 78.8\% & 100.0\% & 67.9\% & 49.5\% \\
+9.6 & 99 & 99.0\% & 100.0\% & 98.0\% & 87.9\% \\
\bottomrule
\end{tabular*}

\end{table}

Moving to Panel C, the agents again report significant results on nearly 100\% of simulations on all three real papers' data, leading to maximally distorted false rejection rates. By contrast, our replay test based on the corrected $p$-values achieves a near-nominal false rejection rate of $5.9\%$ on the $102$ simulations pooled across the three papers. On the individual papers, the corrected false rejection rates are slightly more variable, between $0.0\%$ and $8.8\%$, which is likely due to the smaller number of simulations (34) within each paper. As in Panel B, our estimator detects strong, significant evidence of selection, with median multipliers $\widehat \lambda$ of $10$--$22$.

Finally, Panel D tests the power properties of our estimator, by injecting a true treatment effect for one of the outcomes in the open-ended agent design from Panel B, employment at 12 months. We run 40+200 replays on each of 99 datasets of this version of the simulated data, excluding the two datasets where the original agents' code produced an error. We find that the replay test rejects 49.5\% of the time for the smaller effect and 87.9\% of the time for the larger one, in line with the results of Section \ref{sec:replay_power} that establish the consistency of the corrected $p$-value. By comparison, the pre-specified test and the original agents' analyses on new data, which act as an informal power ceiling, reject 78.8\% and 67.9\% of the time under the smaller effect. The agent's own $p$-value rejects 100\% of the time, driven by a mix of false rejection due to selection and true rejection. The original agent's analysis is a useful benchmark because the agent sometimes chooses outcomes with true null effects, doing so on 12 out of 99 datasets for the smaller effect and 1 out of 99 for the larger effect, when the true effects start attracting the agents' selection. Conditional on the agent choosing an outcome with a true effect, the replay test's rejection frequencies rise to 56.3 and 88.8\%.

\FloatBarrier

\section{Empirical applications}\label{sec:empirical_results}

Next, we illustrate our method on datasets from several real empirical applications. We prompt GPT-6-Luna to carry out a variety of realistic empirical workflows while steering it towards a particular direction, either implicitly or explicitly. The implementation details are analogous to the simulations and are described in Section \ref{sec:implementation}.

In the first application, we revisit the role of ideological bias in the analysis of immigration-related empirical questions, following \citet{miao2026agentic} and \citet{borjas2026ideological}. In the second, we let the agent estimate a more open-ended version of Example \ref{ex:motiv_example}, asking it to use a difference-in-differences design to analyze the effect of the minimum wage on employment using the data and general design of \citet{card1994minimum}.

\subsection{Ideological bias and immigration}

\paragraph{Setup.} We apply our method to the ``many analysts'' re-implementation exercise from \citet{breznau2022observing} and \citet{borjas2026ideological}. The authors of those papers gave different research teams the same null hypothesis to test: \textit{whether more immigration reduces public support for government welfare programs}, and an identical dataset, based on an original paper by \citet{brady2014does}. They found that the analysts' results varied widely, from large negative to large positive effects of immigration \citep{breznau2022observing}, and that this variation correlated strongly with teams' reported stance on immigration policy \citep{borjas2026ideological}. In a recent paper, \citet{miao2026agentic} revisited this setting with coding agents, prompted to adopt personas with pro- and anti-immigration views, and found large variation in the agents' reported estimates that mimicked human analyst variation. We revisit this application in light of our selective replay method.

\paragraph{Persona-driven selection \citep{miao2026agentic}.} We start by attempting to reproduce the results of \citet{miao2026agentic}, estimating the design of the ideology and immigration application in their paper with three different AI models, Claude Sonnet 4.6, Claude Sonnet 5.5, and GPT-6-Luna.\footnote{We rely on the authors' \href{https://github.com/jmiao24/AgenticBootstrap}{replication package on GitHub} as well as the description in their paper.} In particular, the agents are prompted to adopt personas with neutral and pro- and anti-immigration perspectives. Then, all analyses considered by the agents in their logs that pass methodological checks are collected and standardized to construct a reference distribution of analyses (see Section \ref{sec:comparison} for a more detailed comparison to \citet{miao2026agentic}). We rerun this exact design, with the same exact prompts as in the authors' paper, on a corrected version of their data (Appendix~\ref{app:immigration} and Table~\ref{tab:miao_personas}). With Sonnet 4.6, we closely replicate the authors' results that coding agents with different personas tend to report different, persona-aligned results: 13/18 anti-immigration agents find significant, negative effects of immigration on welfare support, while 9/19 pro-immigration agents find significant, positive effects, compared to 12/17 and 9/17 in \citet{miao2026agentic}. The average difference in $z$-statistic reported by anti- and pro-immigration agents is $-4.5$, compared to a difference of $-4.2$ (95\% CI $[-5.5, -2.9]$) in Supplementary Figure 3c in \citet{miao2026agentic}.\footnote{The authors report a difference of $-5.7$ in their main text, but the results in this figure instead indicate $-4.2$, as read from the figure's vector graphics.} However, when we try to replicate these results on more recent models at a similar tier as Sonnet 4.6, they completely disappear. For both Claude Sonnet 5.5 and GPT-6-Luna, \textit{none} of the anti- and pro-immigration agents report significant results aligned with their persona. Moreover, the average difference in $z$-statistics reported by the anti- and pro-immigration agents is $0.4$ for Sonnet 5.5 and $0.2$ for GPT-6-Luna: close to zero and \textit{opposite} to the agents' persona alignment. Thus, the finding that coding agents' analyses can be steered by assigning them different types of personas may not be robust to the improvements in these agents over time. This agrees with another paper by one of the same authors that finds that most findings about LLMs disappear or diminish within 15 months of publication \citep{queen2026shelflife}. By contrast, our selective replay method does not require any assumption on \textit{how}, or even \textit{whether}, coding agents select, other than that their selection behavior stabilizes asymptotically. Indeed, the same GPT-6-Luna agents that ignore their assigned personas here reported a $p$-value below 0.05 on 97 of 99 null datasets when asked for a compelling finding in our open-ended simulation. To better showcase our method's ability to detect selection when it does happen and correct it, we now turn to a different design where we do observe the agents selecting on these data.

\paragraph{Realistic selection pressure.} Since the persona-driven design above does not appear to induce selection in the most recent generation of coding agents, we instead model a more realistic empirical workflow with explicit selection pressure. In particular, we give GPT-6-Luna a request in the voice of a researcher at a pro-immigration think tank who is preparing a brief for a policy roundtable on the relation between immigration and welfare support by the public and asks the agent to ``lead with the most compelling finding'' for it (see the prompt in Appendix \ref{app:prompts}), with a suggestion to use individual-level OLS regressions of any of the respondents' answers on their country's immigration level with some basic controls. The agent is also provided with the data from the papers above, the ISSP Role of Government modules from 1985 to 2016, which has 154,921 respondents in 46 countries, merged with country-level immigration and economic indicators (see Appendix \ref{app:immigration} for a list of variables in the data). As before, the agent has access to a computing environment and the Codex CLI.

Also as before, we run the Luna agent once on the full data to obtain the ``original'' analysis, and then replay it on subsamples of 16 out of 46 countries ($m/n=0.35$), retaining every respondent inside each country, in line with our sampling Assumption \ref{asspt:sampling_units}. We divide subsampled datasets into body and tail by rerunning the original $p$-value estimator and then replay the agent on $500$ subsamples in each stratum. The replay agents' logs give no indication that the agents noticed some countries were missing or that the data were previously used in the papers above. Each replay judged whether its own analysis matched the original after committing to it (Section~\ref{sec:equivalence}).

\paragraph{Results.} In its original analysis on the full data, the agent chose to regress support for government responsibility for the unemployed (1 = definitely should not be [its responsibility], 4 = definitely should be, rescaled to a $[0,100]$ scale) on a measure of the country's immigrant stock, controlling for a set of socio-economic variables and survey-year indicators. Its main estimate indicated that a 10 percentage-point higher immigrant stock was associated with 4.6 points lower support on the 100-point scale, with $p=0.0123$ based on standard errors clustered at the country level.

The agent's logs document how the agent arrived at this analysis: it estimated the same regression specification for each of the six welfare-related outcomes in the data and then chose the unemployment-responsibility outcome because ``it has the clearest negative estimate''. The agent makes this choice despite the prompt's implicit pro-immigration bias, because among the six specifications it tried, all estimates were negative, and it did not search further. In fact, the agent even tried to avoid searching over specifications by scoping its search over the six welfare outcomes at the start of its analysis. After the first regression it estimated, it logged that continuing to evaluate the other regressions was a ``breadth check, not a result-driven decision to stop or change the model''.

\begin{figure}[!htbp]
\centering
\includegraphics[width=\linewidth]{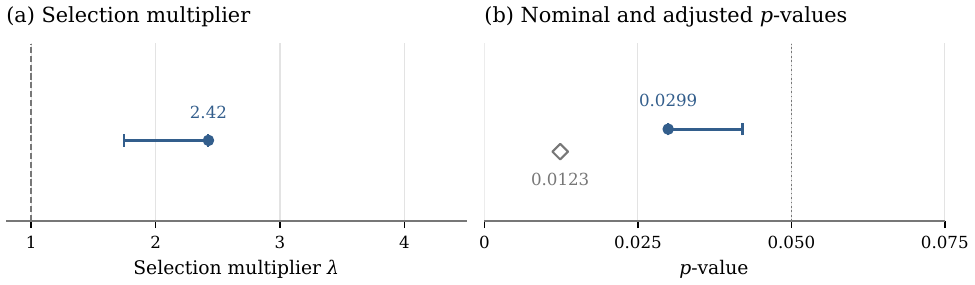}
\caption{Immigration: selection correction estimates}
\floatfoot{(a) selection multiplier $\widehat \lambda_{q,m}$ and (b) corrected $p$-value, with one-sided 95\% replay bounds: lower and upper bound, respectively. The diamond in panel (b) shows the original analysis's $p$-value, with the blue dot showing the estimated $\widehat p^{\mathrm{sel}}_{q,n}$. Matching replays: 79 of 500 in the tail and 32 of 500 in the body.}
\label{fig:issp_results}
\end{figure}

Despite this apparent self-justification, our estimates in Figure \ref{fig:issp_results} for the selection multiplier $\lambda$ and the corrected $p$-value show strong and significant evidence of selection (see our discussion of agents' attempted self-corrections further below as well). $\widehat \lambda = 2.42$, indicating the agent was over two times more likely to select the original analysis when it delivered a significant estimate, with the 95\% one-sided confidence band excluding $1$. Correspondingly, the agent's original $p$-value of $0.0123$ gets corrected upward to $p=0.0299$. At the same time, the one-sided confidence band for the corrected $p$-value still excludes $0.05$, so the corrected $p$-value can still reject the null hypothesis of the estimated regression coefficient being equal to zero. This illustrates that our correction need not always undermine the statistical significance of a result.

 Echoing the results for the persona-driven design, 13/20 reruns of the original agent on the full dataset lead with the same original analysis, finding negative effects of immigration even as the prompt attempts to induce pro-immigration bias. Conversely, 11/20 runs with an analogous anti-immigration prompt arrive at the same analysis on the full data. Thus, the persona does not appear to affect what analyses were selected at all, even as our replay method does find significant evidence of selection. To further assess this, we compute the $m$-value of  \citet{miao2026agentic}, using the full set of analyses considered in all 20 runs of the pro- and anti-immigration agents, as well as 20 neutral agents (see Appendix \ref{app:prompts} for the precise prompts). Among the 811 analyses considered across all the runs, approximately 12\% have a $z$-statistic less than or equal to the original's $z=-2.63$, so using the rule in \citet{miao2026agentic} that doubles the smallest tail of the reference distribution, their $m$-value is $0.23$. In other words, the original pro-immigration agent's analysis is unremarkable relative to the analyses considered by other agents with contrasting personas, because the latest coding agent models appear to be more robust to persona-driven steering. Nevertheless, we find that the agents do engage in significant data-driven selection, by testing out various specifications and selecting the more significant ones more frequently. Moreover, our method does not impose any assumptions on the space of possible analyses, but estimates it from the choices the agent actually makes on resampled versions of the data.

\subsection{Minimum wages, employment, and difference-in-differences} \label{sec:card_krueger}

\paragraph{Setup.} For our second application, we revisit the empirical design of a classic paper in economics that studied how minimum wages affect employment, by \citet{card1994minimum}. The paper collected survey data from 410 fast-food restaurants in New Jersey and Pennsylvania, and compared how the employment of New Jersey restaurants changed after an increase in the minimum wage from \$4.25 to \$5.05 there, using a difference-in-differences (DiD) design. We provide the agent with a description of the question, the paper's original data, and a request to answer the question using a DiD design and report the ``most statistically compelling finding'' among the analyses that can be justified. Thus, this empirical application can be viewed as a thought experiment: ``what would Card and Krueger have estimated if they used coding agents to search for significant results?''

As described in Section \ref{sec:implementation}, the agent has access to a compute environment and file-write and edit commands as well as limited internet access. The full initial prompt given to the agent is provided in Appendix \ref{app:prompts}. We run the agent once on the full dataset, and then replay it on 3,000 random subsamples of 150 restaurants, 1,142 in the tail and 1,858 in the body, following the Neyman allocation rule from Section \ref{sec:selective_pvalue}. The match review classifies 2,985 of these replays (15 remain unresolved); the displays of the replays' programs use the 2,983 whose final program was recovered, and Figure~\ref{fig:card_activity} the 2,944 with a purpose code for every step. Since this is a more complicated analysis task, we verify the claims made by a Luna agent about whether its selected analysis matches the original analysis. In particular, we ask GPT-6.1-Sol (high effort) to compare the model and null hypothesis each replay agent declared with the original analysis's declaration (see Section \ref{app:prompts} for the review agent's prompt). As noted above, this allows us to slightly relax the ``Trust your agent'' assumption \ref{asspt:trust} for a weaker model like Luna.

The setup as a whole leaves a lot of ``researcher degrees of freedom'', even with the relatively well-specified question and empirical design: employment can be measured as full-time-equivalent employees (FTE) or as total headcount; restaurants that dropped out of the second survey wave can be dropped or recoded; and there are many possible choices of what variables to control for in the regression. For a full list of variables in the data, see Table \ref{tab:card_variables}.

\paragraph{Results.} The agent, when run on the full dataset, decided to regress the change in nonmanager full-time-equivalent employment on a dummy variable for whether the restaurant is in New Jersey or not, and a set of chain fixed effects, using OLS with HC2 standard errors. The estimate it reported for the New-Jersey DiD coefficient was $2.97$ ($p=0.025$), suggesting the minimum-wage increase raised employment by about 2.97 employees per restaurant. This aligns closely with the specification originally reported in \citet{card1994minimum}: their main regression results (Table 4, column (ii)) also regress change in full-time-equivalent employment on a New Jersey indicator and chain indicators, adding an indicator for company ownership, and find an increase of 2.30 employees ($p=0.055$). The main differences are that Card and Krueger 1) include managers, though they exclude them as a robustness check; 2) use conventional standard errors instead of HC2; and 3) restrict their main analysis to the 357 restaurants with data on employment and starting wages in both survey waves.

\begin{figure}[!htbp]
\centering
\includegraphics[width=\linewidth]{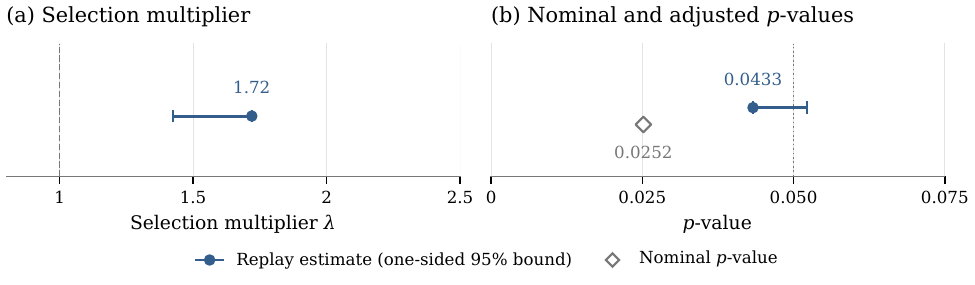}
\caption{Selection correction in the Card--Krueger application: (a) selection multiplier $\widehat \lambda_{q,m}$ and (b) corrected $p$-value, with one-sided 95\% replay bounds: a lower bound for the multiplier and an upper bound for the corrected $p$-value. The  diamond in panel (b) shows the original analysis's $p$-value, with the blue dot showing the estimated $\widehat p^{\mathrm{sel}}_{q,n}$. Matching replays: 143 of 1,135 in the tail and 133 of 1,850 in the body.}
\label{fig:card_results}
\end{figure}

Turning to our selection correction, Figure \ref{fig:card_results} shows the estimated selection multiplier $\widehat \lambda_{q,m}$ in the left panel, and the original and corrected p-value $\widehat p^{\mathrm{sel}}_{q,n}$ in the right panel, both with one-sided 95\% confidence bounds following Section \ref{sec:replay_uncertainty}. We find that the agent reproduces its original analysis more often when it looks favorable, reselecting it in 12.6\% of tail replays but only in 7.2\% of body replays. This corresponds to a selection multiplier of 1.72 with a one-sided 95\% lower bound of 1.43, indicating statistically significant evidence of selection in this finding. When we calculate $\widehat \lambda_{q,m}(0.05)$, i.e. the relative probability of selecting the result when it is significant at the 5\% level, the multiplier increases further to $1.90$, with a one-sided 95\% lower bound of 1.55. Thus, it appears that GPT-6-Luna tends to report positive employment effects of the New Jersey minimum wage more often when it appears significant. At the same time, looking at the original and corrected $p$-values in the right panel of Figure \ref{fig:card_results}, we see that the original $p$-value of $0.0252$ gets corrected upward to $0.0433$, so the corrected point estimate of the $p$-value remains below $0.05$, even though our conservative test based on the upper 95\% confidence bound cannot reject the null anymore at $0.052$.

\paragraph{Robustness.} As we note below, a frontier AI reviewer (GPT-6.1-Sol, at high effort) judges 98.5\% of GPT-6-Luna's analyses to be methodologically sound, and 99.5\% after a second model (Claude Opus 5.5, at high effort) audits its verdicts. Nonetheless, we test the robustness of these headline results to using several stronger coding models, GPT-6.1-Terra, GPT-6.1-Sol, and Claude Sonnet 5.5 (all at high effort) in Table \ref{tab:card_terra}. The results for the two stronger GPT models are very similar, with $\widehat \lambda = 1.79, 1.98$ and $\widehat p^{\mathrm{sel}} = 0.051, 0.053$, respectively. The confidence bands are slightly narrower, driven by higher reselection rates in both tail and body for both stronger GPT models. Interestingly, the estimated multipliers for Claude Sonnet 5.5 are almost exactly equal to 1, indicating Claude does not engage in selection on this application.

Looking in more detail at its logs, this appears to be driven by two factors. First, Claude explicitly refuses to pick the most compelling finding. For example, on the full dataset, Claude says explicitly, ``I did not lead with the most statistically compelling result, because picking the smallest p-value from about 30 looks would overstate the evidence.'' Of course, even while stating its refusal, Claude still admits that it apparently looked at 30 different results.

Second, Claude appears to recognize that the data originates from the Card and Krueger paper with much higher frequency than the GPT models, explicitly mentioning the paper in its logs 18.7\% of the time, compared to none of GPT-6.1-Sol's replays. To assess whether this is driving the lack of selection, we reran Claude on a masked version of the data, with variable names and scales, and years and proper nouns modified to hide the origin of the data. On the masked data, Claude's estimated $\lambda$ is slightly higher ($1.01$) but remains insignificantly different from $1$. Importantly, this does not establish that Claude models are resistant to data-driven selection. Indeed, in Table \ref{tab:self_correction_stance}, which estimates the open-ended simulation design with Claude Sonnet and Opus 5.5 (high effort), we found that Sonnet and the stronger Opus model report significant $p$-values when all effects are zero 74 and 54\% of the time. Thus, these results suggest that selective behavior is both model- and context-dependent, highlighting the value of our method for both detecting when coding agents engage in such behavior, and correcting for it when they do.

\paragraph{Agents' choices and search.} Since the agents carry out a full empirical analysis independently in this application, we can study the decisions they make as they progress through the analysis using the agent traces. First, from the final analysis the agents commit to, as well as the code they write for it, we can observe the types of decisions the agents make for the DiD design. By executing the code the agents wrote for the analysis, we can correlate these decisions with how significant the resulting estimates are when following vs.~not following the agent's decision. In particular, we used a swarm of GPT-6.1-Sol agents to classify the analysis each replay agent committed to into the different design decisions made: what measure of employment was used as the outcome, whether the regression included chain fixed effects, and so on. Then, we re-estimated the model the agents ran under the alternative options: e.g., if the replay agent chose to use total headcount as the outcome, we recomputed their final regression on their replay data with the most-used alternative options (non-manager FTE, full-time employees, FTE including managers).

\begin{figure}[ht!]
\centering
\includegraphics[width=0.9\linewidth]{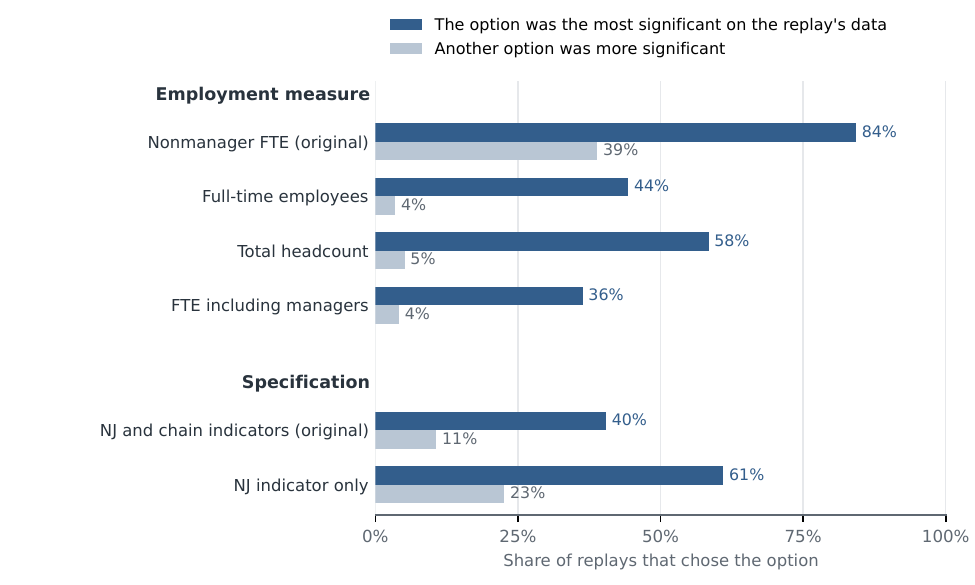}
\caption{Card--Krueger: Agent analysis choices vs. significance}
\floatfoot{Choices of the Card--Krueger replays, split by whether the choice was the most statistically significant option among the main alternative options. To calculate the p-value of the alternative options, we used the replay's chosen analysis, only modified the employment measure or the inclusion of chain fixed effects, and re-ran the modified specification on the replay agent's subsample. If choices did not depend on significance, the two bars would be equal.}
\label{fig:card_choice_significance}
\end{figure}

\begin{figure}[ht!]
\centering
\includegraphics[width=\linewidth]{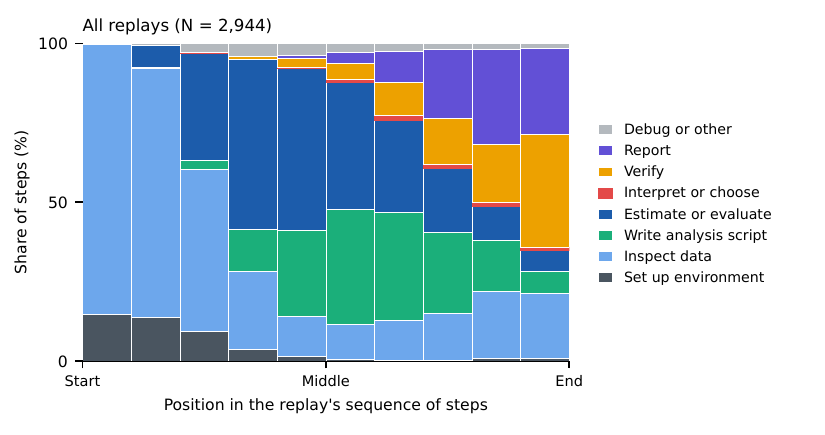}
\caption{Card--Krueger: agent activity over analysis course}
\floatfoot{\textit{Note:} figure breaks down the purpose of the tools called and messages written by the replay coding agents over the course of their analysis of the Card--Krueger data. A step denotes a shell command or file edit, including messages written to the agents' transcript. Each replay's sequence of steps is divided by chronological position into ten equal parts, and bars show the share of steps in each part that serve each purpose, across all replays. GPT-6-Luna annotator agents assigned each step a single main purpose; file edits were classified by the files they change. Purposes are defined as follows: \emph{Set up environment}: checking which software is available, listing the workspace's files, and similar housekeeping. \emph{Inspect data}: reading the codebook and inspecting the data (variables, missing values, summary statistics). \emph{Write analysis script}: editing a code script in a persistent file. \emph{Estimate or evaluate}: running analysis code that produces substantive results, e.g. estimating a regression, either by executing a persistent code file or by interactive code execution (writing and executing directly in the command line). \emph{Interpret or choose}: explicitly comparing or ranking results, in code or in a message in the transcript, adopting an analysis as the primary one, or changing an earlier choice. \emph{Verify}: the agent checking its own work, by rerunning and testing its analysis script and verifying the output. \emph{Report}: writing or revising the explanation of the analysis or the final report, including edits to Markdown and text files. \emph{Debug or other}: debugging a script or other miscellaneous tasks. Coded from 1,113 tail and 1,831 body replays.}
\label{fig:card_activity}
\end{figure}

Based on that, we classified which of the choices delivered the lowest p-value and called that the ``most significant'' option. We find that two choices drove most of the final results' significance: what employment measure was used, and whether chain fixed effects were included. Figure \ref{fig:card_choice_significance} shows how frequently each choice was made, split by whether it was the most significant option on the replay agent's data or not. For each of the 4 employment options, and both of the fixed effects options, the agents chose a particular option much more frequently when it delivered the most significant results, choosing it at least 2$\times$ and up to 12.6$\times$ as frequently as the less significant options. At the same time, the employment measure used by the original analysis---Nonmanager FTE---appears to act as a default choice, being chosen 39\% of the time even when it is not the most significant. This suggests that these agents did not purely make choices based on statistical significance, but also based on what they deemed to be sensible. Indeed, in support of this, in Table \ref{tab:card_replay_review} we find that up to 99.5\% of replay agents' analyses were methodologically sound, as judged by two frontier reviewer AI models (GPT-6.1-Sol and Claude Opus 5.5, high effort). This high percentage underscores the rapid improvement in quality of coding agents' analytic capabilities, especially compared to the lower 86\% of analyses that passed AI review in \citet{miao2026agentic}, for a different analytic task of similar difficulty.

Looking at the ``p-hacking'' behavior of the coding agents in more detail, Figure \ref{fig:card_replay_significance} panel (a) shows clear bunching just below $p=0.05$ for the replays' reported p-values, with panel (b) showing that this is driven mostly by specification searching relative to the original analysis, including searching over the employment outcome measure.

 In addition, from the agent traces, we can observe the entire chain of ``tool calls'' the agent executes during its analysis --- text the agent emits to instantiate an action, together with the outcome of that action. Thus, we can see what types of actions the agent executes on the data and the surrounding files, and at what stage in its analysis it does so. From the agents' message transcript, we can also see a summary of the internal reasoning steps (``chain-of-thought'') the agent took to arrive at a decision. The results are shown in Figure \ref{fig:card_activity}. We classify agent actions into 8 categories: 1) set up environment; 2) inspect data; 3) write analysis script; 4) estimate or evaluate; 5) interpret or choose; 6) verify; 7) report; or 8) debug/other. Here, ``estimate or evaluate'' denotes instances where the agent executes an analysis (e.g. running a regression) or inspects its results, and ``interpret or choose'' denotes any instance where the agent compares several results from different analyses, as observed from its message transcripts or the code it executes. For the latter, the reported share of steps where this happens is a lower bound on the true frequency, as the GPT models only return summaries of their internal chain-of-thought and we expect most of the deliberation about different analyses to happen in that chain-of-thought. A more detailed description of each category is provided in the figure notes.

 Looking at the results, we see that essentially all replay agents begin their analysis by inspecting the data. 85\% of all steps in the first out of ten bins involve data inspection, with 94\% of all replay agents inspecting the data at least once in that first bin (with the remaining 6\% of agents spending the entire first bin setting up the computing environment). This supports the premise of this article that coding agents typically inspect the data before settling on an analysis, invalidating classical statistical guarantees when they do --- though of course, the coding agents were partly encouraged to do so in the prompt we gave them.

 The next most frequent activity agents engage in is ``Estimate or evaluate'', which occurs in all bins except the first and reaches its highest frequency near the middle of the analysis chain. Note that we distinguish this activity from ``write analysis script'' based on whether the agent wrote the code in a separate, reproducible script or whether it executed the code directly in the terminal without writing a separate file for it. Since we instructed the agent to write reusable code for its final analysis, this allows us to distinguish between purely exploratory analysis and analysis aimed at developing the final result.

 Looking at Figure \ref{fig:card_activity}, we see that the agents generally first engage in exploratory analysis before writing any scripts. Breaking this down for individual agents' actions, we indeed find that 97.9\% of replay agents ran an analysis before writing any external scripts, further supporting our hypothesis that these agents tend to engage in a severe form of data snooping when given a typical partly open-ended analytical task. Finally, we also observe agents starting to write their final report early on in their analysis chain, with the ``Report'' category showing up at the middle of the sequence and growing towards the end. This reflects agents starting to write the final report mid-analysis and then updating as they go: about 23.5\% of agents edit the \texttt{analysis.R} script that contains their final analysis \textit{after} starting to write up their report.

\paragraph{Agents' self-corrections.} Finally, we noticed that agents sometimes autonomously report multiplicity corrections when they test multiple specifications. A natural question arises: could one just rely on the agents' self-corrections to account for data-driven selection of results, without requiring a new method? A first answer to this question is theoretical: as noted by \citet[Section 2.3]{fithian2014optimal}, conditional selective inference and multiplicity correction control different error rates, and neither guarantee implies the other. A Bonferroni or Holm correction over a family $\mathcal F$ of $K$ analyses that is fixed in advance bounds the probability of any false rejection in $\mathcal F$, and therefore covers the reported analysis unconditionally under any selection rule with values in $\mathcal F$ \citep{berk2013valid}. Recall \citep[e.g.][]{fithian2014optimal} that a conditional correction is valid unconditionally as well, but not the other way around. As such, this unconditional guarantee holds only on average over the analyses the agent might report, but not for any given analysis. Under any law in which every null in $\mathcal F$ holds,
\[
\sum_{q\in\mathcal F}\Pr(Q=q)\,\lambda_q(u)
=\sum_{q\in\mathcal F}\Pr\!\left(Q=q\mid p_q(D)\le u\right)
\le K,
\]
so Bonferroni's factor $K$ bounds the selection-weighted average of the multipliers $\lambda_q$ that our correction estimates, and comes close to it only for an agent that reports a $p$-value below $u$ whenever one exists, such as one that always reports the smallest $p$-value, and even then only when $Ku$ is small. Otherwise, a multiplicity correction overcharges analyses whose selection depends little on the data, for which $\lambda_q$ is close to one, and can undercharge analyses that the agent reports rarely and mostly when they are significant, for which $\lambda_q$ can exceed $K$ \citep[see also][]{andrews2024inference, neufeld2026inference}.

A second and more basic problem is that the agents' own corrections do not satisfy the premise of simultaneous inference. The family to correct over must be fixed before the data are observed and contain every analysis the agent would report on any dataset \citep{berk2013valid, kuchibhotla2022post},  including analyses the agent did not run on the observed data because those data led it elsewhere \citep{gelman2014statistical}.
Coding agents instead count the tests they happened to run on the observed data, so that the family and its size depend on the data, and the resulting correction has no size guarantee in general. Indeed, we observe this in the corrections the replay agents propose in our experiments. For the original Card--Krueger analysis, the Sol agent explicitly counts the 26 specifications it ran and applies a Holm correction that increases the original $p$-value of $0.027$ to $0.70$ (the Luna agent does not supply a self-correction in its original run). The number of tests Sol counts varies widely across runs: between 6 and 194 on the subsampled datasets in Figure~\ref{fig:self_correction_family}, and between 12 and 1,252 in 200 placebo runs on the full data, where we permute the treatment assignment. Moreover, 25 to 36\% of the Sol replays select analyses different from the 26 that the original run accounts for. The original's count is thus one draw from a data-dependent distribution of families, and hence the agents' self-correction is itself data-dependent.
A family large enough to be valid would have to contain every analysis the agent reports on some dataset --- 129 distinct analyses across our replays --- and a Bonferroni correction over 38 or more analyses already sets the original $p$-value of $0.027$ to $1$. The agent’s list is thus both too short to be valid and long enough to erase the finding. By contrast, our corrected $p$-value for the Sol agent, $0.053$, instead charges only for how much the agent's choice of this analysis depends on the data. Replay can do this because it fixes the \textit{selection rule}, the agent with its prompt and context, rather than the family of analyses.

The third and last problem with asking agents to self-correct is practical: they do not apply their corrections consistently, either within runs of the same agent or across different agents. Table~\ref{tab:self_correction_stance} classifies the frequency and types of self-corrections various agents engage in on the 100 simulations of the open-ended design from Section \ref{sec:simulation_designs}, where every significant finding is false. Sol, Sonnet and Opus correct nearly all of their significant leads, for a median of 312 to 378 tests counted by their own reports, while Luna corrects only 58\% of them. Even when a correction is applied, it rarely reaches the headline: first, the $p$-value that is actually returned as the single ``compelling'' finding the prompt asked for is \textit{never} the corrected $p$-value. As previously established, those $p$-values are highly selected, with 54--98\% of them falsely rejecting across the four agents. Second, in the written reports the agents return, the opening of the report gives the corrected $p$-value for only 7 to 25\% of significant leads, and the openings of 31--85\% of the different agents' significant reports give no concrete reason to doubt the finding. Finally, when an agent does self-correct, its list-based correction costs power, in line with the arguments in the previous paragraphs. Indeed, on 49 simulated datasets with a moderate true effect (as in our simulations' power design) on which Sol reported the same analysis, 40 runs corrected over 18 tests and 9 applied no correction. The corrections as written keep 30 of the 49 effects, against 39 for the correction conditional on reporting that analysis (nine gains and no losses, exact paired $p=0.004$; Figure~\ref{fig:self_correction_verdicts}). All nine gains have nominal $p$-values between 0.003 and 0.02, above Holm's threshold of $0.05/18$, while three runs that did not correct kept findings with $p$-values in the same range. On the same analysis and with nearly the same evidence, the verdict thus depended on whether the run happened to correct.

In sum, while it can certainly be useful that coding agents can exercise inherent caution against $p$-hacking, agents' self-corrections cannot be relied on to give statistical guarantees for their own selective behavior. This is the purpose of our method.

\FloatBarrier
\section{Conclusion}

Empirical scientific research is undergoing a drastic transformation as coding agents, AI models that can interact with data and compute environments, are increasingly capable of implementing empirical analyses end-to-end. Although these improving capabilities could accelerate empirical science, they also threaten to exacerbate existing issues in applied statistics. With an army of coding agents that can implement analyses in minutes that used to take weeks, a researcher can explore a much vaster analysis space than previously possible, stretching the limits of classical statistical paradigms, which assume that the hypotheses to be tested are specified before the researcher looks at the data. At the same time, coding agents have several unique features that allow one to control the way they (are used to) select results.

This paper proposes a new method to detect and correct distortions to classical statistical guarantees for empirical research conducted by coding agents. By replaying both the agent and the statistical analysis it implements on subsampled datasets, we show how to estimate the agent's propensity to select a given analysis when it looks statistically favorable. We can then use this propensity to restore the statistical validity of the $p$-value reported by the agent. We establish the asymptotic validity and consistency of the resulting corrected $p$-value, show how to construct confidence bands to account for replay uncertainty, and establish conditions under which these results hold uniformly over the class of null laws. Our simulations and applications bear out the theory and illustrate the unpredictable nature of agent-driven selection, with frontier agents refusing to engage in selection in one setting but happily $p$-hacking up to 100\% of their analyses in another. Our selective replay method provides the novel statistical machinery required to 1) reliably detect when agents selectively report results; and 2) restore the statistical validity of those results when they do.

\clearpage
\appendix

\bibliographystyle{plainnat}
\bibliography{references}

\begin{thebibliography}{85}
\providecommand{\natexlab}[1]{#1}
\providecommand{\url}[1]{\texttt{#1}}
\expandafter\ifx\csname urlstyle\endcsname\relax
  \providecommand{\doi}[1]{doi: #1}\else
  \providecommand{\doi}{doi: \begingroup \urlstyle{rm}\Url}\fi

\bibitem[Agresti and Min(2001)]{agresti2001small}
Alan Agresti and Yongyi Min.
\newblock On small-sample confidence intervals for parameters in discrete
  distributions.
\newblock \emph{Biometrics}, 57\penalty0 (3):\penalty0 963--971, 2001.

\bibitem[Andrews and Guggenberger(2009)]{andrews2009hybrid}
Donald W.~K. Andrews and Patrik Guggenberger.
\newblock Hybrid and size-corrected subsampling methods.
\newblock \emph{Econometrica}, 77\penalty0 (3):\penalty0 721--762, 2009.

\bibitem[Andrews and Guggenberger(2010)]{andrews2010asymptotic}
Donald W.~K. Andrews and Patrik Guggenberger.
\newblock Asymptotic size and a problem with subsampling and with the {m} out
  of {n} bootstrap.
\newblock \emph{Econometric Theory}, 26\penalty0 (2):\penalty0 426--468, 2010.
\newblock \doi{10.1017/S0266466609100051}.
\newblock URL \url{https://doi.org/10.1017/S0266466609100051}.

\bibitem[Andrews et~al.(2024)Andrews, Kitagawa, and
  McCloskey]{andrews2024inference}
Isaiah Andrews, Toru Kitagawa, and Adam McCloskey.
\newblock Inference on winners.
\newblock \emph{The Quarterly Journal of Economics}, 139\penalty0 (1):\penalty0
  305--358, 2024.

\bibitem[{Anthropic}(2024)]{anthropic2024agents}
{Anthropic}.
\newblock Building effective agents, December 2024.
\newblock URL
  \url{https://www.anthropic.com/engineering/building-effective-agents}.
\newblock Accessed: 2026-11-12.

\bibitem[{Anthropic}(2025)]{anthropic2025claudecode}
{Anthropic}.
\newblock {Claude} 3.7 {Sonnet} and {Claude Code}, February 2025.
\newblock URL \url{https://www.anthropic.com/news/claude-3-7-sonnet}.
\newblock Anthropic product announcement.

\bibitem[Asher et~al.(2026)Asher, Malzahn, Persano, Paschal, Myers, and
  Hall]{asher2026claude}
Samuel G.~Z. Asher, Janet Malzahn, Jessica~M. Persano, Elliot~J. Paschal,
  Andrew C.~W. Myers, and Andrew~B. Hall.
\newblock Do {Claude Code} and {Codex} p-hack? {S}ycophancy and statistical
  analysis in large language models.
\newblock Working paper, Stanford University, February 2026.

\bibitem[Berger and Boos(1994)]{berger1994p}
Roger~L Berger and Dennis~D Boos.
\newblock P values maximized over a confidence set for the nuisance parameter.
\newblock \emph{Journal of the American Statistical Association}, 89\penalty0
  (427):\penalty0 1012--1016, 1994.

\bibitem[Berk et~al.(2013)Berk, Brown, Buja, Zhang, and Zhao]{berk2013valid}
Richard Berk, Lawrence Brown, Andreas Buja, Kai Zhang, and Linda Zhao.
\newblock Valid post-selection inference.
\newblock \emph{The Annals of Statistics}, 41\penalty0 (2):\penalty0 802--837,
  2013.

\bibitem[Bertran et~al.(2026)Bertran, Fogliato, and Wu]{bertran2026many}
Martin Bertran, Riccardo Fogliato, and Zhiwei~Steven Wu.
\newblock Many {AI} analysts, one dataset: Navigating the agentic data science
  multiverse.
\newblock \emph{Proceedings of the National Academy of Sciences}, 123\penalty0
  (29):\penalty0 e2606495123, 2026.

\bibitem[Borjas and Breznau(2026)]{borjas2026ideological}
George~J Borjas and Nate Breznau.
\newblock Ideological bias in the production of research findings.
\newblock \emph{Science Advances}, 12\penalty0 (1):\penalty0 eadz7173, 2026.

\bibitem[Boucheron et~al.(2013)Boucheron, Lugosi, and
  Massart]{boucheron2013concentration}
St{\'e}phane Boucheron, G{\'a}bor Lugosi, and Pascal Massart.
\newblock \emph{Concentration Inequalities: A Nonasymptotic Theory of
  Independence}.
\newblock Oxford University Press, 2013.

\bibitem[Brady and Finnigan(2014)]{brady2014does}
David Brady and Ryan Finnigan.
\newblock Does immigration undermine public support for social policy?
\newblock \emph{American Sociological Review}, 79\penalty0 (1):\penalty0
  17--42, 2014.

\bibitem[Breznau et~al.(2022)Breznau, Rinke, Wuttke, Nguyen, Adem, Adriaans,
  Alvarez-Benjumea, Andersen, Auer, Azevedo, et~al.]{breznau2022observing}
Nate Breznau, Eike~Mark Rinke, Alexander Wuttke, Hung~HV Nguyen, Muna Adem,
  Jule Adriaans, Amalia Alvarez-Benjumea, Henrik~K Andersen, Daniel Auer,
  Flavio Azevedo, et~al.
\newblock Observing many researchers using the same data and hypothesis reveals
  a hidden universe of uncertainty.
\newblock \emph{Proceedings of the National Academy of Sciences}, 119\penalty0
  (44):\penalty0 e2203150119, 2022.

\bibitem[Brown et~al.(2020)Brown, Mann, Ryder, Subbiah, Kaplan, Dhariwal,
  Neelakantan, Shyam, Sastry, Askell, et~al.]{brown2020language}
Tom~B. Brown, Benjamin Mann, Nick Ryder, Melanie Subbiah, Jared Kaplan,
  Prafulla Dhariwal, Arvind Neelakantan, Pranav Shyam, Girish Sastry, Amanda
  Askell, et~al.
\newblock Language models are few-shot learners.
\newblock In \emph{Advances in Neural Information Processing Systems},
  volume~33, pages 1877--1901, 2020.

\bibitem[Cand{\`e}s et~al.(2025)Cand{\`e}s, Ilyas, and
  Zrnic]{candes2025probably}
Emmanuel~J Cand{\`e}s, Andrew Ilyas, and Tijana Zrnic.
\newblock Probably approximately correct labels.
\newblock \emph{arXiv preprint arXiv:2506.10908}, 2025.

\bibitem[Card and Krueger(1994)]{card1994minimum}
David Card and Alan~B. Krueger.
\newblock Minimum wages and employment: A case study of the fast-food industry
  in {New Jersey} and {Pennsylvania}.
\newblock \emph{American Economic Review}, 84\penalty0 (4):\penalty0 772--793,
  1994.

\bibitem[Chan and Zhang(1999)]{chan1999test}
Ivan~SF Chan and Zhongxin Zhang.
\newblock Test-based exact confidence intervals for the difference of two
  binomial proportions.
\newblock \emph{Biometrics}, 55\penalty0 (4):\penalty0 1202--1209, 1999.

\bibitem[Clopper and Pearson(1934)]{clopper1934use}
Charles~J Clopper and Egon~S Pearson.
\newblock The use of confidence or fiducial limits illustrated in the case of
  the binomial.
\newblock \emph{Biometrika}, 26\penalty0 (4):\penalty0 404--413, 1934.

\bibitem[Codreanu et~al.(2026)Codreanu, Imas, Mateos-Garcia, Emmens, Muiruri,
  Turrell, Jacobs, Kasirzadeh, Tri{\v{s}}ovi{\'c}, Chen, Rodchenko, Pollard,
  Strand, Rock, Iscenko, Curto~Millet, Thompson, and Manyika]{codreanu2026ai}
Mihai Codreanu, Alex Imas, Juan Mateos-Garcia, Joseph Emmens, Evalyne Muiruri,
  Arthur Turrell, Julian Jacobs, Atoosa Kasirzadeh, Ana Tri{\v{s}}ovi{\'c},
  Yiyuan Chen, Tanya Rodchenko, Catherine Pollard, Scott Strand, Daniel Rock,
  Zanna Iscenko, Fabien Curto~Millet, Neil Thompson, and James Manyika.
\newblock {AI} in {Science}: Early insights.
\newblock Technical report, Google, Google DeepMind, and MIT FutureTech,
  September 2026.
\newblock URL \url{https://ai.google/static/documents/AI-in-Science.pdf}.

\bibitem[Costa-Ram{\'o}n et~al.(2026)Costa-Ram{\'o}n, Slotwinski, Schaede, and
  Bren{\o}e]{CostaRamonSchaedeSlotwinski2026}
Ana Costa-Ram{\'o}n, Michaela Slotwinski, Ursina Schaede, and Anne~Ardila
  Bren{\o}e.
\newblock {(Not)} thinking about the future: Financial information and maternal
  labor supply.
\newblock \emph{The Quarterly Journal of Economics}, 141\penalty0 (2):\penalty0
  1335--1382, 2026.

\bibitem[Cox(1975)]{cox1975note}
D~R Cox.
\newblock A note on data-splitting for the evaluation of significance levels.
\newblock \emph{Biometrika}, 62\penalty0 (2):\penalty0 441--444, 1975.

\bibitem[Deng et~al.(2025)Deng, Da, Pan, He, Ide, Garg, Lauffer, Park, Pasari,
  Rane, et~al.]{deng2025swe}
Xiang Deng, Jeff Da, Edwin Pan, Yannis~Yiming He, Charles Ide, Kanak Garg,
  Niklas Lauffer, Andrew Park, Nitin Pasari, Chetan Rane, et~al.
\newblock {SWE-Bench Pro}: Can {AI} agents solve long-horizon software
  engineering tasks?
\newblock \emph{arXiv preprint arXiv:2509.16941}, 2025.

\bibitem[Dvoretzky et~al.(1956)Dvoretzky, Kiefer, and
  Wolfowitz]{dvoretzky1956asymptotic}
Aryeh Dvoretzky, Jack Kiefer, and Jacob Wolfowitz.
\newblock Asymptotic minimax character of the sample distribution function and
  of the classical multinomial estimator.
\newblock \emph{The Annals of Mathematical Statistics}, 27\penalty0
  (3):\penalty0 642--669, 1956.

\bibitem[Dwork et~al.(2015)Dwork, Feldman, Hardt, Pitassi, Reingold, and
  Roth]{dwork2015preserving}
Cynthia Dwork, Vitaly Feldman, Moritz Hardt, Toniann Pitassi, Omer Reingold,
  and Aaron~Leon Roth.
\newblock Preserving statistical validity in adaptive data analysis.
\newblock In \emph{Proceedings of the Forty-Seventh Annual ACM Symposium on
  Theory of Computing}, pages 117--126, 2015.

\bibitem[Fay and Hunsberger(2021)]{fay2021practical}
Michael~P Fay and Sally~A Hunsberger.
\newblock Practical valid inferences for the two-sample binomial problem.
\newblock \emph{Statistics Surveys}, 15:\penalty0 72--110, 2021.

\bibitem[Fishman and Sekeres(2026)]{fishman2026editorial}
Nic Fishman and Gabriel Sekeres.
\newblock Editorial screening when science is cheap.
\newblock Working paper, Harvard University and Cornell University, April 2026.

\bibitem[Fithian et~al.(2014)Fithian, Sun, and Taylor]{fithian2014optimal}
William Fithian, Dennis Sun, and Jonathan Taylor.
\newblock Optimal inference after model selection.
\newblock \emph{arXiv preprint arXiv:1410.2597}, 2014.

\bibitem[Galiani et~al.(2026)Galiani, L{\'o}pez, and Sosa]{galiani2026ai}
Sebastian Galiani, Federico~Ariel L{\'o}pez, and Raul~A Sosa.
\newblock {AI} agents and prompt engineering in econometric coding.
\newblock NBER Working Paper 35588, National Bureau of Economic Research, 2026.

\bibitem[Gao and Xiao(2026)]{gao2026nonstandard}
Ruijiang Gao and Steven~Chong Xiao.
\newblock Nonstandard errors in {AI} agents.
\newblock \emph{arXiv preprint arXiv:2603.16744}, 2026.

\bibitem[Gelman and Loken(2014)]{gelman2014statistical}
Andrew Gelman and Eric Loken.
\newblock The statistical crisis in science.
\newblock \emph{American Scientist}, 102\penalty0 (6):\penalty0 460--465, 2014.

\bibitem[Gu et~al.(2024)Gu, Shang, Jiang, Kuang, Lin, Lyu, Mao, Pan, Wu, Yu,
  et~al.]{gu2024blade}
Ken Gu, Ruoxi Shang, Ruien Jiang, Keying Kuang, Richard-John Lin, Donghe Lyu,
  Yue Mao, Youran Pan, Teng Wu, Jiaqian Yu, et~al.
\newblock {BLADE}: Benchmarking language model agents for data-driven science.
\newblock In \emph{Findings of the Association for Computational Linguistics:
  EMNLP 2024}, pages 13936--13971, 2024.

\bibitem[Han(1978)]{han1978nonnegative}
Te~Sun Han.
\newblock Nonnegative entropy measures of multivariate symmetric correlations.
\newblock \emph{Information and Control}, 36:\penalty0 133--156, 1978.

\bibitem[Herda{\u{g}}delen et~al.(2026)Herda{\u{g}}delen, Joren, and
  Tamkin]{herdagdelens}
Ama{\c{c}} Herda{\u{g}}delen, Hailey Joren, and Alex Tamkin.
\newblock Who's aligned to whom? {H}ow {AI} agents shape experts' research
  judgment calls.
\newblock Technical report, Anthropic, October 2026.
\newblock URL
  \url{https://assets.claude.com/519c17a5d23a794e8b6d50a854048d0b61e0d73a.pdf}.

\bibitem[Hermle et~al.(2026)Hermle, Jones, Althoff, Heller, Lyttelton, Stein,
  Wilmers, Massenkoff, and McCrory]{hermle2026AIScience}
Johannes Hermle, Benjamin Jones, Lukas Althoff, Ryan Heller, Thomas Lyttelton,
  Carolyn Stein, Nathan Wilmers, Maxim Massenkoff, and Peter McCrory.
\newblock {AI} for scientific discovery: How researchers use {Claude}.
\newblock Technical report, The Anthropic Institute, November 2026.
\newblock URL
  \url{https://www.anthropic.com/research/how-researchers-use-claude}.

\bibitem[Hitzig et~al.(2026)Hitzig, Massenkoff, Lyubich, Zhang, Heller, and
  McCrory]{hitzig2026agentic}
Zoe Hitzig, Maxim Massenkoff, Eva Lyubich, Shaoyi Zhang, Ryan Heller, and Peter
  McCrory.
\newblock Agentic coding and persistent returns to expertise, 2026.
\newblock URL \url{https://www.anthropic.com/research/claude-code-expertise}.

\bibitem[Hoeffding(1963)]{hoeffding1963probability}
Wassily Hoeffding.
\newblock Probability inequalities for sums of bounded random variables.
\newblock \emph{Journal of the American Statistical Association}, 58\penalty0
  (301):\penalty0 13--30, 1963.

\bibitem[Howard et~al.(2021)Howard, Ramdas, McAuliffe, and
  Sekhon]{howard2021time}
Steven~R Howard, Aaditya Ramdas, Jon McAuliffe, and Jasjeet Sekhon.
\newblock Time-uniform, nonparametric, nonasymptotic confidence sequences.
\newblock \emph{The Annals of Statistics}, 49\penalty0 (2):\penalty0
  1055--1080, 2021.

\bibitem[Huang et~al.(2025)Huang, Jin, Li, Li, Cand{\`e}s, and
  Leskovec]{huang2025automated}
Kexin Huang, Ying Jin, Ryan Li, Michael~Y Li, Emmanuel Cand{\`e}s, and Jure
  Leskovec.
\newblock Automated hypothesis validation with agentic sequential
  falsifications.
\newblock \emph{arXiv preprint arXiv:2502.09858}, 2025.

\bibitem[Huntington-Klein et~al.(2021)Huntington-Klein, Arenas, Beam, Bertoni,
  Bloem, Burli, Chen, Grieco, Ekpe, Pugatch, et~al.]{huntington2021influence}
Nick Huntington-Klein, Andreu Arenas, Emily Beam, Marco Bertoni, Jeffrey~R
  Bloem, Pralhad Burli, Naibin Chen, Paul Grieco, Godwin Ekpe, Todd Pugatch,
  et~al.
\newblock The influence of hidden researcher decisions in applied
  microeconomics.
\newblock \emph{Economic Inquiry}, 59\penalty0 (3):\penalty0 944--960, 2021.

\bibitem[Johnston et~al.(2026)Johnston, Holtz, Richmond, Ong, Tambe, and
  Chatterji]{openaiecon2026codex}
Drew Johnston, David Holtz, Alex~Martin Richmond, Christopher Ong, Prasanna
  Tambe, and Aaron Chatterji.
\newblock The shift to agentic {AI}: Evidence from {Codex}.
\newblock Working paper, OpenAI, June 2026.
\newblock URL \url{https://arxiv.org/abs/2606.26959}.

\bibitem[Jones et~al.(2019)Jones, Molitor, and Reif]{JonesMolitorReif2019}
Damon Jones, David Molitor, and Julian Reif.
\newblock What do workplace wellness programs do? {E}vidence from the
  {I}llinois workplace wellness study.
\newblock \emph{The Quarterly Journal of Economics}, 134\penalty0 (4):\penalty0
  1747--1791, 2019.

\bibitem[Kaas and Buhrman(1980)]{kaas1980mean}
R.~Kaas and J.~M. Buhrman.
\newblock Mean, median and mode in binomial distributions.
\newblock \emph{Statistica Neerlandica}, 34\penalty0 (1):\penalty0 13--18,
  1980.

\bibitem[Kaufmann et~al.(2024)Kaufmann, Andre, and
  K{\H{o}}szegi]{KaufmannAndreKoszegi2024}
Marc Kaufmann, Peter Andre, and Botond K{\H{o}}szegi.
\newblock Understanding markets with socially responsible consumers.
\newblock \emph{The Quarterly Journal of Economics}, 139\penalty0 (3):\penalty0
  1989--2035, 2024.

\bibitem[Korinek(2025)]{korinek2025agents}
Anton Korinek.
\newblock {AI} agents for economic research.
\newblock NBER Working Paper 34202, National Bureau of Economic Research, 2025.

\bibitem[Kuchibhotla et~al.(2022)Kuchibhotla, Kolassa, and
  Kuffner]{kuchibhotla2022post}
Arun~K Kuchibhotla, John~E Kolassa, and Todd~A Kuffner.
\newblock Post-selection inference.
\newblock \emph{Annual Review of Statistics and Its Application}, 9:\penalty0
  505--527, 2022.

\bibitem[Lee et~al.(2016)Lee, Sun, Sun, and Taylor]{lee2016exact}
Jason~D Lee, Dennis~L Sun, Yuekai Sun, and Jonathan~E Taylor.
\newblock Exact post-selection inference, with application to the lasso.
\newblock \emph{The Annals of Statistics}, 44\penalty0 (3):\penalty0 907--927,
  2016.

\bibitem[Lee et~al.(2026)Lee, Ma, Zhao, Nair, Spector, Cohen, and
  Cand{\`e}s]{lee2026fuse}
Joonhyuk Lee, Virginia Ma, Sarah Zhao, Yash Nair, Asher Spector, Regev Cohen,
  and Emmanuel~J Cand{\`e}s.
\newblock {FUSE}: Ensembling verifiers with zero labeled data.
\newblock \emph{arXiv preprint arXiv:2604.18547}, 2026.

\bibitem[Liu and Panigrahi(2025)]{liu2025flexible}
Sifan Liu and Snigdha Panigrahi.
\newblock Flexible selective inference with flow-based transport maps.
\newblock \emph{arXiv preprint arXiv:2506.01150}, 2025.

\bibitem[Liu et~al.(2022)Liu, Markovic-Voronov, and Taylor]{liu2022black}
Sifan Liu, Jelena Markovic-Voronov, and Jonathan Taylor.
\newblock Black-box selective inference via bootstrapping.
\newblock \emph{arXiv preprint arXiv:2203.14504}, 2022.

\bibitem[Lucas and {Open Interpreter
  Contributors}(2023)]{lucas2023openinterpreter}
Killian Lucas and {Open Interpreter Contributors}.
\newblock {Open Interpreter}: A natural language interface for computers,
  September 2023.
\newblock URL \url{https://github.com/openinterpreter/openinterpreter}.
\newblock Open-source software, publicly launched September 5, 2023.

\bibitem[Lyttelton et~al.(2026)Lyttelton, Massenkoff, and
  Wilmers]{lyttelton2026agents}
Thomas Lyttelton, Maxim Massenkoff, and Nathan Wilmers.
\newblock Coding agents in the social sciences.
\newblock Anthropic Research, May 2026.
\newblock URL
  \url{https://www.anthropic.com/research/coding-agents-social-sciences}.

\bibitem[Majumder et~al.(2025)Majumder, Surana, Agarwal, Dalvi~Mishra, Meena,
  Prakhar, Vora, Khot, Sabharwal, and Clark]{majumder2025discoverybench}
Bodhisattwa~Prasad Majumder, Harshit Surana, Dhruv Agarwal, Bhavana
  Dalvi~Mishra, Abhijeetsingh Meena, Aryan Prakhar, Tirth Vora, Tushar Khot,
  Ashish Sabharwal, and Peter Clark.
\newblock {DiscoveryBench}: Towards data-driven discovery with large language
  models.
\newblock In \emph{International Conference on Learning Representations}, 2025.

\bibitem[Markovic et~al.(2019)Markovic, Taylor, and
  Taylor]{markovic2019inference}
Jelena Markovic, Jonathan Taylor, and Jeremy Taylor.
\newblock Inference after black box selection.
\newblock \emph{arXiv preprint arXiv:1901.09973}, 2019.

\bibitem[Massart(1990)]{massart1990tight}
Pascal Massart.
\newblock The tight constant in the {Dvoretzky-Kiefer-Wolfowitz} inequality.
\newblock \emph{The Annals of Probability}, 18\penalty0 (3):\penalty0
  1269--1283, 1990.

\bibitem[McCloskey(2017)]{mccloskey2017bonferroni}
Adam McCloskey.
\newblock Bonferroni-based size-correction for nonstandard testing problems.
\newblock \emph{Journal of Econometrics}, 200\penalty0 (1):\penalty0 17--35,
  2017.

\bibitem[McCloskey and Michaillat(2026)]{mccloskey2026critical}
Adam McCloskey and Pascal Michaillat.
\newblock Critical values robust to p-hacking.
\newblock \emph{Review of Economics and Statistics}, 108\penalty0 (4):\penalty0
  1085--1097, 2026.

\bibitem[McCully(2026)]{mccully2026research}
Brett~A McCully.
\newblock Research choices of agentic artificial intelligence.
\newblock \emph{Economics Letters}, 269:\penalty0 113264, 2026.

\bibitem[{Menlo Ventures}(2026)]{menloventures2026enterprise}
{Menlo Ventures}.
\newblock 2026: The state of enterprise {AI}, 2026.
\newblock URL
  \url{https://menlovc.com/perspective/2026-the-state-of-enterprise-ai/}.

\bibitem[Miao et~al.(2026)Miao, Pritchard, and Zou]{miao2026agentic}
Jiacheng Miao, Jonathan~K Pritchard, and James Zou.
\newblock The agentic garden of forking paths.
\newblock \emph{arXiv preprint arXiv:2607.01507}, 2026.

\bibitem[Neufeld et~al.(2026)Neufeld, Perry, and Witten]{neufeld2026inference}
Anna Neufeld, Ronan Perry, and Daniela Witten.
\newblock Inference conditional on selection: a review.
\newblock \emph{arXiv preprint arXiv:2604.09779}, 2026.

\bibitem[Neyman(1934)]{neyman1992two}
Jerzy Neyman.
\newblock On the two different aspects of the representative method: The method
  of stratified sampling and the method of purposive selection.
\newblock \emph{Journal of the Royal Statistical Society}, 97\penalty0
  (4):\penalty0 558--625, 1934.
\newblock \doi{10.2307/2342192}.

\bibitem[{OpenAI}(2026)]{openai2026reasoning}
{OpenAI}.
\newblock Reasoning models.
\newblock API documentation, 2026.
\newblock URL \url{https://developers.openai.com/api/docs/guides/reasoning}.
\newblock Accessed: 2026-11-12.

\bibitem[Pinsker(1964)]{pinsker1964information}
Mark~S Pinsker.
\newblock \emph{Information and information stability of random variables and
  processes}.
\newblock Holden-Day, San Francisco, 1964.

\bibitem[Politis and Romano(1994)]{politis1994large}
Dimitris~N Politis and Joseph~P Romano.
\newblock Large sample confidence regions based on subsamples under minimal
  assumptions.
\newblock \emph{The Annals of Statistics}, 22\penalty0 (4):\penalty0
  2031--2050, 1994.

\bibitem[Queen and Zou(2026)]{queen2026shelflife}
Owen Queen and James Zou.
\newblock The shelf life of {LLM} findings, November 2026.
\newblock URL \url{https://zou-group.github.io/llm-claims-shelf-life/}.
\newblock Accessed: 2026-11-05.

\bibitem[Romano and Shaikh(2012)]{romano2012uniform}
Joseph~P. Romano and Azeem~M. Shaikh.
\newblock On the uniform asymptotic validity of subsampling and the bootstrap.
\newblock \emph{The Annals of Statistics}, 40\penalty0 (6):\penalty0
  2798--2822, 2012.
\newblock \doi{10.1214/12-AOS1051}.
\newblock URL \url{https://doi.org/10.1214/12-AOS1051}.

\bibitem[R{\"u}schendorf(2009)]{ruschendorf2009distributional}
Ludger R{\"u}schendorf.
\newblock On the distributional transform, {Sklar}'s theorem, and the empirical
  copula process.
\newblock \emph{Journal of Statistical Planning and Inference}, 139\penalty0
  (11):\penalty0 3921--3927, 2009.

\bibitem[Russell and Norvig(2021)]{russell2021artificial}
Stuart Russell and Peter Norvig.
\newblock \emph{Artificial Intelligence: A Modern Approach}.
\newblock Pearson, fourth edition, 2021.

\bibitem[Silberzahn et~al.(2018)Silberzahn, Uhlmann, Martin, Anselmi, Aust,
  Awtrey, Bahn{\'\i}k, Bai, Bannard, Bonnier, et~al.]{silberzahn2018many}
Raphael Silberzahn, Eric~L Uhlmann, Daniel~P Martin, Pasquale Anselmi, Frederik
  Aust, Eli Awtrey, {\v{S}}t{\v{e}}p{\'a}n Bahn{\'\i}k, Feng Bai, Colin
  Bannard, Evelina Bonnier, et~al.
\newblock Many analysts, one data set: Making transparent how variations in
  analytic choices affect results.
\newblock \emph{Advances in Methods and Practices in Psychological Science},
  1\penalty0 (3):\penalty0 337--356, 2018.

\bibitem[Simmons et~al.(2011)Simmons, Nelson, and Simonsohn]{simmons2011false}
Joseph~P Simmons, Leif~D Nelson, and Uri Simonsohn.
\newblock False-positive psychology: Undisclosed flexibility in data collection
  and analysis allows presenting anything as significant.
\newblock \emph{Psychological Science}, 22\penalty0 (11):\penalty0 1359--1366,
  2011.

\bibitem[Sood(2024)]{sood2024selective}
Anav Sood.
\newblock Selective inference is easier with p-values.
\newblock \emph{arXiv preprint arXiv:2411.13764}, 2024.

\bibitem[Tian and Taylor(2018)]{tian2018selective}
Xiaoying Tian and Jonathan Taylor.
\newblock Selective inference with a randomized response.
\newblock \emph{The Annals of Statistics}, 46\penalty0 (2):\penalty0 679--710,
  2018.

\bibitem[Tibshirani et~al.(2016)Tibshirani, Taylor, Lockhart, and
  Tibshirani]{tibshirani2016exact}
Ryan~J Tibshirani, Jonathan Taylor, Richard Lockhart, and Robert Tibshirani.
\newblock Exact post-selection inference for sequential regression procedures.
\newblock \emph{Journal of the American Statistical Association}, 111\penalty0
  (514):\penalty0 600--620, 2016.

\bibitem[Trivedy(2026)]{trivedy2026harness}
Vivek Trivedy.
\newblock The anatomy of an agent harness.
\newblock LangChain blog, March 2026.
\newblock URL
  \url{https://www.langchain.com/blog/the-anatomy-of-an-agent-harness}.
\newblock Accessed: 2026-11-12.

\bibitem[Vaswani et~al.(2017)Vaswani, Shazeer, Parmar, Uszkoreit, Jones, Gomez,
  Kaiser, and Polosukhin]{vaswani2017attention}
Ashish Vaswani, Noam Shazeer, Niki Parmar, Jakob Uszkoreit, Llion Jones,
  Aidan~N Gomez, {\L}ukasz Kaiser, and Illia Polosukhin.
\newblock Attention is all you need.
\newblock \emph{Advances in Neural Information Processing Systems}, 30, 2017.

\bibitem[Wang and Shan(2015)]{wang2015exact}
Weizhen Wang and Guogen Shan.
\newblock Exact confidence intervals for the relative risk and the odds ratio.
\newblock \emph{Biometrics}, 71\penalty0 (4):\penalty0 985--995, 2015.

\bibitem[Waudby-Smith and Ramdas(2024)]{waudbysmith2024estimating}
Ian Waudby-Smith and Aaditya Ramdas.
\newblock Estimating means of bounded random variables by betting.
\newblock \emph{Journal of the Royal Statistical Society Series B: Statistical
  Methodology}, 86\penalty0 (1):\penalty0 1--27, 2024.

\bibitem[Wei et~al.(2022)Wei, Wang, Schuurmans, Bosma, Ichter, Xia, Chi, Le,
  and Zhou]{wei2022chain}
Jason Wei, Xuezhi Wang, Dale Schuurmans, Maarten Bosma, Brian Ichter, Fei Xia,
  Ed~H. Chi, Quoc~V. Le, and Denny Zhou.
\newblock Chain-of-thought prompting elicits reasoning in large language
  models.
\newblock In \emph{Advances in Neural Information Processing Systems},
  volume~35, pages 24824--24837, 2022.

\bibitem[White(2000)]{white2000reality}
Halbert White.
\newblock A reality check for data snooping.
\newblock \emph{Econometrica}, 68\penalty0 (5):\penalty0 1097--1126, 2000.

\bibitem[Willison(2025)]{willison2025agents}
Simon Willison.
\newblock I think ``agent'' may finally have a widely enough agreed upon
  definition to be useful jargon now, September 2025.
\newblock URL \url{https://simonwillison.net/2025/Sep/18/agents/}.
\newblock Accessed: 2026-11-12.

\bibitem[Yang et~al.(2024)Yang, Jimenez, Wettig, Lieret, Yao, Narasimhan, and
  Press]{yang2024sweagent}
John Yang, Carlos~E. Jimenez, Alexander Wettig, Kilian Lieret, Shunyu Yao,
  Karthik Narasimhan, and Ofir Press.
\newblock {SWE}-agent: Agent-computer interfaces enable automated software
  engineering.
\newblock In \emph{Advances in Neural Information Processing Systems},
  volume~37, pages 50528--50652, 2024.

\bibitem[Yao et~al.(2023)Yao, Zhao, Yu, Du, Shafran, Narasimhan, and
  Cao]{yao2023react}
Shunyu Yao, Jeffrey Zhao, Dian Yu, Nan Du, Izhak Shafran, Karthik Narasimhan,
  and Yuan Cao.
\newblock {ReAct}: Synergizing reasoning and acting in language models.
\newblock In \emph{International Conference on Learning Representations}, 2023.

\bibitem[Zheng et~al.(2023)Zheng, Chiang, Sheng, Zhuang, Wu, Zhuang, Lin, Li,
  Li, Xing, et~al.]{zheng2023judging}
Lianmin Zheng, Wei-Lin Chiang, Ying Sheng, Siyuan Zhuang, Zhanghao Wu, Yonghao
  Zhuang, Zi~Lin, Zhuohan Li, Dacheng Li, Eric Xing, et~al.
\newblock Judging {LLM}-as-a-judge with {MT-Bench} and {Chatbot Arena}.
\newblock \emph{Advances in Neural Information Processing Systems},
  36:\penalty0 46595--46623, 2023.

\bibitem[Zrnic and Jordan(2023)]{zrnic2023post}
Tijana Zrnic and Michael~I Jordan.
\newblock Post-selection inference via algorithmic stability.
\newblock \emph{The Annals of Statistics}, 51\penalty0 (4):\penalty0
  1666--1691, 2023.

\end{thebibliography}

\clearpage
\section{Proofs}

\begin{proof}[Proof of Proposition \ref{prop:subsampling_selection}.]
Let $\mu\coloneqq\mathcal L(D_n\mid Q=q)$ and for $S\subseteq\{1,\dots,n\}$ with $|S|=m$, let $\mu_S$ denote the marginal law of $D_S$ under $\mu$. So for every measurable set $B$ of size-$m$ datasets,
\[
\mu_S(B) =\mu\bigl(\{d:d_S\in B\}\bigr) =\Pr(D_S\in B\mid Q=q).
\]
Defining $s(x)\coloneqq\Pr(Q=q\mid D_n=x)\in[0,1]$, by Bayes' rule and conditioning on $D_n$, for every measurable set $A$ of size-$n$ datasets we have
\[
\mu(A)=\Pr(D_n\in A\mid Q=q)=\frac{\Pr(D_n\in A,\ Q=q)}{\Pr(Q=q)}
=\frac{1}{\rho_{q,n}}\int_A s(x)\,P^n(dx).
\]
This means that $\mu$ is absolutely continuous with respect
to $P^n$ with density
\[
\frac{d\mu}{dP^n}(x)=\frac{s(x)}{\rho_{q,n}}
\qquad P^n\text{-almost everywhere}.
\]
Using this density and the fact that $s(x)\le 1$, we have
\[
\mathrm{KL}(\mu\|P^n)=\int \log\!\left(\frac{d\mu}{dP^n}(x)\right)\,\mu(dx)
=\int \log s(x)\,\mu(dx)+\log\frac{1}{\rho_{q,n}}\le \log\frac{1}{\rho_{q,n}}.
\]
This bounds the selection-induced divergence of the full dataset.
We next use Han's inequality to control the average divergence
of its size-$m$ coordinate marginals in terms of this full-dataset
divergence. In fact, since $P^n$ is a product measure, Han's inequality for relative entropy
\citetext{\citealp{han1978nonnegative}; \citealp[Theorem~4.9]{boucheron2013concentration}} gives, for every $k\le n$ and every
$T$ with $|T|=k$, $k^{-1}\sum_{i\in T}\mathrm{KL}(\mu_{T\setminus\{i\}}\|P^{k-1})
\le\frac{k-1}{k}\mathrm{KL}(\mu_T\|P^k)$. Averaging over $T$ and iterating
from $k=n$ down to $k=m+1$ yields
\[
\binom{n}{m}^{-1}\sum_{|S|=m}\mathrm{KL}\big(\mu_S\,\|\,P^m\big)
\le\frac{m}{n}\,\mathrm{KL}\big(\mu\,\|\,P^n\big).
\]
Because $S_m$ is independent of $(D_n,\Xi)$, and hence of $Q$, each subset is
drawn with probability $\binom{n}{m}^{-1}$ regardless of whether $q$ is
selected, so $\mathcal L(D_{S_m}\mid Q=q)=\binom{n}{m}^{-1}\sum_{|S|=m}\mu_S$.
Hence
\begin{align*}
\mathrm{KL}\big(\mathcal L(D_{S_m}\mid Q=q)\,\|\,P^m\big)
&\le\binom{n}{m}^{-1}\sum_{|S|=m}\mathrm{KL}\big(\mu_S\,\|\,P^m\big)
&&\text{(convexity of relative entropy)}\\
&\le\frac{m}{n}\,\mathrm{KL}\big(\mu\,\|\,P^n\big)
&&\text{(Han's inequality)}\\
&\le\frac{m}{n}\log\frac{1}{\rho_{q,n}},
\end{align*}
and Pinsker's inequality \citetext{\citealp{pinsker1964information}; \citealp[Theorem~4.19]{boucheron2013concentration}} gives the result.
\end{proof}

\section{Proofs for statistical results in Section~\ref{sec:statistical_results}}

\label{app:statistical_proofs}
\begin{proof}[Proof of Proposition~\ref{prop:selected_rank_general}]
Conditional on $Q_n=q$, $H_{q,n}$ is the cdf of $U_{q,n}$. For any cdf $H$, let $X\sim H$ and let $V\sim\mathrm{Unif}(0,1)$ be independent of $X$. The randomized probability integral transform
\[
T=H(X^-)+V\bigl[H(X)-H(X^-)\bigr],
\]
where $H(x^-)=\lim_{y\uparrow x}H(y)$, is uniformly distributed on $[0,1]$ \citep[Prop. 2.1]{ruschendorf2009distributional}.
Writing
\[
\Delta_H\coloneqq\sup_x\bigl[H(x)-H(x^-)\bigr]
\]
for the largest jump of $H$, we have
$ 0\le H(X)-T =(1-V)\bigl[H(X)-H(X^-)\bigr] \le\Delta_H.$
In our setting, every such jump is bounded by
\[
\Pr_P(U_{q,n}=u\mid Q_n=q)
\le\frac{F_{q,n}(u)-F_{q,n}(u^-)}{\rho_{q,n}}
\le \frac{2}{\rho_{q, n}} \sup _{v \in[0,1]}\left|F_{q, n}(v)-v\right| \longrightarrow 0,
\]
by Assumptions~\ref{asspt:base_calibration} and \ref{asspt:recurrence}.
As a result, for $V_{q,n}=H_{q,n}(U_{q,n})$,
\begin{equation}\label{eq:oracle_PIT_error}
0\le \alpha-\Pr_P(V_{q,n}\le\alpha\mid Q_n=q)
\to 0.
\end{equation}
\end{proof}

\begin{lemma}[Subsampling cdf bound]
\label{lem:uniform_subsampling_cdf}
Let $T$ be any measurable statistic of a dataset of size $m$ which may depend on the ordering of its observations and on auxiliary randomness $\Xi$. Let $G$ be its cdf under $P^m$ and $G_n^S$ its conditional cdf under a uniformly drawn and randomly ordered size-$m$ subset of $D_n$, that is,
\[
G_n^S(t):=
\Pr\left(T(D_{S_m},\Xi')\le t\mid D_n\right).
\]
Under Assumption~\ref{asspt:sampling_units}, with
$K=\lfloor n/m\rfloor$,
\[
\Pr_P\bigl(\|G_n^S-G\|_\infty>x\bigr)\le 3e^{-Kx^2}.
\]
Conditional on $Q_n=q$, this probability is at most $3e^{-Kx^2}/\rho_{q,n}$.
\end{lemma}

\begin{proof}
Randomly partition the $n$ observations into $K$ disjoint blocks of size $m$. Let $\widehat G_K$ be the empirical cdf of $T$
on these blocks, using the independence of $\Xi$ across blocks. Unconditionally,
the block statistics are i.i.d.\ with cdf $G$. As a result, the
Dvoretzky--Kiefer--Wolfowitz--Massart inequality
\citep{dvoretzky1956asymptotic, massart1990tight} gives
\[
\Pr_P\bigl(\|\widehat G_K-G\|_\infty>x\bigr)
\le 2e^{-2Kx^2}.
\]
Since each block is marginally a uniformly drawn subset of size $m$,
$G_n^S=\E(\widehat G_K\mid D_n)$, and we have
\[
\begin{aligned}
\E\exp\{K\|G_n^S-G\|_\infty^2\}
&\le \E\exp\{K\|\widehat G_K-G\|_\infty^2\}\\
&=1+\int_0^\infty e^t
\Pr\!\left(K\|\widehat G_K-G\|_\infty^2>t\right)\,dt\\
&\le 1+2\int_0^\infty e^{-t}\,dt=3,
\end{aligned}
\]
by Jensen's inequality, the layer-cake representation, and the tail bound above, respectively. Then, Markov's inequality gives the first claim. The second claim follows by noting that
$\Pr_P(E\mid Q_n=q)\le \Pr_P(E)/\rho_{q,n}$ for any event $E$.
\end{proof}

\begin{lemma}[Stability of selected-rank distributions]
\label{lem:random_threshold_stability}
Under Assumptions~\ref{asspt:base_calibration},
\ref{asspt:lambda_stability}(a), and~\ref{asspt:recurrence}, for every
$\varepsilon\in(0,1/2)$,
\[
\sup_{u\in[\varepsilon,1-\varepsilon]}
[H_{q,n}(u)-H_{q,m_n}(u)]_+\longrightarrow0.
\]
If Assumption \ref{asspt:lambda_stability}(b) holds, the conclusion here also holds with an absolute value.
\end{lemma}

\begin{proof}
Write $\kappa_{q,k}=\sup_u|F_{q,k}(u)-u|$, which tends to zero
for both $k=n,m_n$. For any $0\le a<b\le1$ and either sample size $k=n, m_n$,
\[
H_{q,k}(b)-H_{q,k}(a)
\le \frac{F_{q,k}(b)-F_{q,k}(a)}{\rho_{q,k}}
\le \frac{b-a+2\kappa_{q,k}}{\rho_\star}.
\]
At each fixed $u>0$, the identity $H_{q,k}(u)=F_{q,k}(u)\lambda_{q,k}(u)$ and
Assumptions \ref{asspt:base_calibration} and \ref{asspt:lambda_stability} give
\[
[H_{q,n}(u)-H_{q,m_n}(u)]_+
\le [\lambda_{q,n}(u)-\lambda_{q,m_n}(u)]_+
+\frac{\kappa_{q,n}+\kappa_{q,m_n}}{\rho_\star}
\longrightarrow0.
\]
Then, fix a finite grid $\varepsilon=t_0<\cdots<t_L=1-\varepsilon$
with mesh at most $h$. By monotonicity of the cdfs,
for any $u\in[t_j,t_{j+1}]$,
\begin{equation} \label{eq:pointwise_H_q}
\begin{aligned}
H_{q,n}(u)-H_{q,m_n}(u)
&\le H_{q,n}(t_{j+1})-H_{q,m_n}(t_j)\\
&\le H_{q,n}(t_{j+1})-H_{q,m_n}(t_{j+1})
+\frac{h+2\kappa_{q,m_n}}{\rho_\star},
\end{aligned}
\end{equation}
where the second inequality follows from the first bound above. As a result,
\[
\begin{aligned}
&\limsup_{n \to \infty} \sup_{u\in[\varepsilon,1-\varepsilon]}
[H_{q,n}(u)-H_{q,m_n}(u)]_+\\
&\le
\limsup_{n \to \infty} \left\{ \max_{j=0,\ldots,L}
[H_{q,n}(t_j)-H_{q,m_n}(t_j)]_+
+\frac{h+2\kappa_{q,m_n}}{\rho_\star} \right\} = \frac{h}{\rho_\star} ,
\end{aligned}
\]
where the maximum vanishes by the pointwise convergence shown above since we work on a finite grid, and $\kappa_{q,m_n}\to0$ by Assumption~\ref{asspt:base_calibration}.
Since $h>0$ is arbitrary, the result follows. The case with an absolute value also follows by the same argument with $n$ and $m_n$ interchanged.
\end{proof}

\begin{proof}[Proof of Theorem~\ref{thm:asymptotic_validity}]
Fix $q$ and $P\in\mathcal P_{0,q}$. Write $u_n=U_{q,n}$,
$m=m_n$, and $V_n=H_{q,n}(u_n)$.

We first establish consistency of the replay selection probabilities $\widehat a_{1,q}, \widehat a_{0,q}$. To this end, we show that subsampling consistently approximates the joint distribution of the original p-value and the agent's
selection event. Apply Lemma~\ref{lem:uniform_subsampling_cdf}
to $U_{q,m}$ and the auxiliary variable
\[
Z_{q,m}:=
\begin{cases}
U_{q,m}, & Q_m=q,\\
2, & Q_m\ne q.
\end{cases}
\]
For $u\in[0,1]$, their cdfs identify, respectively,
\[
F_{q,m}(u)
\quad\text{and}\quad
A_{q,m}(u):=\Pr_P(U_{q,m}\le u,Q_m=q),
\]
while $\Pr_P(Z_{q,m}\le1)=\rho_{q,m}$.

Define the conditional subsampling distributions
\[
\begin{aligned}
F_{q,n,m}^S(u)
&:=\Pr\left(p_q(D_{S_m})\le u\mid D_n\right),\\
A_{q,n,m}^S(u)
&:=\Pr\left(p_q(D_{S_m})\le u,\,
Q_{S_m}=q\mid D_n\right),
\end{aligned}
\]
where $Q_{S_m}$ denotes the analysis selected by a
fresh agent on dataset $D_{S_m}$. Both probabilities here are taken over
uniformly drawn size-$m$ subsamples $S_m$ of $D_n$ and
the agent's randomness, conditional on $D_n$.

By Lemma~\ref{lem:uniform_subsampling_cdf}, applied to
$U_{q,m}$ and $Z_{q,m}$, and since $m/n\to0$,
\[
\sup_{u\in[0,1]}
\left(
|F_{q,n,m}^S(u)-F_{q,m}(u)|
+
|A_{q,n,m}^S(u)-A_{q,m}(u)|
\right)
\stackrel{p}{\longrightarrow}0
\]
conditional on $Q_n=q$. Evaluating the
second term at $u=1$ also gives convergence of the
subsampling selection probability to $\rho_{q,m}$.
Importantly, the uniformity of these results also gives convergence of these cdfs when evaluated at the random threshold $u_n = p_q(D_n)$.

Then, for $\varepsilon\in(0,1/2)$,
let $E_{n,\varepsilon}=\{\varepsilon\le u_n\le1-\varepsilon\}$. On $E_{n,\varepsilon}$, Assumption~\ref{asspt:base_calibration}
implies that
\[
\min\{F_{q,m}(u_n),1-F_{q,m}(u_n)\}
\ge \varepsilon-\kappa_{q,m},
\]
where $\kappa_{q,k}=\sup_u|F_{q,k}(u)-u|$.
Hence, by the preceding uniform convergence result,
\[
\begin{aligned}
&\min\{F_{q,n,m}^S(u_n),
       1-F_{q,n,m}^S(u_n)\}\\
& \quad \geq \min \left\{F_{q, m}\left(u_n\right), 1-F_{q, m}\left(u_n\right)\right\} -\left|F_{q,n,m}^S\left(u_n\right)-F_{q, m}\left(u_n\right)\right| \\
&\quad\ge \varepsilon-\kappa_{q,m}
-\sup_{u\in[0,1]}
|F_{q,n,m}^S(u)-F_{q,m}(u)|\\
&\quad=\varepsilon-o_p(1).
\end{aligned}
\]
Thus, with probability approaching one, a subsample falls in each of the tail and body strata with probability at least $\varepsilon/2$ conditional on $D_n$, so that the draws fill both strata after finitely many steps almost surely.

Define the selection probabilities within each stratum under the subsampling law given $D_n$ as
\[
\begin{aligned}
b_{1,n}
&:=\frac{A_{q,n,m}^S(u_n)}
{F_{q,n,m}^S(u_n)},\\
b_{0,n}
&:=\frac{A_{q,n,m}^S(1)
-A_{q,n,m}^S(u_n)}
{1-F_{q,n,m}^S(u_n)}.
\end{aligned}
\]
These are the stratum selection probabilities $b_{s,n}$ of Section~\ref{sec:replay_uncertainty}. Recall that the corresponding population selection
probabilities satisfy
\[
\begin{aligned}
a_{1,q,m}(u_n)
&=\frac{A_{q,m}(u_n)}{F_{q,m}(u_n)},\\
a_{0,q,m}(u_n)
&=\frac{\rho_{q,m}-A_{q,m}(u_n)}
{1-F_{q,m}(u_n)}.
\end{aligned}
\]
Since both denominators are at least $\varepsilon-o_p(1)$ on $E_{n,\varepsilon}$, as shown above, the preceding uniform convergence result then implies
\[
\mathbf 1_{E_{n,\varepsilon}}
\left|b_{s,n}
-a_{s,q,m}(u_n)\right|
\stackrel{p}{\longrightarrow}0,
\qquad s=0,1,
\]
conditional on $Q_n=q$.

It remains to show that the actual replay frequencies
consistently estimate these subsampling probabilities. Fix a stratum
$s\in\{0,1\}$, and let $D_{s,1},D_{s,2},\ldots$ denote the subsamples that fall in stratum $s$, in the order in which they are drawn. Conditional on $D_n$, these are i.i.d.\ draws from the subsampling law given $D_n$ restricted to stratum $s$, by Assumption~\ref{asspt:sampling_units} and because stratum membership is a function of the subsample. After $j$ replays, the
selection frequency is
\[
\widehat a_{s,q}^{(j)}(u_n)
=\frac{1}{j}\sum_{i=1}^j
\mathbf{1}\{
\mathcal A_r(D_{s,i},\Xi_{s,i})=q
\},
\]
where the $\Xi_{s,i}$ are independent draws of the
agent's randomness. Conditional on $D_n$, the summands are therefore i.i.d.\ Bernoulli variables with mean $b_{s,n}$, and Hoeffding's inequality
\citep{hoeffding1963probability} gives
\[
\Pr\left(
\left|\widehat a_{s,q}^{(j)}(u_n)
-b_{s,n}\right|>\delta
\mid D_n
\right)
\le 2e^{-2j\delta^2}.
\]

The same bound accommodates adaptive allocation between strata, because the $i$-th replay in stratum $s$ uses the $i$-th subsample to fall in that stratum, whatever the replay counts.
Indeed, for any integer $M\ge1$ and $\delta>0$,
\[
\begin{aligned}
&\Pr\left(
E_{n,\varepsilon}\cap
\left\{
\left|\widehat a_{s,q}^{(B_{s,n})}(u_n)
-b_{s,n}\right|>\delta
\right\}
\mid Q_n=q
\right)\\
&\quad\le
\Pr(B_{s,n}<M\mid Q_n=q)
+\sum_{j=M}^{\infty}2e^{-2j\delta^2}\\
&\quad=
\Pr(B_{s,n}<M\mid Q_n=q)
+\frac{2e^{-2M\delta^2}}
{1-e^{-2\delta^2}}.
\end{aligned}
\]
Since $B_{s,n}\to\infty$ in probability conditional
on $Q_n=q$, letting $n\to\infty$ and then
$M\to\infty$ gives
\[
\mathbf1_{E_{n,\varepsilon}}
\left|\widehat a_{s,q}(u_n)
-b_{s,n}\right|
\stackrel{p}{\longrightarrow}0
\]
conditional on $Q_n=q$.

Combining the two convergence results by the
triangle inequality finally gives
\[
\mathbf 1_{E_{n,\varepsilon}}
\left|\widehat a_{s,q}(u_n)
-a_{s,q,m}(u_n)\right|
\stackrel{p}{\longrightarrow}0,
\qquad s=0,1,
\]
conditional on $Q_n=q$.

Next, we relate the population counterpart of our estimator
to the selected-rank cdf. Define
\[
\widetilde\rho_{q,m}(u)
:=u\,a_{1,q,m}(u)+(1-u)a_{0,q,m}(u),
\qquad
R_{q,m}(u)
:=\frac{u\,a_{1,q,m}(u)}{\widetilde\rho_{q,m}(u)}.
\]
By the law of total probability,
\[
\widetilde\rho_{q,m}(u)-\rho_{q,m}
=
[u-F_{q,m}(u)]
[a_{1,q,m}(u)-a_{0,q,m}(u)],
\]
so that
$|\widetilde\rho_{q,m}(u)-\rho_{q,m}|
\le\kappa_{q,m}$.
Recurrence ensures that both denominators are bounded away
from zero for sufficiently large $n$. Since
$R_{q,m}(u)\le1$ and $\lambda_{q,m}(u)\le1/\rho_\star$,
\[
\begin{aligned}
|R_{q,m}(u)-u\lambda_{q,m}(u)|
&\le\frac{\kappa_{q,m}}{\rho_\star},\\
|u\lambda_{q,m}(u)-H_{q,m}(u)|
&\le\frac{\kappa_{q,m}}{\rho_\star}.
\end{aligned}
\]

Hence, uniformly over $u$,
\[
|R_{q,m}(u)-H_{q,m}(u)|
\le\frac{2\kappa_{q,m}}{\rho_\star}
\longrightarrow0.
\]
Moreover, consistency of the replay frequencies at
$u_n$, together with the denominator bound above
and continuity of the ratio map, which allows application of the continuous mapping theorem, yields
\[
\mathbf1_{E_{n,\varepsilon}}
|\widehat p_{q,n}^{\mathrm{sel}}
-R_{q,m}(u_n)|
\stackrel{p}{\longrightarrow}0
\]
conditional on $Q_n=q$. Evaluating the uniform
bound at $u_n$ and applying the triangle
inequality therefore gives
\[
\mathbf1_{E_{n,\varepsilon}}
|\widehat p_{q,n}^{\mathrm{sel}}
-H_{q,m}(u_n)|
\stackrel{p}{\longrightarrow}0.
\]

Lemma~\ref{lem:random_threshold_stability} then gives,
\[
\mathbf1_{E_{n,\varepsilon}}
[H_{q,n}(u_{n})-\widehat p_{q,n}^{\mathrm{sel}}]_+
\stackrel{p}{\longrightarrow}0
\]
conditional on $Q_n=q$.
Moreover, calibration (Assumption~\ref{asspt:base_calibration}) and recurrence (Assumption~\ref{asspt:recurrence}) imply
\[
\Pr_P(E_{n,\varepsilon}^c\mid Q_n=q)
\le\frac{2\varepsilon+2\kappa_{q,n}}{\rho_\star}.
\]
Taking $n\to\infty$ and then $\varepsilon\downarrow0$
thus gives
\[
[H_{q,n}(u_n)-\widehat p_{q,n}^{\mathrm{sel}}]_+
\stackrel{p}{\longrightarrow}0
\qquad\text{conditional on }Q_n=q.
\]

Finally, since $H_{q,n}(u_n)$ is conditionally super-uniform by
Proposition~\ref{prop:selected_rank_general}, we have, by the union bound and the preceding convergence result, for every $\eta \in(0,1-\alpha)$,
\[
\begin{aligned}
&\Pr_P(\widehat p_{q,n}^{\mathrm{sel}}\le\alpha\mid Q_n=q)\\
&\quad\le
\Pr_P(H_{q,n}(u_n)\le\alpha+\eta\mid Q_n=q)\\
&\qquad+
\Pr_P(H_{q,n}(u_n)-\widehat p_{q,n}^{\mathrm{sel}}>\eta\mid Q_n=q)\\
&\quad\le\alpha+\eta+o(1).
\end{aligned}
\]
Letting $\eta\downarrow0$ proves the result. Under two-sided
pointwise stability, the same argument yields
$|V_n-\widehat p_{q,n}^{\mathrm{sel}}|
\stackrel{p}{\longrightarrow}0$.
Asymptotic exactness then follows from
Proposition~\ref{prop:selected_rank_general}.
For the multiplier, on $u_n\ge\varepsilon$,
\[
\begin{aligned}
&\left|\widehat\lambda_q(u_n)
-\lambda_{q,n}(u_n)\right|\\
&\quad\le
\frac{\left|\widehat p_{q,n}^{\mathrm{sel}}
-H_{q,n}(u_n)\right|}{\varepsilon}
+\frac{\kappa_{q,n}}{\varepsilon\rho_\star}
\stackrel{p}{\longrightarrow}0.
\end{aligned}
\]
Since
\[
\Pr_P(u_n<\varepsilon\mid Q_n=q)
\le\frac{\varepsilon+\kappa_{q,n}}{\rho_\star},
\]
letting $\varepsilon\downarrow0$ proves the
second claim in \eqref{eq:oracle_equivalence}.

\end{proof}

\begin{proof}[Proof of Corollary \ref{cor:uniform_validity}]
Throughout, we write $m=m_n$, $u_n=U_{q,n}$, and $V_n^P=H^P_{q,n}(u_n)$. All suprema over $P$ below are taken over the set of null distributions $\mathfrak P_{0,q}$. For $\varepsilon\in(0,1/2)$, we define
\[
I_\varepsilon=[\varepsilon,1-\varepsilon],
\quad
E_{n,\varepsilon}=\{u_n\in I_\varepsilon\}.
\]

We first make the estimation argument in the proof of Theorem~\ref{thm:asymptotic_validity} uniform over $P$. The subsampling bounds used there are distribution-free. Conditioning on $Q_n=q$ increases an unconditional probability bound by at most a factor $1/\rho_\star$. Since $\lfloor n/m_n\rfloor\to\infty$, the corresponding approximation errors vanish uniformly in conditional probability over $\mathfrak P_{0,q}$.

For adaptive replay counts, the union bound over replay prefixes in the same proof bounds the probability that either replay frequency is unavailable or differs from its stratum selection probability $b_{s,n}$ under the subsampling law by more than $\delta>0$, uniformly over $P$, by
\[
\sup_P\Pr_P\!\left(
\min_{s\in\{0,1\}}B_{s,n}<M
\,\middle|\,Q_n=q
\right)
+\frac{4e^{-2M\delta^2}}{1-e^{-2\delta^2}}.
\]
Letting $n\to\infty$ and then $M\to\infty$, this bound vanishes under Assumption~\ref{asspt:uniform_replay_growth}.

On $E_{n,\varepsilon}$, the uniform calibration bounds both population stratum probabilities below by $\varepsilon-\kappa_n$. The population correction denominator is at least $\rho_\star-\kappa_n$, by the denominator calculation in the proof of Theorem~\ref{thm:asymptotic_validity}. The preceding uniform consistency bounds ensure that the corresponding subsampling denominators are also bounded away from zero with probability tending uniformly to one on this event. Thus, the ratio and calibration arguments in that proof give, for every $\delta>0$,
\[
\sup_P\Pr_P\!\left(
E_{n,\varepsilon}\cap
\left\{
\left|\widehat p^{\mathrm{sel}}_{q,n}
-H^P_{q,m}(u_n)\right|>\delta
\right\}
\,\middle|\,Q_n=q
\right)\to0.
\]

Next, we make the selected-cdf stability argument uniform. The increment bound in Lemma~\ref{lem:random_threshold_stability} becomes
\[
H^P_{q,k}(b)-H^P_{q,k}(a)
\le\frac{b-a+2\kappa_n}{\rho_\star},
\quad k\in\{n,m\},\quad 0\le a<b\le1.
\]
Thus, for a fixed finite grid $\mathcal T_h$ that covers $I_\varepsilon$ with a mesh size of at most $h$, the finite-grid argument of that lemma gives
\[
\sup_P\sup_{u\in I_\varepsilon}
[H^P_{q,n}(u)-H^P_{q,m}(u)]_+
\le
\max_{t\in\mathcal T_h}
\sup_P[\lambda^P_{q,n}(t)-\lambda^P_{q,m}(t)]_+
+\frac{h+4\kappa_n}{\rho_\star}.
\]
The maximum tends to zero by Assumption~\ref{asspt:uniform_lambda_stability}(a), since the grid is finite and independent of $P$. Letting $n\to\infty$ and then $h\downarrow0$ proves uniform one-sided cdf stability on $I_\varepsilon$. Under Assumption~\ref{asspt:uniform_lambda_stability}(b), the same argument gives the corresponding absolute-value bound.

Combining estimation and stability on $E_{n,\varepsilon}$, and using the endpoint bound from the proof of Theorem~\ref{thm:asymptotic_validity},
\[
\sup_P\Pr_P(E_{n,\varepsilon}^c\mid Q_n=q)
\le\frac{2\varepsilon+2\kappa_n}{\rho_\star},
\]
we obtain, by first letting $n\to\infty$ and then $\varepsilon\downarrow0$, that for every $\eta>0$,
\[
r_n(\eta):=
\sup_P\Pr_P\!\left(
V_n^P-\widehat p^{\mathrm{sel}}_{q,n}>\eta
\,\middle|\,Q_n=q
\right)\to0.
\]
Under two-sided stability, the same conclusion holds for
$r_n^{\mathrm{abs}}(\eta)$, defined by replacing the discrepancy inside this probability with its absolute value.

Finally, the oracle super-uniformity from Proposition~\ref{prop:selected_rank_general} gives
\[
\sup_P\sup_{\alpha\in[0,1]}
\left[
\Pr_P\!\left(
\widehat p^{\mathrm{sel}}_{q,n}\le\alpha
\,\middle|\,Q_n=q
\right)-\alpha
\right]_+
\le\eta+r_n(\eta).
\]
For exactness, the jump bound and randomized probability integral transform used in the proof of that proposition yield
\[
\sup_P\sup_{\alpha\in[0,1]}
\left|
\Pr_P(V_n^P\le\alpha\mid Q_n=q)-\alpha
\right|
\le\frac{2\kappa_n}{\rho_\star}.
\]
Under the two-sided stability assumption, comparison with the oracle at thresholds $\alpha-\eta$ and $\alpha+\eta$ therefore gives
\[
\sup_P\sup_{\alpha\in[0,1]}
\left|
\Pr_P\!\left(
\widehat p^{\mathrm{sel}}_{q,n}\le\alpha
\,\middle|\,Q_n=q
\right)-\alpha
\right|
\le
\eta+r_n^{\mathrm{abs}}(\eta)
+\frac{2\kappa_n}{\rho_\star}.
\]
Any shifted thresholds outside $[0,1]$ are handled by the trivial probability bounds zero and one. Letting $n\to\infty$ and then $\eta\downarrow0$ proves both conclusions.
\end{proof}

\begin{proof}[Proof of Proposition~\ref{prop:replay_power}]
For part (a), write
$V_n^0=H_{q,n}^{P_0}(U_{q,n})$.
For every $\eta>0$, the probability that the
feasible and oracle tests disagree satisfies
\[
\begin{aligned}
&\Pr_{P_0}\left(
\mathbf{1}\{\widehat p_{q,n}^{\mathrm{sel}}\le\alpha\}
\neq\mathbf{1}\{V_n^0\le\alpha\}
\mid Q_n=q\right)\\
&\quad\le
\Pr_{P_0}\left(
|\widehat p_{q,n}^{\mathrm{sel}}-V_n^0|>\eta
\mid Q_n=q\right)\\
&\qquad+
\Pr_{P_0}\left(
|V_n^0-\alpha|\le\eta
\mid Q_n=q\right).
\end{aligned}
\]
The first term converges to zero by the
two-sided conclusion of
Theorem~\ref{thm:asymptotic_validity}.
By Proposition~\ref{prop:selected_rank_general},
the limsup of the second term is at most $2\eta$.
Letting $\eta\downarrow0$ establishes that the
disagreement probability vanishes under $P_0$,
conditional on $Q_n=q$.

Let
\[
E_n=
\left\{
\mathbf{1}\{\widehat p_{q,n}^{\mathrm{sel}}\le\alpha\}
\neq
\mathbf{1}\{V_n^0\le\alpha\},
\ Q_n=q
\right\}.
\]
Define
\[
f_n(d):=\Pr(E_n\mid D_n=d),
\]
where the probability is taken over the original
agent and replay randomizations. Since these
randomizations have the same conditional law
given $D_n$ under $P_0$ and $P_n$, the function
$f_n$ is common to both experiments.

By the preceding argument,
\[
\E_{P_0^n}[f_n(D_n)]
=\Pr_{P_0}(E_n)\longrightarrow0,
\]
where $P_0^n$ is the sampling distribution of $D_n$ under the null law $P_0$.
Thus, for every $\varepsilon>0$, by Markov's inequality,
\[
P_0^n(f_n(D_n)>\varepsilon)
\longrightarrow0,
\]
and by contiguity,
\[
P_n^n(f_n(D_n)>\varepsilon)
\longrightarrow0.
\]

Since $0\le f_n\le1$, we have
\[
\begin{aligned}
\Pr_{P_n}(E_n)
&=\E_{P_n^n}[f_n(D_n)]\\
&=\E_{P_n^n}\left[
f_n(D_n)\mathbf{1}\{f_n(D_n)\le\varepsilon\}
\right]\\
&\quad+\E_{P_n^n}\left[
f_n(D_n)\mathbf{1}\{f_n(D_n)>\varepsilon\}
\right]\\
&\le\varepsilon+
P_n^n(f_n(D_n)>\varepsilon).
\end{aligned}
\]
By contiguity, the second term converges to zero.
Letting $n\to\infty$ and then
$\varepsilon\downarrow0$ gives
$\Pr_{P_n}(E_n)\to0$.
 Hence,
\[
\Pr_{P_n}(E_n\mid Q_n=q)
=
\frac{\Pr_{P_n}(E_n)}
{\Pr_{P_n}(Q_n=q)}
\longrightarrow0,
\]
since $\liminf_n\Pr_{P_n}(Q_n=q)>0$. This proves part (a).

For part (b), write $u_n=U_{q,n}$ and define
\[
R_n=\frac{u_n}{\widehat a_{0,q}(u_n)},
\]
with $R_n=+\infty$ whenever the denominator
is zero. Since
$0\le\widehat a_{0,q}(u_n)\le1$,
$R_n\stackrel{p}{\longrightarrow}0$ implies
\[
u_n\stackrel{p}{\longrightarrow}0,
\qquad
\Pr\left(\widehat a_{0,q}(u_n)=0
\mid Q_n=q\right)\longrightarrow0.
\]
Whenever both strata have a replay, $\widehat a_{0,q}(u_n)>0$ and
$u_n<1$, using
$0\le\widehat a_{1,q}(u_n)\le1$ gives
\[
\begin{aligned}
0\le\widehat p_{q,n}^{\mathrm{sel}}
&=
\frac{u_n\widehat a_{1,q}(u_n)}
{u_n\widehat a_{1,q}(u_n)
+(1-u_n)\widehat a_{0,q}(u_n)}\\
&\le
\frac{u_n}
{(1-u_n)\widehat a_{0,q}(u_n)}\\
&=\frac{R_n}{1-u_n}
\stackrel{p}{\longrightarrow}0.
\end{aligned}
\]
The complementary event has probability
tending to zero conditional on $Q_n=q$ by Assumption~\ref{asspt:replay_availability} and the assumption that $R_n \stackrel{p}{\longrightarrow} 0$.
That proves part (b).
\end{proof}

For the proofs of Lemma \ref{lem:exact_ci} and Theorem \ref{thm:replay_confidence}, we need the following definitions.

\label{app:ci_definition}
With the conventions $0\log0=0$ and $\log0=-\infty$, let the log-likelihood of the two independent binomials be,
\[
\ell(a;X)=X_1\log a_1+(B_1-X_1)\log(1-a_1)+X_0\log a_0+(B_0-X_0)\log(1-a_0),
\qquad
\widehat a=(X_1/B_1,\,X_0/B_0),
\]
and let $\partial_t=\{a\in[0,1]^2:g_{u_n}(a)=t\}$: for $t\in(0,1)$ the ray $a_1=r_ta_0$ with $r_t=t(1-u_n)/\{(1-t)u_n\}$, while $\partial_0=\{a_1=0,\,a_0>0\}$ and $\partial_1=\{a_0=0,\,a_1>0\}$. The signed root of the likelihood-ratio statistic for the constraint $g_{u_n}(a)=t$ is
\[
S_t(X)=\operatorname{sign}\!\big(g_{u_n}(\widehat a)-t\big)\sqrt{D_t(X)},
\qquad
D_t(X)=2\Big[\ell(\widehat a;X)-\sup_{a\in\partial_t}\ell(a;X)\Big]\in[0,\infty],
\]
with $g_{u_n}(\widehat a)=1$ when $X_0=0<X_1$ and $S_t(0,0)=0$ for every $t$. The inner supremum is a one-dimensional concave maximization along the ray. The validity argument below does not depend on this choice: any statistic $S_t$ yields valid bounds. Because $S_t$ varies with $t$, the acceptance sets in \eqref{eq:ci_bounds} need not be intervals, which is why the bounds are defined as their hull.

\begin{proof}[Proof of Lemma~\ref{lem:exact_ci}]
First, we confirm that the binomial counts $(X_1, X_0)$ are jointly binomial conditional on the data $D_n$. Condition on $D_n$, so $u_n$ is fixed, and write
$\mathcal L_s:=\mathcal L\big(D_{S_m}\mid D_n,\ \mathbf 1\{p_q(D_{S_m})\le u_n\}=s\big)$ for the law of a uniform size-$m$ subsample of $D_n$ conditional on falling in stratum $s$, so that $b_{s,n}=\Pr(\mathcal A_r(D,\Xi)=q)$ for $D\sim\mathcal L_s$ and a newly sampled $\Xi$. By Assumption~\ref{asspt:sampling_units} the subsampled datasets are i.i.d.\ draws from $\mathcal L(D_{S_m}\mid D_n)$, and their stratum labels are functions of them, so the subsamples that fall in stratum $s$, in the order in which they are drawn, are i.i.d.\ draws from $\mathcal L_s$, independently across strata. The replayed subsamples in stratum $s$ are the first $B_s$ of these, with $B_s$ fixed given $D_n$. Since each replay uses fresh agent randomness, the indicators $\mathbf 1\{\mathcal A_r(D,\Xi)=q\}$ returned by the matching step (Assumption~\ref{asspt:trust}(iii)) are i.i.d.\ Bernoulli$(b_{s,n})$ within stratum $s$ and independent across strata. Hence, conditional on $D_n$, $(X_1,X_0)\sim\Pr_b$ with $b=(b_{1,n},b_{0,n})$.

 Fix $a$ with $u_na_1+(1-u_n)a_0>0$ and let $t=g_{u_n}(a)$, so that $a\in\Theta_-(t)$. Let $q_a(X)=\Pr_a(S_t(X')\ge S_t(X))$. As the upper-tail probability of $S_t(X)$ under its own law, $q_a$ satisfies $\Pr_a(q_a(X)<c)\le c$ for every $c\in[0,1]$. On the event $\{a\in C(X)\}$ the point $a$ belongs to the set over which the supremum in $p_-(t;X)$ is taken, so $p_-(t;X)\ge\min\{1,\beta+q_a(X)\}$, and $p_-(t;X)<\gamma/2$ forces $q_a(X)<\gamma/2-\beta$. The two Clopper--Pearson intervals for $a_0$, $a_1$ have coverage of at least $1-\beta/2$ each and the two counts are independent, so $\Pr_a(a\notin C(X))\le1-(1-\beta/2)^2\le\beta$, and therefore
\[
\Pr_a(p_-(t;X)<\gamma/2)
\le\Pr_a(a\notin C(X))+\Pr_a(q_a(X)<\gamma/2-\beta)
\le\beta+(\gamma/2-\beta)=\gamma/2 .
\]
Since $\underline p_n$ is the infimum of the acceptance set, $\underline p_n>t$ implies $p_-(t;X)<\gamma/2$, which gives $\Pr_a(\underline p_n>g_{u_n}(a))\le\gamma/2$. The bound for $\overline p_n$ follows in the same way from $a\in\Theta_+(t)$ and the lower tail. This gives Eqs. \eqref{eq:finite-sample-validity}.

Finally, apply both bounds at $a=(b_{1,n}, b_{0,n})$, the true selection probabilities, for which $g_{u_n}(b)=w_n$, and combine them with a union bound to obtain the validity of the confidence band in \eqref{eq:finite-sample-coverage}.
\end{proof}

\begin{proof}[Proof of Theorem~\ref{thm:replay_confidence}]
\emph{Notation.} Write $u_n=U_{q,n}$, $m=m_n$, $V_n=H_{q,n}(u_n)$, $W_n=u_n\lambda_{q,n}(u_n)$, $\kappa_{q,k}=\sup_u|F_{q,k}(u)-u|$, $b=(b_{1,n},b_{0,n})$, $d_n=u_nb_{1,n}+(1-u_n)b_{0,n}$, and $\widehat a=(X_1/B_1,X_0/B_0)$. Recall that $g_u\left(a_1, a_0\right)=\frac{u a_1}{u a_1+(1-u) a_0}$, $w_n:=g_{u_n}\left(b_{1, n}, b_{0, n}\right)$ as defined in the main text. Fix $\varepsilon\in(0,1/2)$ and let $E_{n,\varepsilon}=\{\varepsilon\le u_n\le1-\varepsilon\}$. For a fresh dataset of size $m$, let $a_{1,q,m}(u)=\Pr_P(Q_m=q\mid U_{q,m}\le u)$ and $a_{0,q,m}(u)=\Pr_P(Q_m=q\mid U_{q,m}>u)$ be the population stratum selection probabilities, and $R_{q,m}(u)=g_u(a_{1,q,m}(u),a_{0,q,m}(u))$ the resulting population correction. The subsample quantities $b_{s,n}$ are the same conditional probabilities computed under the subsampling law given $D_n$. All limits are conditional on $Q_n=q$. From the proof of Theorem~\ref{thm:asymptotic_validity} we use three facts: (F1) $\Pr_P(E_{n,\varepsilon}^c\mid Q_n=q)\le(2\varepsilon+2\kappa_{q,n})/\rho_\star$; (F2) the denominator of $R_{q,m}(u)$ is at least $\rho_\star-\kappa_{q,m}$ and $\sup_u|R_{q,m}(u)-H_{q,m}(u)|\le2\kappa_{q,m}/\rho_\star$; (F3) by Lemma~\ref{lem:uniform_subsampling_cdf}, the joint conditional subsampling distribution of $(p_q(D_{S_m}),\mathbf 1\{\mathcal A_r(D_{S_m},\Xi')=q\})$ given $D_n$ converges uniformly in probability to its counterpart for a new dataset of size $m$, $D_m$ sampled from the dataset law $P$.

\emph{Step 1: $w_n$ approximates the targets.}
Decompose
\begin{equation}\label{eq:ci_three_terms}
V_n-w_n
=\underbrace{\{H_{q,n}(u_n)-H_{q,m}(u_n)\}}_{\text{stability}}
+\underbrace{\{H_{q,m}(u_n)-R_{q,m}(u_n)\}}_{\text{calibration}}
+\underbrace{\{R_{q,m}(u_n)-w_n\}}_{\text{subsampling}} .
\end{equation}
First, the calibration term is at most $2\kappa_{q,m}/\rho_\star\to0$ by (F2). Second, for the subsampling term, Assumption~\ref{asspt:base_calibration} gives $\min\{F_{q,m}(u_n),1-F_{q,m}(u_n)\}\ge\varepsilon-\kappa_{q,m}$ on $E_{n,\varepsilon}$, so the conditional probabilities defining $a_{s,q,m}(u_n)$, and by (F3) those defining $b_{s,n}$, have denominators bounded away from zero with probability tending to one, and (F3) yields $\mathbf 1_{E_{n,\varepsilon}}|b_{s,n}-a_{s,q,m}(u_n)|\stackrel{p}{\longrightarrow}0$ for $s=0,1$. Since the denominator of $g_{u_n}$ at $(a_{1,q,m}(u_n),a_{0,q,m}(u_n))$ is at least $\rho_\star-\kappa_{q,m}$ by (F2), it follows that $d_n\ge\rho_\star/2$ on $E_{n,\varepsilon}$ with probability tending to one, and continuity of $g_u$ away from a vanishing denominator gives $\mathbf 1_{E_{n,\varepsilon}}|R_{q,m}(u_n)-w_n|\stackrel{p}{\longrightarrow}0$. Third, the stability term is controlled by Lemma~\ref{lem:random_threshold_stability}: uniformly on $[\varepsilon,1-\varepsilon]$, its positive part vanishes under Assumption~\ref{asspt:lambda_stability}(a) and its absolute value under (b). Putting the three together, we obtain,
\[
\mathbf 1_{E_{n,\varepsilon}}(V_n-w_n)_+\stackrel{p}{\longrightarrow}0
\ \text{under (a)},
\qquad
\mathbf 1_{E_{n,\varepsilon}}|V_n-w_n|\stackrel{p}{\longrightarrow}0
\ \text{under (b)}.
\]
Since $V_n=F_{q,n}(u_n)\lambda_{q,n}(u_n)$ and $\lambda_{q,n}\le1/\rho_\star$, we have $|V_n-W_n|\le\kappa_{q,n}/\rho_\star\to0$, so both statements hold with $V_n$ replaced by $t_+=\max\{V_n,W_n\}$, and the second also with $t_-=\min\{V_n,W_n\}$. Finally, Step~2 below requires $t_+\le1$. We know that $V_n\le1$ and $V_n$ is asymptotically uniform by Proposition~\ref{prop:selected_rank_general}, so $\Pr_P(t_+>1\mid Q_n=q)\le\Pr_P(V_n>1-\kappa_{q,n}/\rho_\star\mid Q_n=q)\to0$.

\emph{Step 2: local coverage.}
Lemma~\ref{lem:exact_ci} covers $w_n$, whereas we need coverage of $t_\pm$, which differ from $w_n$ by an $o_p(1)$ term. We therefore bound the probability that an endpoint of the interval falls on the wrong side of a point $t$. Fix $t\in[0,1]$, suppose $d_n>0$, and consider the pair with overall reselection probability $d_n$ and tail share $t$,
\[
x=\frac{d_nt}{u_n},\qquad y=\frac{d_n(1-t)}{1-u_n},
\]
so that $u_nx+(1-u_n)y=d_n$ and $g_{u_n}(x,y)=u_nx/d_n=t$. This pair need not lie in $[0,1]^2$, so set $c=\max\{1,x,y\}$ and $b^\dagger=(x,y)/c$. Then $b^\dagger\in[0,1]^2$, and $g_{u_n}(b^\dagger)=t$ because $g_{u_n}$ is invariant to rescaling its argument. Because $b_{1,n}=d_nw_n/u_n$ and $b_{0,n}=d_n(1-w_n)/(1-u_n)$ with $d_n\le1$, on $E_{n,\varepsilon}$ we have $|x-b_{1,n}|\vee|y-b_{0,n}|\le|t-w_n|/\varepsilon$, and since $b_{s,n}\le1$ the rescaling by $c$ moves each coordinate by at most $c-1\le\max\{x-b_{1,n},y-b_{0,n}\}_+\le|t-w_n|/\varepsilon$. Hence, $\|b^\dagger-b\|_\infty\le2|t-w_n|/\varepsilon$. Generate the $B_s$ Bernoulli variables of stratum $s$ as $\mathbf 1\{U_i\le b_{s,n}\}$ and $\mathbf 1\{U_i\le b^\dagger_s\}$ from the same independent uniform random variables $U_i$, so that the first set of counts has law $\Pr_b$ and the second $\Pr_{b^\dagger}$. Then, the two count vectors differ only if some $U_i$ falls between the two thresholds, an event of probability at most $(B_1+B_0)\|b^\dagger-b\|_\infty$. By the coupling inequality $d_{\mathrm{TV}}(\Pr_b,\Pr_{b^\dagger})$ is bounded from above by this probability. Applying the one-sided validity in Lemma~\ref{lem:exact_ci} at $a=b^\dagger$ and transferring to $\Pr_b$ through this total variation bound, we obtain, on $E_{n,\varepsilon}$ and conditional on $D_n$,
\begin{equation}\label{eq:ci_nearby}
\Pr_b(\overline p_n<t)\ \vee\ \Pr_b(\underline p_n>t)
\le\frac{\gamma}{2}+\min\Big\{1,\frac{2(B_1+B_0)|t-w_n|}{\varepsilon}\Big\}.
\end{equation}

\emph{Step 3: (a) and (b).}
Both upper bounds in (a) hold exactly when $\overline p_n\ge t_+$, since $W_n\le\overline p_n$ is equivalent to $\lambda_{q,n}(u_n)\le\overline p_n/u_n$ by definition of $W_n, V_n$. Given $D_n$, both $t_+$ and $w_n$ are fixed. If $t_+\le w_n$, then $\{\overline p_n<t_+\}\subseteq\{\overline p_n<w_n\}$, whose conditional probability is at most $\gamma/2$ by Lemma~\ref{lem:exact_ci}. If $t_+>w_n$, apply \eqref{eq:ci_nearby} at $t=t_+$. In either case, on $E_{n,\varepsilon}\cap\{t_+\le1\}$,
\[
\Pr(\overline p_n<t_+\mid D_n)
\le\frac{\gamma}{2}+\min\Big\{1,\frac{2(B_1+B_0)(t_+-w_n)_+}{\varepsilon}\Big\}.
\]
Taking expectations given $Q_n=q$, the second term is bounded and converges to zero in probability by Step 1, so its expectation vanishes, while the complementary event $E_{n,\varepsilon}^c\cup\{t_+>1\}$ has conditional probability at most $(2\varepsilon+2\kappa_{q,n})/\rho_\star+o(1)$ by (F1) and Step 1. Letting $n\to\infty$ and then $\varepsilon\downarrow0$ proves (a). For (b), the same argument at $t=t_-$ bounds $\Pr(\underline p_n>t_-)$ using $|t_--w_n|\stackrel{p}{\longrightarrow}0$, and a union bound gives joint coverage $1-\gamma$.

\emph{Step 4: part (c).}
We show that, when the interval can be computed and $\gamma\le1/2$,
\begin{equation}\label{eq:ci_containment}
\inf g_{u_n}(C(X))\ \le\ \underline p_n\ \le\ g_{u_n}(\widehat a)\ \le\ \overline p_n\ \le\ \sup g_{u_n}(C(X)),
\end{equation}
so that the interval is contained in the image of the Clopper--Pearson rectangle. Then, we show that this image shrinks as $n$ and $B_1, B_0 \to \infty$. For the outer inequalities, if $t<\inf g_{u_n}(C(X))$ then $C(X)\cap\Theta_-(t)=\emptyset$, so $p_-(t;X)=\beta<\gamma/2$ and $t$ is rejected. An analogous argument establishes the upper bound. For the inner inequalities, let $t^*=g_{u_n}(\widehat a)$ be the corrected $p$-value at the observed frequencies, and consider the ordering statistic $S_t$ at $t=t^*$. Since $\widehat a\in\partial_{t^*}$ by construction and $\widehat a$ maximizes $\ell(\cdot;X)$ without constraint, the constrained and unconstrained maxima in $D_{t^*}(X)$ coincide, so $S_{t^*}(X)=0$. Then, note that $\widehat a\in C(X)\cap\Theta_-(t^*)$ because Clopper--Pearson intervals contain the sample proportion. Since $g_{u_n}$ is increasing in $a_1$ and decreasing in $a_0$, $\{X_1'\ge X_1,\,X_0'\le X_0\}\subseteq\{S_{t^*}(X')\ge0\}$, and under $\Pr_{\widehat a}$ each coordinate is binomial with integer mean $X_s$, which is then also its median \citep{kaas1980mean}, so each of the two events has probability at least $1/2$. Hence $p_-(t^*;X)\ge\min\{1,\beta+1/4\}\ge\gamma/2$, so $t^*$ is accepted and $\underline p_n\le t^*$; symmetrically $\overline p_n\ge t^*$.

Turning to the width of $g_{u_n}(C(X))$, write $B_{\min}=\min\{B_1,B_0\}$ and $z_B=\sqrt{\log(4/\beta)/(2B_{\min})}$. The Clopper--Pearson interval at level $1-\beta/2$ is contained in the Hoeffding interval at the same level, so every point of $C(X)$ lies within sup-norm distance $z_B$ of $\widehat a$, and by Hoeffding's inequality $\|\widehat a-b\|_\infty\le\delta$ except with conditional probability $4e^{-2B_{\min}\delta^2}$. Since the map $a\mapsto u_na_1+(1-u_n)a_0$ is $1$-Lipschitz in sup norm, on $E_{n,\varepsilon}\cap\{d_n\ge\rho_\star/2\}$ and whenever $z_B+\delta\le\rho_\star/4$, the rectangle $C(X)$ is contained in the convex set $\{a:u_na_1+(1-u_n)a_0\ge\rho_\star/4\}$, on which both partial derivatives of $g_{u_n}$ are bounded by $4/\rho_\star$. Thus $g_{u_n}$ and $g_{u_n}/u_n$ are Lipschitz on $C(X)$ in sup norm with constants $8/\rho_\star$ and $8/(\rho_\star\varepsilon)$, and by \eqref{eq:ci_containment} the two intervals have diameters at most $16z_B/\rho_\star$ and $16z_B/(\rho_\star\varepsilon)$, both of order $B_{\min}^{-1/2}$.

It remains to control the events excluded above. The interval is unavailable only if a stratum has probability zero under the subsampling law given $D_n$, so that it never fills, or if $X_1=X_0=0$. The former has probability vanishing as $n\to\infty$, because, as shown in the proof of Theorem~\ref{thm:asymptotic_validity}, both stratum probabilities are at least $\varepsilon/2$ on $E_{n,\varepsilon}$ with probability tending to one; the latter has conditional probability $(1-b_{1,n})^{B_1}(1-b_{0,n})^{B_0}\le e^{-B_{\min}(b_{1,n}+b_{0,n})}\le e^{-B_{\min}d_n}$, which vanishes as $B_{\min}\to\infty$ on $\{d_n\ge\rho_\star/2\}$. The event $E_{n,\varepsilon}\cap\{d_n<\rho_\star/2\}$ has vanishing probability by Step 1, and $E_{n,\varepsilon}^c$ has probability at most $(2\varepsilon+2\kappa_{q,n})/\rho_\star$ by (F1). Fix $\varepsilon'>0$ and choose $\delta$ with $\delta\le\rho_\star/8$. For $B_{\min}$ large enough that $z_B\le\rho_\star/8$ and $16z_B/(\rho_\star\varepsilon)\le\varepsilon'$, the diameter of either interval exceeds $\varepsilon'$ only on the union of the excluded events, so
\[
\limsup_{n\to\infty}\Pr_P\!\left(\max\{\operatorname{diam}\mathcal I^p_{n,B},\operatorname{diam}\mathcal I^\lambda_{n,B}\}>\varepsilon'\mid Q_n=q\right)
\le\frac{2\varepsilon}{\rho_\star}+4e^{-2B_{\min}\delta^2}+e^{-B_{\min}\rho_\star/2}.
\]
Letting $B_{\min}\to\infty$ and then $\varepsilon\downarrow0$ proves \eqref{eq:replay_ci_precision}.
\end{proof}

\section{Additional results}\label{app:additional_results}

\subsection{Definitions} \label{app:definitions}

This section defines the terms from the AI literature that the paper uses and relates them to the objects of Section~\ref{sec:setup}. The definitions are meant to give a concise understanding of the concepts involved and are by no means exhaustive. See \citet{korinek2025agents} for an introduction to AI agents for economic research.

\paragraph{Large language model.} A large language model (LLM) is a neural network, typically a transformer \citep{vaswani2017attention}, trained on large amounts of text to predict the next \emph{token} (a word or word fragment) given the preceding ones \citep{brown2020language}. It generates text by drawing a token from this predicted distribution of tokens, appending it to its input, and repeating. The input is called the \emph{context}, and the part of the context that contains the user's instructions is called the \emph{prompt}. Input and output are text (character strings) in a broad sense that includes prose, code, and shell commands. Many current models first generate intermediate \emph{reasoning} tokens that are fed back into the model with the aim of improving the accuracy of the final answer, called a chain of thought \citep{wei2022chain}. The user can specify how much of this reasoning the models are required to perform (the ``reasoning effort''; \citealp{openai2026reasoning}). Two properties of these models are important for this paper: first, because tokens are drawn at random, the same context (input) can produce different outputs; second, the model's parameters do not change during use, so the model does not retain any memory from one conversation to the next (not counting renewed ingestion of external inputs as ``memory'').

\paragraph{Agent.} In the classical definition in the field of AI, an agent is anything that can perceive its environment and intervene in it \citep[Chapter~2]{russell2021artificial}. An LLM becomes an agent through \emph{tool calling}: the model emits text in a prescribed format that names an action (a ``tool'', e.g. running a shell command) and its arguments, a computer program then executes the prescribed action in the embedded computing environment and appends the result to the context, and the expanded context is fed back to the model so that it receives environmental feedback from its action \citep{yao2023react}. Current usage of the term ``agent'' largely reserves it for systems in which the model itself decides which actions to take and in what order, as opposed to workflows in which a programmer fixes the sequence of model calls in advance, which are encompassed by the broader classical definition \citep{anthropic2024agents}. A common short definition is that an LLM agent ``runs tools in a loop to achieve a goal'' \citep{willison2025agents}.

\paragraph{Harness.} A program that executes an agentic loop is called a \emph{harness}. In one common formulation, the harness denotes every part of an agentic system other than the model itself \citep{trivedy2026harness}. It assembles the context, including a \emph{system prompt} with standing instructions and descriptions of the available tools, sends it to the model, and executes the tool calls that the model requests, subject to the permissions granted by the user. The model is usually hosted by a provider on a server in a data center and queried through an API over the internet, while the harness runs in a separate compute environment. Claude Code \citep{anthropic2025claudecode} and the Codex CLI, used in this paper, each combine a harness with a command-line interface.

\paragraph{Coding agent.} A coding agent is an LLM agent whose available tools let it read and write files and execute programs on a computer \citep{lucas2023openinterpreter, yang2024sweagent}. Statistical software can be run programmatically, usually through the command line, so such an agent can load a dataset, write code to estimate a statistical model, execute that code, read the output, and revise its code in response. We call the folder that the agent can read and modify its \emph{workspace}, one execution of the loop from the initial prompt to the agent's final answer a \emph{run} or session, and the record of all messages, tool calls, and results emitted by a coding agent and its program executions in a run its \emph{trace} or log.

\paragraph{Relation to the formal setup.} In the notation of Section~\ref{sec:setup}, $\mathcal A_r$ comprises the base model, the harness, the tools, and their settings, such as the reasoning effort; $r$ collects the instructions (the prompt) and the files other than the data that are available to the agent at the start of a run; and $\Xi$ collects intrinsic model randomness (due to the token draws and other nondeterminism in LLM inference) and any additional randomness in the tool results. Because the model does not retain any context across runs, a new run that starts from the same $r$ in a fresh copy of the workspace, with the model and harness held fixed, applies the same mapping with a fresh draw of $\Xi$, which is the sense in which a coding agent can be reset and replayed. Finally, some harnesses provide a way for the model to store notes that persist across runs. Since these are stored outside the model and enter a run by being reinjected into the model's context, holding them fixed is part of holding $r$ fixed.

\subsection{Simulation details}\label{app:simulation_details}

\paragraph{Open-ended design.} Each dataset has $n = 2{,}000$ participants, generated as follows, with all error terms drawn independently. An individual is drawn from site $S \in \{1, 2, 3, 4\}$ with the respective probabilities $(0.30, 0.25, 0.25, 0.20)$ and labour-market strengths $\ell(S) \in \{0.25, 0, -0.10, -0.30\}$. Age $A = \mathrm{round}(\mathcal{N}(37, 11^2))$ clipped to $[18, 64]$; female $F \sim \mathrm{Bernoulli}(0.56)$; parent $P \sim \mathrm{Bernoulli}(\Lambda(-0.6 + 0.6F + 0.03(A - 35)))$ with $\Lambda$ the logistic function; education $E \in \{1, 2, 3, 4\}$ with probabilities $(0.16, 0.40, 0.30, 0.14)$; household size $H = 1 + \mathrm{Poisson}(0.8) + P\,(1 + \mathrm{Poisson}(0.7))$ clipped to $[1, 8]$. The two latent traits are
\begin{align*}
C &= 0.35\,(E - 2.4) - 0.015\,|A - 40| + \ell(S) + \varepsilon, & \varepsilon &\sim \mathcal{N}(0, 1), \\
W &= 0.4\,C + \eta, & \eta &\sim \mathcal{N}(0, 1).
\end{align*}
Baseline employment is $D_0 = \mathbf{1}\{C + \mathcal{N}(0, 0.9^2) > -0.1\}$, baseline earnings $Y_0 = D_0\,\mathrm{round}_{10}\bigl(\exp(7.7 + 0.25\,C + \mathcal{N}(0, 0.45^2))\bigr)$, with $\mathrm{round}_{10}$ rounding to ten dollars, and months unemployed $M_0 = (1 - D_0)\,\mathrm{round}(1 + \mathrm{Exponential}(8))$ clipped to $[1, 60]$. Treatment $T$ equals one for exactly half of each site's participants, chosen uniformly at random, and does not appear in any other equation. The traits at the follow-ups are $C_6 = C + \mathcal{N}(0, 0.5^2)$, $C_{12} = 0.7\,C_6 + 0.3\,C + \mathcal{N}(0, 0.5^2)$, $W_6 = W + \mathcal{N}(0, 0.6^2)$ and $W_{12} = 0.7\,W_6 + 0.3\,W + \mathcal{N}(0, 0.6^2)$. For $t \in \{6, 12\}$, writing $e = C_t$ and $w = W_t$, the outcomes are
\begin{align*}
\text{employed}_t &= \mathbf{1}\{e + 0.8\,D_0 - 0.4 + \mathcal{N}(0, 0.8^2) > 0\}, \\
\text{hours}_t &= \text{employed}_t \cdot \mathrm{round}(\mathcal{N}(31 + 3e, 8^2)) \text{ clipped to } [4, 60], \\
\text{earnings}_t &= \text{employed}_t \cdot \mathrm{round}_{10}\bigl(\exp(7.6 + 0.3e + 0.25\log(1 + Y_0)/8 + \mathcal{N}(0, 0.5^2))\bigr), \\
\text{financial security}_t &= \mathrm{round}(10 + 3e + 1.5w + \mathcal{N}(0, 3^2)) \text{ clipped to } [0, 20], \\
\text{food insecure}_t &= \mathbf{1}\{-(0.7e + 0.3w) + 0.25\,(H - 2.5) + \mathcal{N}(0, 1) > 0.9\}, \\
\text{debt in collections}_t &= \mathbf{1}\{-0.6e + \mathcal{N}(0, 1) > 0.8\}, \\
\text{distress}_t &= \mathrm{round}(8 - 2.5w - 0.8e + \mathcal{N}(0, 3.5^2)) \text{ clipped to } [0, 24], \\
\text{health}_t &= \mathrm{round}(3.2 + 0.45w + 0.15e - 0.02\,(A - 37) + \mathcal{N}(0, 0.8^2)) \text{ clipped to } [1, 5], \\
\text{training}_t &= \mathbf{1}\{-0.6\,\text{employed}_t + 0.2\,(E - 2.4) - 0.01\,(A - 37) + \mathcal{N}(0, 1) > 0.9\}.
\end{align*}
 Since $T$ does not enter into any equation, every treatment effect is exactly zero. For the power results, the equation for $\text{employed}_{12}$ gains the term $\tau T$ inside the indicator, with $\tau=0.28$ on 100 datasets and $\tau=0.43$ on the other 100. All other equations are left unchanged.

\subsection{Ideological bias and immigration}\label{app:immigration}

This Appendix section describes our replication of the results in \citet{miao2026agentic} in some more detail and reports the corresponding results on the faulty data file included in their replication package.

\paragraph{Replication.} The design involves rerunning the coding agents on the full datasets, as opposed to on subsamples as in our method, and changing the agents' prompts to induce diverse perspectives in their analysis. Any given analysis reported by one single agent is then tested against the resulting reference distribution of analyses for how extreme it is, similar to a two-sided p-value. As discussed in Section~\ref{sec:comparison}, the authors call this an ``m-value'', and suggest that it captures ``the probability that an analysis path would produce a claim at least as extreme as the reported one''. Table \ref{tab:miao_personas} reports the results, for the AI model used by the authors (Claude Sonnet 4.6, through the Claude Agent SDK) as well as a more recent version of the same  model (Claude Sonnet 5.5) and GPT-6-Luna (through the Codex CLI). As in the original paper, we ran 20 pro- and 20 anti-immigration agents on the same data, and subjected the agents' outputs to a three-stage methodological review to filter out incorrect analyses. Any remaining small differences from the authors' results are likely due to LLM stochasticity and the fact that we used a more recent model as the methodology judge (GPT-6.1-Sol instead of GPT-5.4).

\begin{table}[!htbp]
\centering
\small
\caption{Immigration: persona design of \citet{miao2026agentic}}
\floatfoot{\textit{Notes:} table replicates the findings of \citet{miao2026agentic} on three coding agents, Claude Sonnet 4.6 (used in the original paper) and two more recent models, Claude Sonnet 5.5 and GPT-6-Luna. \textit{Passed} indicates how many out of the 40 agents' analyses passed methodological review by GPT-6.1-Sol; \textit{Anti/Pro} indicate how many of the estimates of immigration's effect on welfare support were significant in the direction of the agent's assigned persona (negative for anti-immigration agent, positive for pro) with \textit{Share} computing the overall share together with a Clopper-Pearson CI; \textit{Mean $z$} gives the average $z$-statistic (reported estimate / SE) for both types of agents and the difference $\Delta$ between the average anti- and pro- $z$-statistics. }
\label{tab:miao_personas}
\setlength{\tabcolsep}{3.5pt}
\begin{tabular}{lrrrcrrc}
\toprule
 & & \multicolumn{3}{c}{Significant in the persona\textquoteright s direction} & \multicolumn{3}{c}{Mean $z$} \\
\cmidrule(lr){3-5}\cmidrule(lr){6-8}
Model & Passed & Anti & Pro & Share [95\% CI] & Anti & Pro & $\Delta$ [95\% CI] \\
\midrule
Claude Sonnet 4.6 & 37/40 & 13/18 & 9/19 & 59\% [42, 75] & $-2.6$ & $+1.9$ & $-4.5$ [$-5.3$, $-3.7$] \\
Claude Sonnet 5.5 & 40/40 & 0/20 & 0/20 & 0\% [0, 9] & $+0.4$ & $0.0$ & $+0.4$ [$+0.1$, $+0.8$] \\
GPT-6-Luna & 35/40 & 0/20 & 0/15 & 0\% [0, 10] & $-0.3$ & $-0.5$ & $+0.2$ [$0.0$, $+0.5$] \\
\bottomrule
\end{tabular}

\end{table}

\paragraph{Design.} We refer to \citet{miao2026agentic} and the accompanying \href{https://github.com/jmiao24/AgenticBootstrap}{replication package} for the details on their design, and just highlight some more details on how we deviated from it here. The Claude runs use the harness settings of \citet{miao2026agentic} in the Claude Agent SDK, which use the default high effort setting. Since the SDK runs Claude Code CLI under the hood, for the GPT-6-Luna runs, we use the analogous Codex CLI at high reasoning effort. We pass the persona prompts as the system prompt in both harnesses, replacing the default system prompt. Following the paper, each final result produced by an agent undergoes an ``inference repair'' where a separate, independent reviewer agent (same model as the original) reviews and fixes any methodological errors it can find in the code. Finally, each analysis undergoes a methodological review of a structured summary of the repaired specification, run three times with the review prompt provided in the replication package. A result is kept if at least two of the three verdicts pass it, and its $z$-statistic is recovered from the repaired two-sided $p$-value. As noted above, summary and review use GPT-6.1-Sol whereas theirs use GPT-5.4. Since the authors did not publish their summary agent's prompt, we construct it from their example summaries.

\paragraph{Data.} We noticed several data mismatches in the file distributed by \citet{miao2026agentic}, with respondents being assigned the wrong country and outcome columns holding survey questions, likely due to a faulty data merge. We rebuilt the file ourselves from the ISSP archive releases and the replication package of \citet{breznau2022observing} and use the corrected file for our analyses. Our results with Claude Sonnet 4.6 are close to the authors' on both files and, if anything, closer on the corrected file (Tables~\ref{tab:miao_personas} and~\ref{tab:miao_personas_original}: a gap in mean $z$ of $-4.5$, against $-3.8$ on their file and $-4.2$ in their paper), so we suspect the faulty merge arose in the replication package rather than in the analysis behind the paper. Table \ref{tab:issp_variables} lists all the variables in the data that were given to the agent and hence available to choose for its analysis. For full transparency, Table \ref{tab:miao_personas_original} reports our replication results on the faulty data file, which still gives highly similar results.

\begin{table}[!htbp]
\centering
\footnotesize
\caption{Immigration: variables in the data given to the agent}
\floatfoot{\textit{Notes:} One row per respondent in the ISSP Role of Government modules of 1985, 1990, 1996, 2006 and 2016: 154,921 respondents in 46 countries and 107 country-waves.  Missing values are empty; no survey weights are included.}
\label{tab:issp_variables}
\setlength{\tabcolsep}{4pt}
\begin{tabular}{@{}p{0.26\linewidth}p{0.52\linewidth}p{0.16\linewidth}@{}}
\toprule
Variable & Description & Coding \\
\midrule
\multicolumn{3}{@{}p{0.96\linewidth}}{\textit{Outcomes: ``should it be the government\textquoteright s responsibility to \ldots''; 1 definitely should be, 2/3 probably should be, 1/3 probably should not be, 0 definitely should not be}} \\
\raggedright \texttt{jobs} & \raggedright provide a job for everyone who wants one & 0--1 \\
\raggedright \texttt{healthcare} & \raggedright provide health care for the sick & 0--1 \\
\raggedright \texttt{old\_age} & \raggedright provide a decent standard of living for the old & 0--1 \\
\raggedright \texttt{unemployed} & \raggedright provide a decent standard of living for the unemployed & 0--1 \\
\raggedright \texttt{income\_diff} & \raggedright reduce income differences between the rich and the poor & 0--1 \\
\raggedright \texttt{housing} & \raggedright provide decent housing for those who can\textquoteright t afford it (not asked in 1985) & 0--1 \\
\midrule
\multicolumn{3}{@{}p{0.96\linewidth}}{\textit{Immigration (country, module year)}} \\
\raggedright \texttt{migstock} & \raggedright Immigrant stock: World Bank where available, otherwise UN & \% of population \\
\raggedright \texttt{migstock\_wb}, \texttt{migstock\_un}, \texttt{migstock\_oecd} & \raggedright Immigrant stock from the World Bank (1985--2006), the UN (1990--2016) and the OECD (members only) & \% of population \\
\raggedright \texttt{mignet\_un} & \raggedright Net migration rate, UN (none for 2016) & per 1,000 per year \\
\raggedright \texttt{migflow\_pct} & \raggedright The same rate in percent & \% per year \\
\midrule
\multicolumn{3}{@{}p{0.96\linewidth}}{\textit{Respondent}} \\
\raggedright \texttt{age} & \raggedright Age & years \\
\raggedright \texttt{female} & \raggedright Female & 0/1 \\
\raggedright \texttt{employed} & \raggedright In paid work & 0/1 \\
\raggedright \texttt{educ} & \raggedright Highest completed education, harmonized (1996--2016) & 0--5 \\
\raggedright \texttt{topbot} & \raggedright Self-placement from the bottom to the top of society (2006--2016) & 1--10 \\
\midrule
\multicolumn{3}{@{}p{0.96\linewidth}}{\textit{Country controls (module year)}} \\
\raggedright \texttt{gdp\_wb}, \texttt{log\_gdp} & \raggedright GDP per capita and its logarithm & current US\$ \\
\raggedright \texttt{pop\_wb} & \raggedright Population & persons \\
\raggedright \texttt{ginid\_solt} & \raggedright Gini index of disposable household income (SWIID) & 0--1 \\
\raggedright \texttt{gini\_wb} & \raggedright Gini index (World Bank; sparse) & 0--100 \\
\raggedright \texttt{wdi\_unempilo} & \raggedright Unemployment rate (ILO; from 1991) & \% of labor force \\
\raggedright \texttt{socx\_oecd} & \raggedright Public social expenditure (OECD members) & \% of GDP \\
\raggedright \texttt{al\_ethnic} & \raggedright Ethnic fractionalization & 0--1 \\
\midrule
\multicolumn{3}{@{}p{0.96\linewidth}}{\textit{Identifiers}} \\
\raggedright \texttt{wave}, \texttt{year} & \raggedright Survey module: 1985, 1990, 1996, 2006, 2016 &  \\
\raggedright \texttt{iso\_country}, \texttt{cntry\_wave} & \raggedright Country (ISO code) and country-wave &  \\
\bottomrule
\end{tabular}

\end{table}

\begin{table}[!htbp]
\centering
\small
\caption{Immigration: persona design of \citet{miao2026agentic} on their distributed data file}
\floatfoot{\textit{Notes:} results equivalent to Table \ref{tab:miao_personas}, but estimated on the faulty data file provided with the authors' replication package. On their file, the newer models were run as a check before we rebuilt the data: 10 runs per persona for Claude Sonnet 5.5 and 4 for GPT-6-Luna. The last row reports the runs plotted in Supplementary Figure~3c of \citet{miao2026agentic}, with $z$-statistics read from the figure's vector graphics.}
\label{tab:miao_personas_original}
\setlength{\tabcolsep}{3.5pt}
\begin{tabular}{lrrrcrrc}
\toprule
 & & \multicolumn{3}{c}{Significant in the persona\textquoteright s direction} & \multicolumn{3}{c}{Mean $z$} \\
\cmidrule(lr){3-5}\cmidrule(lr){6-8}
Model & Passed & Anti & Pro & Share [95\% CI] & Anti & Pro & $\Delta$ [95\% CI] \\
\midrule
Claude Sonnet 4.6 & 35/40 & 12/17 & 8/18 & 57\% [39, 74] & $-2.6$ & $+1.2$ & $-3.8$ [$-5.1$, $-2.7$] \\
Claude Sonnet 5.5 & 20/20 & 0/10 & 0/10 & 0\% [0, 17] & $-1.0$ & $-1.1$ & $+0.1$ [$-0.2$, $+0.5$] \\
GPT-6-Luna & 8/8 & 0/4 & 0/4 & 0\% [0, 37] & $-0.1$ & $-0.7$ & $+0.6$ [$-0.1$, $+0.9$] \\
\midrule
\citet{miao2026agentic} & 34/40 & 12/17 & 9/17 & 62\% [44, 78] & $-2.3$ & $+1.9$ & $-4.2$ [$-5.5$, $-2.9$] \\
\bottomrule
\end{tabular}

\end{table}

\paragraph{Reference distribution for the m-value.} The m-value of the application's original uses 60 full-data GPT-6-Luna sessions with the original's harness and data: 20 with the original request, 20 with the team described as pushing for lower immigration to protect a strong safety net, and 20 with a nonpartisan team. Their logs hold 811 estimated analyses, nearly all with standard errors clustered by country. The m-value pools them with equal weight, as in the calculator of \citet{miao2026agentic}; weighting sessions equally gives 0.25.

\subsubsection{Known selection rules on a fixed menu}\label{app:immigration_all}

As a complement to the fixed simulation design in Section \ref{sec:simulation_designs}, we also considered a design for the immigration application where the agents were told to implement a known selection rule on the data. This allows us to compare our estimated $p$-values to the oracle $p$-values, which are known exactly because we fixed the selection rule.  We gave GPT-6-Luna (high effort) a menu of 108 OLS regressions of country-level support for the six welfare-related outcomes in the ISSP data on the foreign-born share, varying the item, the source of the immigrant-stock measure, the countries covered and the controls, with HC3 standard errors. An anti- and a pro-immigration analyst both gave the agent one of three instructions: report a specification named in advance whatever its result (``fixed model''); report the specification with the most negative/positive $t$-statistic (``frontier''); or report the named specification if it is negative/positive and significant, and otherwise report the most negative/positive alternative (``client pressure'').

The named specification for the anti-immigration analyst regresses support for government responsibility for the unemployed on a measure of the share of immigrants in the country without controls. It is the most negative specification on the full dataset ($p = 0.026$), so all three workflows report it as the ``original'' analysis, differing only in what they report on the subsampled datasets. Then, we replayed each workflow on 100-200 subsampled datasets per stratum (tail vs. body) and computed the oracle corrected $p$-value by applying the mechanical selection rules to an enlarged bank of $20{,}000$ subsampled datasets. Every replay agent followed the rule it was assigned exactly.

Table~\ref{tab:immigration_all} shows that the corrected $p$-values computed from the agent replays are very close to their oracle counterparts. As expected, the rule that instructed the agent to always pick the same analysis has an estimated multiplier of exactly $1$, while for the anti-immigration analyst, the other rules estimate $\lambda$s larger than $1$, and more strongly so for the most selective rule, which discards the original analysis whenever it is not significant.

Interestingly, we find no evidence of strong selection under these rules for the pro-immigration analyst, which appears to be driven by the fact that it fails to find significant results in its original analysis to begin with ($p=0.755$ and $0.919$).

\begin{table}[!htbp]
\centering
\small
\caption{Immigration: known selection rules on a fixed menu}
\floatfoot{\textit{Notes:} \textit{Tail} and \textit{Body}: replays selecting the original analysis. The\textit{ multiplier} and \textit{corrected $p$} are agentic replay estimates, one-sided 95\% bound in parentheses; the \textit{exact $p$} applies each rule programmatically (not using coding agents) to all 20,000 subsamples.}
\label{tab:immigration_all}
\setlength{\tabcolsep}{4pt}
\begin{tabular}{@{}lcccccc@{}}
\toprule
Workflow & Nominal $p$ & Tail & Body & Multiplier & Corrected $p$ & Exact $p$ \\
\midrule
\multicolumn{7}{@{}l}{\emph{Analyst expects a negative association}} \\
Fixed model & 0.026 & 100/100 & 100/100 & 1.00 ($\geq$ 0.97) & 0.026 ($\leq$ 0.026) & 0.026 \\
Frontier & 0.026 & 32/199 & 23/200 & 1.38 ($\geq$ 0.91) & 0.035 ($\leq$ 0.054) & 0.038 \\
Client pressure & 0.026 & 198/200 & 9/200 & 14.31 ($\geq$ 9.81) & 0.366 ($\leq$ 0.524) & 0.350 \\
\midrule
\multicolumn{7}{@{}l}{\emph{Analyst expects a positive association}} \\
Fixed model & 0.755 & 100/100 & 100/100 & 1.00 ($\geq$ 0.99) & 0.755 ($\leq$ 0.760) & 0.755 \\
Frontier & 0.755 & 20/200 & 25/200 & 0.94 ($\geq$ 0.80) & 0.711 ($\leq$ 0.800) & 0.758 \\
Client pressure & 0.919 & 7/199 & 8/200 & 0.99 ($\geq$ 0.87) & 0.909 ($\leq$ 0.961) & 0.887 \\
\bottomrule
\end{tabular}

\end{table}

\subsection{Card--Krueger Application} \label{app:card_krueger}

\begin{table}[!htbp]
\centering
\footnotesize
\caption{Card--Krueger: variables in the data given to the agent}
\floatfoot{\textit{Notes:} one row in the data is one restaurant, for 410 restaurants total surveyed in February and March 1992, before New Jersey's minimum wage rose from \$4.25 to \$5.05 on April 1, and again in November and December 1992. Employee counts and closure codes are given as recorded, without recoding; missing entries are NA.}
\label{tab:card_variables}
\setlength{\tabcolsep}{4pt}
\begin{tabular}{@{}p{0.26\linewidth}p{0.52\linewidth}p{0.16\linewidth}@{}}
\toprule
Variable & Description & Coding \\
\midrule
\multicolumn{3}{@{}p{0.96\linewidth}}{\textit{Store}} \\
\raggedright \texttt{store\_id}, \texttt{sheet} & \raggedright Store identifier (state and survey sheet number) &  \\
\raggedright \texttt{chain} & \raggedright Burger King, KFC, Roy Rogers, Wendy\textquoteright s & 1--4 \\
\raggedright \texttt{co\_owned} & \raggedright Company-owned & 0/1 \\
\raggedright \texttt{state} & \raggedright New Jersey (1) or Pennsylvania (0) & 0/1 \\
\raggedright \texttt{southj}, \texttt{centralj}, \texttt{northj}, \texttt{shore} & \raggedright Region of New Jersey; shore location & 0/1 \\
\raggedright \texttt{pa1}, \texttt{pa2} & \raggedright Region of Pennsylvania & 0/1 \\
\midrule
\multicolumn{3}{@{}p{0.96\linewidth}}{\textit{Employment (wave 1 / wave 2)}} \\
\raggedright \texttt{empft / empft2} & \raggedright Full-time employees, excluding managers & count \\
\raggedright \texttt{emppt / emppt2} & \raggedright Part-time employees, excluding managers & count \\
\raggedright \texttt{nmgrs / nmgrs2} & \raggedright Managers and assistant managers & count \\
\midrule
\multicolumn{3}{@{}p{0.96\linewidth}}{\textit{Wages and benefits (wave 1 / wave 2)}} \\
\raggedright \texttt{wage\_st / wage\_st2} & \raggedright Starting wage & \$ per hour \\
\raggedright \texttt{inctime / inctime2} & \raggedright Months to the usual first raise & months \\
\raggedright \texttt{firstinc / firstin2} & \raggedright Usual first raise & \$ per hour \\
\raggedright \texttt{bonus / special2} & \raggedright Cash bounty (wave 1) or special program (wave 2) for new workers & 0/1 \\
\raggedright \texttt{pctaff} & \raggedright Share of employees affected by the new minimum wage (wave 1) & \% \\
\raggedright \texttt{meals / meals2} & \raggedright Meal benefit: none, free, reduced-price, both & 0--3 \\
\midrule
\multicolumn{3}{@{}p{0.96\linewidth}}{\textit{Prices and operations (wave 1 / wave 2)}} \\
\raggedright \texttt{psoda / psoda2} & \raggedright Price of a medium soda, including tax & \$ \\
\raggedright \texttt{pfry / pfry2} & \raggedright Price of small fries, including tax & \$ \\
\raggedright \texttt{pentree / pentree2} & \raggedright Price of an entree, including tax & \$ \\
\raggedright \texttt{open / open2r} & \raggedright Opening hour & hour \\
\raggedright \texttt{hrsopen / hrsopen2} & \raggedright Hours open per day & hours \\
\raggedright \texttt{nregs / nregs2} & \raggedright Cash registers & count \\
\raggedright \texttt{nregs11 / nregs112} & \raggedright Cash registers open at 11 am & count \\
\midrule
\multicolumn{3}{@{}p{0.96\linewidth}}{\textit{Survey (wave 1 / wave 2)}} \\
\raggedright \texttt{ncalls / ncalls2} & \raggedright Call-backs before contact & count \\
\raggedright \texttt{type2} & \raggedright Wave-2 interview by phone (1) or in person (2) & 1--2 \\
\raggedright \texttt{status2} & \raggedright Wave-2 status: refused, answered, or closed (renovation, permanently, highway construction, mall fire) & 0--5 \\
\raggedright \texttt{date2} & \raggedright Wave-2 interview date & MMDDYY \\
\bottomrule
\end{tabular}

\end{table}

\begin{table}[!htbp]
\centering
\small
\caption{Card--Krueger: methodological review of replay agents' analyses}
\floatfoot{\textit{Notes}: table shows percentage of Card--Krueger replay agents' analyses that pass methodological review by a stronger model. \emph{Reviewer} column: percentage of replays passed by GPT-6.1-Sol on high effort after it read the replay's declared model and null hypothesis and its final \texttt{analysis.R} file. Comments in code, estimates, and $p$-values were removed to not bias the review. The review agent was asked to judge if the analysis contains a major methodological error that would invalidate it as a reliable answer to the research question. \emph{Reviewer + audit}: percentage of replays passed after a separate audit agent from a different model family (Claude Opus 5.5, high effort) was given the same inputs as well as the reviewer agent's verdict and asked to audit the reviews.}
\label{tab:card_replay_review}
\begin{tabular}{@{}lrrr@{}}
\toprule
& & \multicolumn{2}{c}{Pass (\%)} \\
\cmidrule(l){3-4}
 & Replays & Reviewer & Reviewer + audit \\
\midrule
All replays & 2,983 & 98.5 & 99.5 \\
Tail datasets & 1,132 & 98.7 & 99.2 \\
Body datasets & 1,851 & 98.3 & 99.7 \\
\bottomrule
\end{tabular}

\end{table}

\begin{figure}[!h]
\centering
\includegraphics[width=\linewidth]{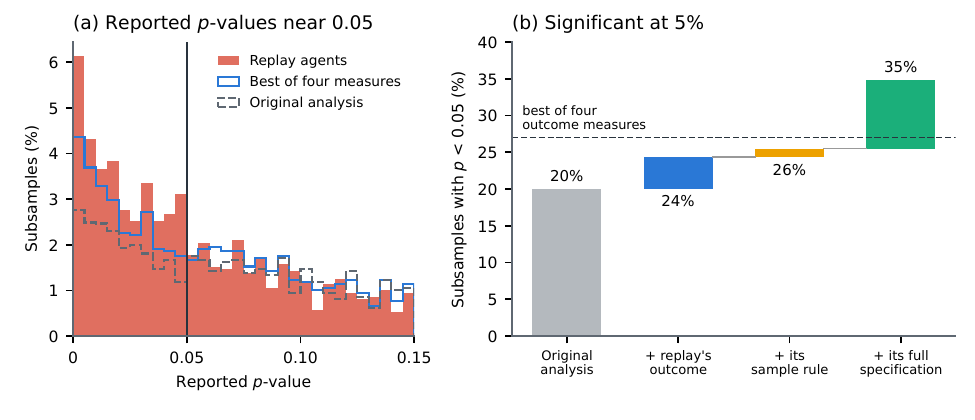}
\caption{Card--Krueger: agentic p-hacking}
\floatfoot{\textit{Note}: (a) histogram of p-values reported by replay agents (red bars), p-values produced by original analysis on the subsampled data (dashed line), and p-values produced by picking the employment outcome measure that makes the original's analysis most significant on the subsampled data (solid line).
Based on 2,983 classified replays. (b) Share of analyses significant (y-axis) on a replay agent's subsampled data when starting from running the original analysis (chosen by the agent on the full data: OLS on the store-level change with HC2 standard errors) on the subsampled data and progressively introducing choices made by the replay agent for its analysis on the subsampled data. Dashed line: share significant when using original analysis with the most significant of the four employment outcome measures.}
\label{fig:card_replay_significance}
\end{figure}

\begin{table}[!h]
\centering
\small
\caption{Card--Krueger: headline results with stronger coding agents}
\floatfoot{\textit{Note}: table compares headline results in Card--Krueger application (estimated $\lambda$ and corrected p-value) using GPT-6-Luna (high effort) against results using two stronger models from the same family (GPT-6.1-Terra and GPT-6.1-Sol, high effort) and a stronger model from a different family (Claude Sonnet 5.5). Setup for all agents is identical, and replays for all agents draw their subsamples from the same pool, with tail and body counts set by each cohort's own pilot. A separate reviewing agent (GPT-6.1-Sol, high reasoning effort) judged the declared analysis matches in the three GPT cohorts. The same agent judged the matches in both Sonnet cohorts. The masked-data row reruns Sonnet, original and replays, on the same stores with neutral variable names, the states, chains and survey details described generically, the dates moved to 2019 and dollar amounts rescaled. By a keyword rule, Sonnet declines the prompt's request to lead with the most statistically compelling finding in its visible reasoning or messages in 97.1\% of replays (masked data: 97.9\%), against 3 of GPT-6-Luna's 3,000 and none of GPT-6.1-Sol's, and names Card and Krueger somewhere in the session in 18.7\% (masked data: 10.6\%), against 1 of GPT-6-Luna's and none of GPT-6.1-Sol's. Parentheses give one-sided 95\% bounds: a lower bound for $\lambda$ and an upper bound for the corrected $p$-value.}
\label{tab:card_terra}
\begin{tabular}{@{}lccccc@{}}
\toprule
 & Original & Tail & Body & Multiplier & Corrected \\
Agent & $p$-value & matches & matches & $\lambda$ & $p$-value \\
\midrule
GPT-6-Luna & 0.025 & 143/1,135 (12.6\%) & 133/1,850 (7.2\%) & 1.72 ($\geq$ 1.43) & 0.043 ($\leq$ 0.052) \\
GPT-6.1-Terra & 0.028 & 251/1,073 (23.4\%) & 243/1,909 (12.7\%) & 1.79 ($\geq$ 1.57) & 0.051 ($\leq$ 0.058) \\
GPT-6.1-Sol & 0.027 & 270/1,493 (18.1\%) & 133/1,493 (8.9\%) & 1.98 ($\geq$ 1.68) & 0.053 ($\leq$ 0.062) \\
Claude Sonnet 5.5 & 0.041 & 1,348/1,494 (90.2\%) & 1,351/1,494 (90.4\%) & 1.00 ($\geq$ 0.98) & 0.041 ($\leq$ 0.042) \\
\quad masked data & 0.040 & 1,155/1,487 (77.7\%) & 1,151/1,490 (77.2\%) & 1.01 ($\geq$ 0.97) & 0.041 ($\leq$ 0.042) \\
\bottomrule
\end{tabular}

\end{table}

\begin{table}[!htbp]
\centering
\caption{Agents' self-correction}
\floatfoot{\textit{Notes:} table summarizes runs on the 100 null datasets of the open-ended design (Panel B) where a given coding agent autonomously stated a correction for its main estimate's $p$-value; one report per agent and dataset; usable reports (no errors or refusals) 99, 100, 100 and 97. \textit{Leads with $p\le0.05$}: number of agents' reports whose lead finding has $p\le0.05$; all other columns in \% of these leads. \textit{Disclaimer in the opening}: blind read of the report's first 800 characters with statistics masked by GPT-6.1-Sol (high effort). \textit{None}: no concrete reason given to doubt the finding (calling it exploratory or a hypothesis does not count); \textit{Warning}: a concrete reason provided for why the finding needs to be interpreted with caution; \textit{Disclaimer}: the finding is said to fail, e.g.\ under a correction or as consistent with chance. \textit{Own multiplicity correction}: columns give statistics on the multiplicity corrections the agents reported autonomously. \textit{Tests stated} and \textit{Applied}: second blind read of the full report by GPT-6.1-Sol (high effort). Tests stated counts how many tests the agent says it ran and is correcting for. Applied counts whether the agent actually applied any Bonferroni, Holm, permutation or other multiplicity correction to the lead finding. \textit{Given in the opening}: corrected $p$-value and the correction's name in the first 800 characters, by text search.}
\label{tab:self_correction_stance}
\footnotesize
\begin{tabular}{@{}lrrrrrrr@{}}
\toprule
 & & \multicolumn{3}{c}{Disclaimer in the opening (\%)} & \multicolumn{3}{c}{Own multiplicity correction} \\
\cmidrule(lr){3-5}\cmidrule(l){6-8}
Agent & \begin{tabular}[b]{@{}c@{}}Leads with\\$p\le0.05$\end{tabular} & None & Warning & Disclaimer & \begin{tabular}[b]{@{}c@{}}Tests stated\\(median)\end{tabular} & \begin{tabular}[b]{@{}c@{}}Applied\\(\% of leads)\end{tabular} & \begin{tabular}[b]{@{}c@{}}Given in the\\opening (\% of leads)\end{tabular} \\
\midrule
GPT-6-Luna & 97 & 85 & 14 & 1 & 90 & 58 & 7 \\
GPT-6.1-Sol & 88 & 70 & 11 & 18 & 312 & 100 & 25 \\
Claude Sonnet 5.5 & 74 & 31 & 4 & 65 & 342 & 100 & 7 \\
Claude Opus 5.5 & 52 & 35 & 12 & 54 & 378 & 96 & 10 \\
\bottomrule
\end{tabular}

\end{table}

\begin{figure}[!htbp]
\centering
\includegraphics[width=.78\linewidth]{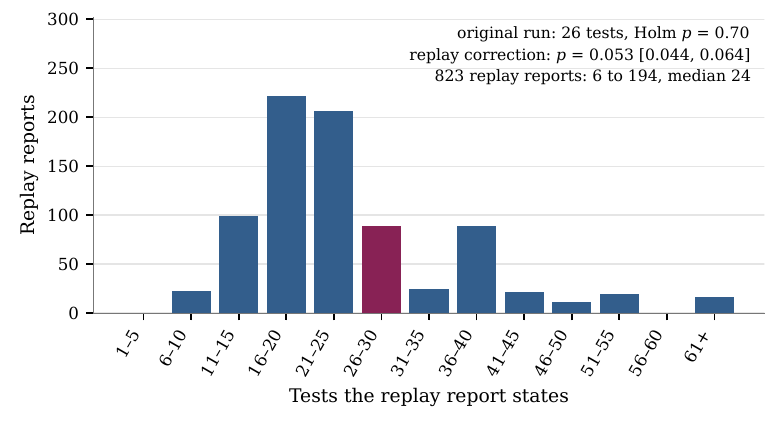}
\caption{The family an agent self-corrects over is a draw from its procedure}
\floatfoot{\textit{Notes:} Number of tests stated in GPT-6.1-Sol's replay reports on 150-store subsamples of the Card--Krueger data, classified by GPT-6.1-Sol (high effort) through blind reads of 831 reports, 823 of which state a count. Shaded bin: contains 26, the count stated by the original run on all 410 stores. Holm $p$: the original's correction of its selected analysis (nominal $p=0.027$) over those 26 tests. Replay correction: 2,986 classified replays, with 95\% replay interval. Not shown: 200 full-size placebo runs (state labels permuted across all 410 stores, same prompt and agent) state 12 to 1,252 tests; quartiles 24 and 36, median 30.}
\label{fig:self_correction_family}
\end{figure}

\begin{figure}[!htbp]
\centering
\includegraphics[width=.8\linewidth]{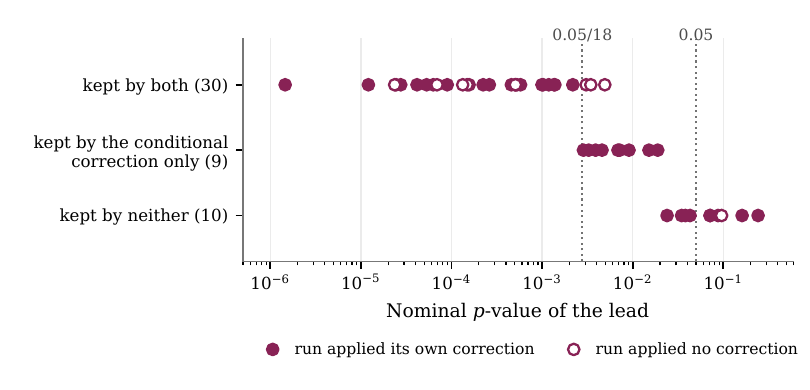}
\caption{Agents' self-corrections lose power compared to the conditional correction}
\floatfoot{\textit{Notes:} 49 datasets of the open-ended simulation design with a moderate true effect on which GPT-6.1-Sol, prompted for one compelling finding without being directed to search, reported the same analysis: employment at 12 months, all participants, site controls. One marker per dataset, at that analysis's nominal $p$-value; rows by verdict at the 5\% level under the agent's own correction as written and under the conditional correction. No dataset is kept by the own correction only. Filled: agent estimated a Holm or Bonferroni correction over a stated family of 18 (40 runs); hollow: no correction, nominal $p$ as written (9 runs). Dotted lines at $0.05/18$ and $0.05$. Conditional correction: rank of the nominal $p$-value among the 51 of 100 fresh null runs that reported the same analysis.}
\label{fig:self_correction_verdicts}
\end{figure}

\FloatBarrier

\section{Implementation Details}

\FloatBarrier
\subsection{Prompts}\label{app:prompts}
\newcommand{\promptbox}[1]{\par\nopagebreak\smallskip\noindent\includegraphics[page=#1,width=\linewidth]{proposals/prompt-displays/prompts-outlined.pdf}\par\smallskip}
We reproduce the instructions that defined each analysis task. The original prompts also asked the agents to save their analysis as a program that can be rerun on another dataset, and contained instructions on file names, output formats and logging that we added for our own record keeping; we omit these and mark omissions with [\ldots]. Every agent ran in a workspace containing the data and supporting documentation such as a codebook.

\paragraph{Open-ended simulation design.} Prompt:
\promptbox{1}
The codebook describes the nine outcomes and states:
\promptbox{2}

\paragraph{Power design.} The power design uses the research prompt and codebook above unchanged.

\paragraph{Real data designs.} The research prompt is the one above without the words ``of the Pathways to Work employment-support program'', so that one prompt serves all three studies:
\promptbox{18}
Each study has its own codebook. It opens with a description of the study, reproduced below, then defines the design and baseline variables and the outcomes, and ends with the analysis convention above, with block indicators in place of site indicators. Every codebook describes the randomization as our placebo carries it out, and none mentions that the treatment was reassigned. The wellness and teacher codebooks describe the real intervention. The consumer survey randomly assigned respondents to the topics of its questions but had no treatment of this kind, so its codebook describes an invented information module, as the simulation's codebook describes an invented program; its description of the sample, the eight randomly assigned consumption topics and the four social topics follows the real survey.

\emph{Wellness:}
\promptbox{19}

\emph{Teachers:}
\promptbox{20}

\emph{Consumers:}
\promptbox{21}

\paragraph{Open-ended immigration analyst.} Prompt:
\promptbox{13}
The two other versions differ only in the second sentence.

\emph{Anti-immigration version:} ``Our team works with groups that push for lower immigration to protect a strong safety net, and the argument we keep making is that immigration erodes public support for social programs.''

\emph{Nonpartisan version:} ``Our team is nonpartisan and works with groups on all sides of the immigration debate, and an argument we keep hearing is that immigration erodes public support for social programs.''

\paragraph{Card--Krueger application.} Prompt:
\promptbox{9}

\paragraph{Commitment prompt.} After its analysis, every original agent and replay received the following prompt in the same conversation. In the open-ended simulation design, a longer version also requested the numerical results; only the model and null description entered the comparison.
\promptbox{11}

\paragraph{Comparison prompt.} Each replay then received the following prompt, again in the same conversation, with the original analysis's commitment in place of [original commitment].
\promptbox{12}
The omitted passages clarify further details about the setting and how to determine a match, e.g. that differently measured outcomes remain different even when they play symmetric roles, that identical restrictions may be paraphrased in a variety of ways, that realized sample sizes and estimates are irrelevant, and that the agent may use its own earlier statements about its final analysis to resolve any abbreviations in its commitment.

In the Card--Krueger application, the reported verdicts come from a separate reviewing agent. It received the two declarations, the codebook and a short description of the study design, and was instructed to compare the declared models, nulls, outcomes, populations and sample restrictions, to ignore differences in estimators, inference procedures, code and presentation, and to mark a comparison unresolved rather than infer missing details from code or numerical results. Its prompt is given below, with the study description, the codebook and the declarations in place of the bracketed placeholders.
\promptbox{22}

\end{document}